\documentclass[twocolumn]{aastex63}
\usepackage[utf8]{inputenc}
\usepackage{amsmath}
\usepackage{natbib}
\usepackage{bm}
\usepackage{multirow}

\newcommand{\p}{\partial}

\newcommand{\ii}{\mathrm{i}}

\newcommand{\dd}{\delta}
\newcommand{\real}[1]{\operatorname{Re}\left(#1\right)}

\newcommand{\Hdust}{H_\mathrm{d}}
\newcommand{\Hgas}{H_\mathrm{g}}

\newcommand{\hgas}{h_\mathrm{g}}

\newcommand{\rhod}{\rho_\mathrm{d}}
\newcommand{\rhog}{\rho_\mathrm{g}}

\newcommand{\taus}{\tau_\mathrm{s}}
\newcommand{\st}{\mathrm{St}}

\newcommand{\fdust}{f_\mathrm{d}}
\newcommand{\fgas}{f_\mathrm{g}}

\newcommand{\Hd}{H_\mathrm{d}}

\newcommand{\re}{\operatorname{Re}}

\newcommand{\OmK}{\Omega_{\rm K}}
\newcommand{\OmKref}{\Omega_0}
\newcommand{\etatot}{\eta_\mathrm{tot}}
\newcommand{\etatotzero}{\eta_\mathrm{tot,0}}
\newcommand{\etatilde}{\widetilde{\eta}}
\newcommand{\hgref}{h_\mathrm{g0}}
\newcommand{\rhogref}{\rho_\mathrm{g0}}
\newcommand{\alphass}{\alpha_\mathrm{SS}}

\shorttitle{Magnetized streaming instabilities}
\shortauthors{M.-K. Lin and C.-Y. Hsu}

\begin{document}

\title{Streaming instabilities in accreting and magnetized laminar protoplanetary disks} 

\correspondingauthor{Min-Kai Lin}
\email{mklin@asiaa.sinica.edu.tw}

\author[0000-0002-8597-4386]{Min-Kai Lin}
\altaffiliation{Both authors contributed equally to this work.}
\affiliation{Institute of Astronomy and Astrophysics, Academia Sinica, Taipei 10617, Taiwan}
\affiliation{Physics Division, National Center for Theoretical Sciences, Taipei 10617, Taiwan}

\author{Chun-Yen Hsu}
\altaffiliation{Both authors contributed equally to this work.}
\affiliation{Institute of Astronomy and Astrophysics, Academia Sinica, Taipei 10617, Taiwan}

\begin{abstract}
The streaming instability is one of the most promising pathways to the formation of planetesimals from pebbles. Understanding how this instability operates under realistic conditions expected in protoplanetary disks is therefore crucial to assess the efficiency of planet formation. Contemporary models of protoplanetary disks show that magnetic fields are key to driving gas accretion through large-scale, laminar magnetic stresses. However, the effect of such magnetic fields on the streaming instability has not been examined in detail. To this end, we study the stability of dusty, magnetized gas in a protoplanetary disk. We find the streaming instability can be enhanced by passive magnetic torques and even persist in the absence of a global radial pressure gradient{. In this case,} instability is attributed to the azimuthal drift between dust and gas, unlike the classical streaming instability{, which is driven by radial drift.}  This suggests that the streaming instability can remain effective inside dust-trapping pressure bumps in accreting disks. When a live vertical field is considered, we find the magneto-rotational instability can be damped by dust feedback, while the classic streaming instability can be stabilized by magnetic perturbations. We also find that Alfv\'{e}n waves can be destabilized by dust-gas drift, but this instability requires nearly ideal conditions. We discuss the possible implications of these results for dust dynamics and planetesimal formation in protoplanetary disks. 
\end{abstract}

\section{Introduction}\label{intro}

The formation of 1--100 km-sized planetesimals is a key stage in the core accretion theory of planet formation \citep{chiang10,johansen14,birnstiel16} but is still not well understood.  Specifically, the growth of solids from micron-sized grains through sticking is limited to mm to cm-sized pebbles, beyond which collisions result in bouncing or fragmentation  \citep{blum08,blum18}. Furthermore, gas drag can lead to a rapid inwards drift of solids \citep{whipple72, weiden77}, which introduces a radial drift barrier 
\citep{birnstiel10,birnstiel12}. 

One way to circumvent these growth barriers is the collective, self-gravitational collapse of a particle swarm directly into planetesimals \citep{goldreich73,youdin02}. To do so, the particle swarm must attain a dust-to-gas mass density ratio $\epsilon$ well above unity \citep{shi13}, which should be compared to the typical value of $\epsilon\sim 0.01$ in the interstellar medium and is a reasonable expectation in protoplanetary disks \citep[PPDs,][]{testi14}, at least initially. Thus, some other mechanism is needed to enhance the local dust-to-gas ratio. 

To this end, the streaming instability \citep[SI,][]{youdin05,youdin07,johansen07,bai10,bai10c,kowalik13,yang14,carrera15,yang17,schreiber18,li18,flock21,li21} is a leading candidate for raising the dust-to-gas ratio in PPDs to the point of gravitational collapse  \citep{johansen09,simon16,simon17,schafer17,li19,abod19}. 

The SI is a linear instability in rotating flows of dust and gas when the two components interact through mutual drag \citep{jacquet11,lin17,pan20a,pan20b,pan21,jaupart20,squire20} and is powered by their relative radial drift, which itself is usually driven by a global radial pressure gradient that offsets the gas from Keplerian rotation. More recently, the SI was shown to be a member of a broader class of `Resonant Drag Instabilities' in dusty-gas \citep[RDI,][]{squire18a,squire18b,zhuravlev19}, where instability arises from a resonance between neutral waves in the gas and the relative motion between dust and gas. 

There have been several extensions to the classic SI of \cite{youdin05}, or more generally that of dust-gas  interaction, including the effect of external turbulence \citep{chen20,umurhan20,gole20,schafer20}, multiple grain sizes \citep{schaffer18,schaffer21,llambay19,krapp19,zhu20,paardekooper20,mcnally21}, pressure bumps  \citep{taki16,onishi17,auffinger18,carrera21a,carrera21b}, and disk stratification \citep{ishitsu09,lin21}. These efforts are necessary to determine the efficiency of the SI in realistic PPDs. 

On the other hand, the influence of a magnetic field on the SI has been less well-explored, but PPDs are expected to be magnetized, albeit subject to non-ideal magneto-hydrodynamic (MHD) effects \citep{lesur20}. Early studies focused on the impact of MHD turbulence sustained by the magneto-rotational instability \citep[MRI,][]{balbus91}
on dust dynamics, including their vertical settling and radial diffusion and migration \citep{johansen05,fromang05,fromang06,johansen06}. Subsequent simulations do find dust clumping in MHD-turbulent disks, which has been attributed to the SI \citep{johansen07b,balsara09,tilley10}, weak radial diffusion in resistive disks \citep{yang18}, or concentration by zonal flows or pressure bumps induced by the MRI  \citep{johansen09b,johansen11,dittrich13,xu21}. The latter two effects may facilitate the SI as it requires $\epsilon\gtrsim 1$ for dynamical growth, although pressure bumps present a dilemma as the classic SI does not formally operate without a radial pressure gradient.

In addition to driving small-scale (perhaps weak) turbulence, magnetic fields are also expected to regulate the large-scale gas dynamics of PPDs. Modern numerical simulations often find magnetized disk winds that drive laminar accretion  \citep{gressel15,bethune17,bai17,wang19,gressel20}. These windy PPDs are the new paradigm for planet formation and evolution. In some of these models, namely those including the Hall effect with the field and disk rotation being aligned, large-scale horizontal magnetic stresses can develop in the disk midplane, which leads to radial gas accretion. Such inflows can strongly modify the orbital migration of protoplanets compared to conventional, alpha-type viscous disk models \citep{mcnally17,kimmig20}. 

A natural question is how do planets form in these laminar, accreting PPDs in the first place? Related to this issue is the formation of rings or pressure bumps commonly seen in such simulations  \citep{bethune16,suriano17,suriano18,suriano19,riols19,cui21b}, which can act as dust traps \citep{krapp18,riols20} and may be sites of preferential planetesimal formation, for example through the SI.  These pressure bumps may also explain dust rings observed in bright PPDs  \citep[e.g.][]{andrews18,long18}. 

In this work, we examine the basic properties of dusty-gas dynamics in a magnetized PPD. Motivated by the aforementioned studies, we are particularly interested in the effect of a magnetically-driven accretion flow on the SI, as well as how the MRI and SI interact, and their implications for planetesimal formation. 
We find that a laminar gas accretion flow can enhance the SI and { lead to} instability even without radial pressure gradients, which suggests that pressure bumps remain feasible sites for the SI. We also find dust-loading reduces MRI growth rates, while magnetic perturbations { can be}  effective in stabilizing the classic SI. Finally, we { demonstrate} an RDI unique to dusty and magnetized gas, in which Alfv\'{e}n waves are rendered unstable by dust-gas drift, although it may have limited relevance to realistic PPDs.  

This paper is organized as follows. In \S\ref{basic_eqs} we list the basic equations for a dusty, magnetized PPD and specify the physical setups under consideration. We describe the linear problem in \S\ref{Theoretical_study} { and give an overview of analytic results, some of which are developed in the appendices. We present numerical results in \S\ref{results}}. In \S\ref{nonlinear} we verify the main findings from our linear stability analyses with direct integration of the dusty MHD equations. We discuss implications of our results to dust dynamics in PPDs in \S\ref{discussion} before summarizing in \S\ref{summary}.

\section{Basic equations} \label{basic_eqs}
We consider a three-dimensional (3D) PPD comprised of gas and a single species of uncharged dust grains in orbit around a central star of mass $M_*$. Cylindrical coordinates $(R,\phi, z)$ are centered on the star. The gas component has density, pressure, and velocity fields ($\rhog, P, \bm{V}$), respectively, and is threaded { by} a magnetic field $\bm{B}$. We consider small grains (defined below) and approximate the dust population as a pressureless fluid with density and velocity fields $(\rhod,\bm{W})$, respectively. The frictional drag between dust and gas is characterized by the stopping time $\taus$. 

We make several simplifications for tractable analyses. We consider a strictly isothermal gas so that $P=C_s^2\rhog$, where $C_s = \Hgas\OmK$ is a constant sound-speed. Here $\Hgas$ is the pressure scale height,  $\OmK\equiv\sqrt{GM_*/R^3}$ is the Keplerian frequency, and $G$ is the gravitational constant. { We neglect gas viscosity and dust diffusion \citep{dubruelle95} except in some calculations for regularization and comparison purposes.} As a proxy for non-ideal MHD effects, we consider Ohmic resistivity with a constant diffusion coefficient $\eta_O$. For the dust, we assume a constant Stokes number $\st \equiv \taus \OmK$. Then the fluid approximation applies to 
small grains with $\st\ll 1$ \citep{jacquet11}{, which we assume throughout.} We focus on dynamics close to the disk midplane ($z=0$) and thus neglect the vertical component of stellar gravity, so our models are unstratified ($\p_z = 0$ in equilibrium). We also { impose} axisymmetry { from the outset} ($\p_\phi \equiv 0$). 

Under these approximations, the governing equations for the magnetized gas are:
\begin{align}
&\frac{\partial \rho_g }{\partial t} + \nabla \cdot (\rho_g { V}) = {\rm 0}, \label{gas_mass_global}\\
&\frac{\partial { V} }{\partial t} + { V} \cdot \nabla { V}  = 
- R\OmK^2\hat{\bm{R}} - \frac{1}{\rho_g} \nabla P \notag \\ 
&\phantom{\frac{\partial { V} }{\partial t} + { V} \cdot \nabla { V}  =}
+\frac{1}{\rho_g\mu_0}\left(\nabla\times\bm{B}\right)\times\bm{B} +  \frac{\epsilon}{\taus}({ W}-{ V}) , \label{gas_mom_global}\\
&\frac{\partial { B} }{\partial t}  =  \nabla \times \left( { V} \times { B } - \eta_{O} \nabla\times\bm{B}\right),  \label{Faraday_global}
\end{align}
together with $\nabla\cdot\bm{B} = 0$, $\mu_0$ is the magnetic permeability, and $\epsilon \equiv \rhod/\rhog$ is the dust-to-gas ratio. 
The dust equations are 
\begin{align}
&\frac{\partial \rho_d }{\partial t} + \nabla \cdot (\rho_d { W}) = 0, \label{dust_mass_global}\\
&\frac{\partial { W} }{\partial t} + { W} \cdot \nabla { W}  = - R\OmK^2\hat{\bm{R}} - \frac{1}{\taus}({ W}-{ V}).  \label{dust_mom_global}
\end{align}

\subsection{Steady state drift with a magnetic field}
We consider axisymmetric{ ($\p_\phi=0$), unstratified ($\p_z=0$)} basic states with vanishing vertical velocities ($V_z = W_z = 0$). { The vertical component of the magnetic field $B_z$ is taken to be a constant, which does not affect the equilibrium solutions below.} The solenoidal condition implies $B_R\propto R^{-1}$. Hence we write 
\begin{align}
    B_R(R) = B_{R0}\left(\frac{R_0}{R}\right),\label{eqm_br}
\end{align}
where $R_0$ is a reference radius and $B_{R0}$ is a constant. For a thin disk we expect $\bm{V}\simeq R\OmK\hat{\bm{\phi}}$. Inserting this into the induction equation, we find  
\begin{align}
    B_\phi(R) = -B_{\phi0}\sqrt{\frac{R_0}{R}} \quad\text{with } B_{\phi0}\equiv \frac{2R_0^2\OmKref B_{R0}}{\eta_O},\label{eqm_bphi}
\end{align}
where $\OmKref = \OmK(R_0)$, see also \cite{mcnally17}.

We next seek approximate solutions to the horizontal velocity fields given the above field configuration. We write $\bm{V} = R\OmK\hat{\bm{\phi}} + \bm{V}^\prime$ and assume the disk is nearly Keplerian so that $|\bm{V}^\prime|\ll R\OmK$, and similarly for $\bm{W}$. Neglecting terms quadratic in the primed variables, the equilibrium equations become
\begin{align}
     -2\OmK V_\phi^\prime &= F_R  + 2\eta R\OmK^2 + \frac{\epsilon}{\taus}\left(W_R^\prime - V_R^\prime\right),\label{eqm_gasr}\\
   \frac{1}{2}\OmK V_R^\prime &= F_\phi + \frac{\epsilon}{\taus}\left(W_\phi^\prime - V_\phi^\prime\right),\\
    -2\OmK W_\phi^\prime &= - \frac{1}{\taus}\left(W_R^\prime - V_R^\prime\right),\\
      \frac{1}{2}\OmK W_R^\prime &= - \frac{1}{\taus}\left(W_\phi^\prime - V_\phi^\prime\right),\label{eqm_dustphi}
\end{align}
where $F_{R,\phi}$ are the radial and azimuthal components of the Lorentz force:  
\begin{align}
    &F_R \equiv -\frac{B_\phi^2}{2\mu_0R\rhog}, \quad F_\phi \equiv \frac{B_R B_\phi}{2\mu_0R\rhog}\label{lorentz_force},
\end{align}
with $B_{R,\phi}$ given by Eqs. \ref{eqm_br}--\ref{eqm_bphi}, and 
\begin{align}
    \eta \equiv - \frac{1}{2R\OmK^2\rhog}\frac{\p P}{\p R}\label{eta_def}
\end{align}
is a dimensionless measure of the radial gas pressure gradient. 
Solving Eqs. \ref{eqm_gasr}--\ref{eqm_dustphi} give 
\begin{align}
    & V_{R}^\prime = \frac{2 \epsilon \st}{\Delta^2} \etatot R \OmK + \frac{2\left(\st^2 + \epsilon +1\right)}{\Delta^2}\frac{F_\phi}{\OmK}, \label{eqm_vr}\\
& V_{\phi}^\prime = - \frac{\left(\st^2 + \epsilon +1\right)}{\Delta^2} \etatot R \OmK +\frac{\epsilon \st}{\Delta^2}\frac{F_\phi}{\OmK}, \label{eqm_vphi}\\
& W_{R}^\prime = - \frac{2 \st}{\Delta^2}\etatot R \OmK + \frac{ 2(\epsilon +1)}{\Delta^2}\frac{F_\phi}{\OmK},  \label{eqm_wr}  \\
& W_{\phi}^\prime = - \frac{\left(\epsilon +1\right)}{\Delta^2}\etatot R \OmK - \frac{\st}{\Delta^2}\frac{F_\phi}{\OmK}\label{eqm_wphi},
\end{align}
where $\Delta^2 \equiv \st^2 + (1 + \epsilon)^2$, and 
\begin{align}
    \eta_\mathrm{tot} = \eta + \frac{F_R}{2R\OmK^2}\label{etatot} 
\end{align}
is a dimensionless measure of the total (gas plus magnetic) radial pressure gradient. Eqs. \ref{eqm_vr}--\ref{eqm_wphi} generalizes the standard, two-fluid steady-state drift solutions for a dusty gas \citep{nakagawa86} to include { the effect of} horizontal magnetic fields in a resistive disk{, where magnetic torques drive gas accretion.} See \cite{umurhan20} for a similar set of equations that account for { viscous gas accretion.} 

We remark that the above velocity fields do not produce global steady states for arbitrary density fields. This requires the mass fluxes $RV_R^\prime\rhog$ and $RW_R^\prime\rhod$ to be global constants, which constrains the density profiles. We explore this in Appendix \ref{global_density} for the case of constant $\st$ and $\epsilon$. However, small-scale instabilities such as the SI are not expected to be sensitive to the global density profile. Indeed, it is possible to model them in a local disk model, in which case a strict steady state can be obtained for constant densities, velocities, pressure gradients, and magnetic torques, as we do below.

\subsection{Dust drift indirectly induced by magnetic fields}\label{dust_drift_magnetized}
In the limit $\epsilon \to 0$, or negligible dust feedback, we find
\begin{align}
    V_R^\prime \to \frac{2F_\phi}{\OmK}, \quad V_\phi^\prime \to - \etatot R\OmK, \label{vr_nodust}
\end{align}
so the gas rotates at a sub-Keplerian speed (assuming $\etatot>0$) and drifts inward since $F_\phi < 0$. The gas accretion rate is a constant: 
\begin{align}
    \dot{M}_\mathrm{g} = -R \rhog V_R^\prime = \frac{B_{\phi 0}^2\eta_O}{2\mu_0\OmKref^2 R_0^2}.\label{mgdot} 
\end{align} 
In this limit the dust radial drift is  
\begin{align}
     W_R^\prime \to - \frac{2\st}{1 + \st^2}\etatot R\OmK + \frac{2}{1 + \st^2}\frac{F_\phi}{\OmK}, \notag \\ & 
\end{align}
Thus the inwards drift of dust is enhanced by the magnetic torque. This is expected as the dust is partially coupled to the inwardly-accreting gas through drag. The dust, therefore, feels the magnetic field indirectly. 

\subsection{Shearing box approximation} \label{shear_box}
We study the local dynamics of the above system by focusing on a small patch of the disk using the shearing box framework \citep{goldreich65}. The box is anchored at a fiducial point  $(R_0,\phi_0,0)$ that rotates around the star at an angular frequency of $\OmKref$, so $\phi_0 = \OmKref  t $. Cartesian coordinates $(x,y,z)$ in the box correspond to the radial, azimuthal, and vertical directions in the global disk. For a sufficiently small box size ($\ll R_0$) we can ignore curvature effects and approximate Keplerian rotation as  $\bm{U}_\mathrm{K}=-q x\OmKref \hat{\bm{y}}$ with $q=3/2$. We then define $\bm{v}$ and $\bm{w}$ as the dust and gas velocities in the shearing box relative to this linear shear flow.  
The total gravitational and centrifugal force in the box is $3 x \OmKref^2 \hat{\bm{x}}$. 

In terms of velocity deviations, the axisymmetric shearing box equations read
\begin{align}
&\frac{\partial \rhog }{\partial t} + \nabla \cdot (\rhog \bm{v}) = 0 \label{gas_mass_local},\\
&\frac{\partial \bm{v} }{\partial t} + \bm{v}\cdot \nabla \bm{v} = 2 v_y \Omega_0 \hat{\bm{x}} - v_x \frac{\Omega_0}{2} \hat{\bm{y}} - \frac{1}{\rhog} \nabla P \notag \\ 
&\phantom{\frac{\partial \bm{v} }{\partial t} + \bm{v}\cdot \nabla \bm{v} =}
 + 2\etatotzero R_0\OmKref^2\hat{\bm{x}} + F_{\phi 0}\hat{\bm{y}} \notag\\
 & \phantom{\frac{\partial \bm{v} }{\partial t} + \bm{v}\cdot \nabla \bm{v} =}
 + \frac{1}{\rhog\mu_0}\left(\nabla\times\bm{B}\right)\times \bm{B} + \frac{\epsilon}{\taus}(\bm{ w}-\bm{v}),\label{gas_mom_local}\\
&\frac{\partial \bm{B}}{\partial t}  =  \nabla \times (\bm{v} \times \bm{B}) + \nabla \times (\bm{U}_\mathrm{K} \times \bm{B}) + \eta_{O} \nabla^2 \bm{B}  \label{Faraday_local},\\
&\frac{\partial \rho_d }{\partial t} + \nabla \cdot (\rho_d \bm{w}) = 0, \label{dust_mass_local}\\
&\frac{\partial \bm{w} }{\partial t} + \bm{w} \cdot \nabla \bm{w} = 2 w_y \Omega_0 {\hat x} - w_x \frac{\Omega_0}{2} {\hat y} - \frac{1}{\taus}(\bm{w}-\bm{v})   \label{dust_mom_local}
\end{align}
\citep[e.g.][]{yang18}, where subscript $0$ here and below denotes evaluation at the reference radius in steady state.
In Eq. \ref{gas_mom_local}, we include the effect of a global gas-plus-magnetic pressure gradient through the term $\propto \etatotzero$ and that from a large-scale horizontal magnetic torque through $F_{\phi0}$. In the shearing box approximation these are modeled as constant forcing terms that do not respond to the  dynamics in the box. Note that $P$ here refers to pressure fluctuations and is zero in equilibrium. An  exact equilibrium state consists of constant densities and velocities, with the latter given by Eqs. \ref{eqm_vr}--\ref{eqm_wphi} evaluated at $R_0$. That is, $\bm{v} = \bm{V}^\prime(R_0)$ and $\bm{w} = \bm{W}^\prime(R_0)$. 

\subsection{Problem specification}
 We study two specific roles of magnetic fields and simplify Eqs. \ref{gas_mass_local}--\ref{dust_mom_local} accordingly:
\begin{enumerate}
    \item { Case I}: The effect of gas accretion flows induced by large-scale, horizontal fields. Here, the magnetic field is only included in the equilibrium disk and is neglected in the perturbed state. We  thus set the Lorentz force to zero in Eq. \ref{gas_mom_local} and drop the induction equation. This is expected to be valid for gas poorly coupled to the field, as expected in PPDs, in which case the magnetic field will remain effectively unperturbed. See \cite{mcnally17} for a similar treatment. 
    
    \item { Case II}: The effect of a live vertical magnetic field, which enables the MRI and other magnetic modes.
    { A vertical field does not modify the equilibrium from the hydrodynamic limit. However, magnetic forces are fully active in the perturbed state, i.e. we include Lorentz forces and the induction equation. Note that a live radial field cannot be included initially as it would be sheared apart by differential rotation so no equilibrium can be constructed. For simplicity we also neglect background horizontal fields, thus $\etatotzero \to \eta_0$ and $F_{\phi 0} \to 0$, and there is no magnetically-induced gas accretion.  
    } 
\end{enumerate}

\subsection{Physical parameters}\label{Physical_parameters}

Our magnetized, dusty disks are characterized by several parameters, some of which are specific to Case I and II. For clarity, we drop the subscript 0 notation with the understanding that all variables below are evaluated at the reference radius in the equilibrium state.

The disk opening angle at the reference radius 
\begin{align}
\hgas \equiv \frac{\Hgas}{R} 
\end{align}
is a measure of the disk temperature. In this work we fix $\hgas=0.05$. The global pressure gradient $\etatot$ is typically of $O(\hgas^2)$ and is also a measure of gas compressibility when considering the SI \citep{youdin05,youdin07}. 

For convenience we also define the reduced pressure gradient parameter 
\begin{align}
    \etatilde \equiv \frac{\eta_\mathrm{tot}}{\hgas}. 
\end{align}
We take $\etatilde\geq 0$ unless otherwise stated. We will consider small values of $\etatilde$ to explore how the SI behaves around pressure extrema.  

For the magnetic field, we define the azimuthal and vertical Alfv\'{e}n speeds 
\begin{align}
C_{A\phi,z} \equiv \frac{\left|B_{\phi,z}\right|}{\sqrt{\mu_0 \rhog}},  
\end{align}
{ and the plasma beta parameters}
\begin{align}
\beta_{\phi,z} \equiv \frac{C_s^2}{C_{A\phi,z}^2},
\end{align}
to quantify the (inverse) strength of the magnetic field in Cases I and II, respectively. 
We use the Elsasser numbers  
\begin{align}
&\Lambda_{\phi,z} \equiv \frac{C_{A\phi, z}^2}{\eta_O\Omega} = \frac{C_s^2}{\eta_O\beta_{\phi,z}\Omega} \label{def_elsa}
\end{align}
to quantify the effect of Ohmic resistivity.

A given disk model is parameterized by $\etatilde$, $\st$, $\epsilon$, $\beta_{\phi, z}$, and $\Lambda_{\phi, z}$. However, for Case I the magnetic parameters only appear in the azimuthal force $F_\phi$, which physically represents Maxwell stresses from the background disk. It is therefore convenient to define the dimensionless stress 
\begin{align}
    \alpha_M \equiv -\frac{B_RB_\phi}{\mu_0 P} = \frac{\hgas^2}{2\Lambda_\phi\beta_\phi^2}, \label{alphaMdef}
\end{align}
 so $F_\phi = -\alpha_MC_s^2/2R$. { Our results for Case I do not depend on
  values of $\beta_\phi$ and $\Lambda_\phi$ separately, but only on the magnetic stress $\alpha_M$. 
 } Note that $\alpha_M$ should \emph{not} be interpreted as the viscous $\alphass$ parameter of \cite{shakura73} that attempts to mimic small-scale turbulence. Here, $\alpha_M$ characterizes radial angular momentum transported by large-scale, { horizontal} magnetic torques, although our results are also applicable to gas accretion driven by other means  (see \S\ref{discuss_pressure_bump}). 

\subsection{Radial and azimuthal drifts}\label{dimensionless_drifts}
We define a dimensionless measure of the relative radial dust-gas drift as 
\begin{align}
    \zeta_x \equiv \frac{w_x - v_x}{C_s} = -\frac{\st}{\Delta^2}\left[ 2(1+\epsilon)\etatilde -\alpha_M\hgas\st\right].
    \label{dimensionless_xdrift}
\end{align}
The first term in the square brackets of Eq. \ref{dimensionless_xdrift} is usually of $O(\hgas)$. It is thus larger than the second term by a factor of  $O\left(\frac{1}{\alpha_M\st}\right) \gg 1$ for typical disk parameters. That is, radial drift is generally attributed to radial pressure gradients. 

However, near a pressure extremum, $\zeta_x$ can vanish as the inwards drift due to pressure gradients is cancelled out by the magnetic torque. For $\etatilde \lesssim \alpha_M\hgas\st$, the magnetic torque dominates and dust drifts \emph{outwards relative to the gas}. For $\hgas=0.05$, $\st=0.1$, and $\alpha_M=0.01$, the critical $\etatilde$ is $O(10^{-5})$.  


Similarly, we define a dimensionless measure of the azimuthal drift as
\begin{align}
    \zeta_y \equiv \frac{w_y - v_y}{C_s} = \frac{\st}{\Delta^2}\left[\st \etatilde + \frac{1}{2}\alpha_M\hgas(1+\epsilon)\right]. 
    \label{dimensionless_ydrift}
\end{align}
Both pressure gradients and the magnetic torque leads to a positive azimuthal drift. For $\etatilde \lesssim \alpha_M\hgas/\st$, the azimuthal drift is attributed to the magnetic torque. For the same parameters as above, the critical $\etatilde $ is $O(10^{-3})$, implying that as one approaches a pressure extremum 
the magnetically-induced azimuthal drift dominates over that due to pressure gradients, before the same occurs for the radial drift. 

We remark that while the radial pressure gradient causes an $O(\st)$ radial drift and an $O(\st^2)$ azimuthal drift; the magnetic torque has the opposite effect in driving an $O(\st)$ azimuthal drift and an $O(\st^2)$ radial drift. When $\etatilde = 0$, the azimuthal drift is larger than radial drift by a factor of $O(\st^{-1})$. 
 

\section{Linear { theory}} \label{Theoretical_study}
We consider axisymmetric Eulerian perturbations for any variable $A$ such that  
\begin{align}
A\to A + \re \sum_{k_x, k_z} \delta A \exp{\left[\ii(k_{x}x+k_{z}z) + \sigma t \right]},  \label{def_linearized}
\end{align}
where $\delta A$ is a complex amplitude; $k_{x}$ and $k_{z}$ are real radial and vertical wavenumbers, respectively; the complex growth rate $\sigma = s - \ii \omega$, where $s$ is the real growth rate and 
$\omega$ is the oscillation frequency. { For the linear problem, we take $k_{x,z}>0$ without loss of generality.} Explicit expressions of the linearized equations are given in Appendix \ref{linear_eqns}, where we also reproduce the standard MRI and SI. 

The linearized system constitutes the eigenvalue problem 
\begin{align}
& \bm{M}\bm{q} = \sigma \bm{q} \label{ex_eigen}
\end{align}
where $\bm{M}$ is the matrix representation of the right-hand side of Eqs. (\ref{lin_gas_mass}) -- (\ref{lin_dust_mom_z}) and $\bm{q}$ is the eigenvector of the complex amplitudes. Note that for Case I, $\bm{M}$ is a $8\times 8$ matrix and $\bm{q} = \left[\delta\rhog, \delta\rhod, \delta\bm{v},\dd\bm{w} \right]^T$ since the induction equation is dropped. For Case II, $\bm{M}$ is an $11\times 11$ matrix and $\bm{q}$ additionally  includes magnetic field perturbations. We solve Eq. \ref{ex_eigen} using standard numerical methods in \textsc{matlab}. In addition to the physical parameters described in \S\ref{Physical_parameters}, each calculation also depends on the wavenumbers $k_{x,z}$. 

{
\subsection{Streaming instabilities driven by azimuthal drift}\label{azi_SI}
The classic SI of \cite{youdin05} is driven by the radial drift between dust and gas, $\zeta_x$.  There also exists an azimuthal drift $\zeta_y$ between dust and gas (Eq. \ref{dimensionless_ydrift}), but this is sub-dominant to the radial drift provided that magnetic torques are weak compared to pressure gradients, which is usually the case. 

However, as discussed in \S\ref{dimensionless_drifts}, azimuthal drifts can dominate over radial drifts near pressure extrema. Indeed, for Case I and sufficiently small $\etatilde$, we find the SI can still operate, but is now related to the azimuthal drift, and thereby to the magnetically-induced accretion flow. In Appendix \ref{azi_si} we present a simplified model of this new form of SI. We find they have growth rates 
\begin{align}
    s = \sqrt{\frac{\epsilon K_x \zeta_y}{1+\epsilon}}\Omega,\label{azimuthal_si_growth}
\end{align}
for $K_z=0$, where $K_{x,z} = k_{x,z}\Hgas$ are normalized wavenumbers. This is distinct from the classic SI that requires $K_z\neq0$ \citep{youdin05}. This azimuthal drift-driven SI also differs from the \emph{Resonant Drag Instabilities} described below, which would necessarily require non-axisymmetric disturbances. 
}

\subsection{Resonant drag instabilities} \label{Resonance}

A powerful description of instabilities driven by mutual dust-gas drag is the RDI theory developed by \cite{squire18a,squire18b}. This family of instabilities arise from a resonance between waves in the gas and the relative drift between dust and gas when $\omega_\mathrm{gas}(\bm{k}) = \bm{k}\cdot(\bm{w}-\bm{v})$, where $\omega_\mathrm{gas}$ is the neutral frequency of a wave mode in the gas when there is no dust, and $\bm{k}$ is the wavevector. 

In our unstratified, axisymmetric shearing box the resonance condition is  
\begin{align}
    \omega_\mathrm{gas}(k_x, k_z) = k_x(w_x - v_x).\label{rdi_cond}
\end{align}
 Eq. \ref{rdi_cond} gives the relation between $k_x$ and $k_z$ that maximizes growth rates. Since a variety of waves can be supported in a gas, RDIs are generic so that dusty gas is generally unstable. Note that the azimuthal drift ($w_y-v_y$) cannot cause an RDI in axisymmetric disks. 

\subsubsection{Classic streaming instabilities}\label{rdi_si}
{ When $\epsilon\ll 1$ the classic SI is an RDI in which $\omega_\mathrm{gas}$ corresponds to an inertial wave \citep{squire18b}}, 
\begin{align}
    \omega_\mathrm{gas}^2 = \frac{k_z^2}{k_x^2 + k_z^2}\Omega^2 \label{inertial_wave}
\end{align}
\citep[e.g.][]{balbus03}. { Using Eq. \ref{inertial_wave} and 
 the radial drift given by Eq. \ref{dimensionless_xdrift}, the resonance condition Eq. \ref{rdi_cond} becomes}
\begin{align}
    K_z^2 = \frac{K_x^4\zeta_x^2}{1 - K_x^2\zeta_x^2}\label{rdi_si_condition}
\end{align}
\citep[see also][]{umurhan20}. 

Although RDI theory formally applies to the limit $\epsilon  \ll 1$, we find Eq. \ref{rdi_si_condition} is still a useful guide in identifying classic SI modes even when $\epsilon > 1$. Note also that { RDIs do not} distinguish between different origins of dust-gas drift. Thus for example, at a pressure extremum, $\etatilde=0$ the radial drift caused by the magnetic torque alone can, in principle, also drive { RDIs}. { However, in practice, we find the azimuthal drift dominates in this limit and leads to the non-RDI instability described in \S\ref{azi_SI}}.

{
\subsubsection{Magnetic fields and the classic SI}
When a live magnetic field is present, for example in Case II, we find that magnetic perturbations can stabilize the classic SI, most notably in dust-poor disks, even with large resistivities. In Appendix \ref{mag_onefluid} we present a toy model based on a one-fluid description of a dusty, magnetized gas. For $K_x/K_z,\, \epsilon,\,\Lambda_z \ll 1$, Eq. \ref{magSI_onefluid_disp_simple} gives the dispersion relation 
\begin{align}
&\hat{\sigma}^3 + \left(\epsilon\st - 2\ii K_x \st \etatilde\right)\hat{\sigma}^2 + \hat{\sigma} -2\ii K_x \st \etatilde\notag\\
&+ 2\hat{\sigma}\Lambda_z\left(\epsilon\st\hat{\sigma}^2 + \hat{\sigma} - 2\ii K_x \st \etatilde\right)=0 
\end{align}
to leading order in $\Lambda_z$, where $\hat{\sigma}=\sigma/\Omega$. One can then show that at  $K_x=1/2\st\etatilde$ ($=1/\left|\zeta_x\right|$ for $\epsilon\ll 1$), where the resonant $K_z\to\infty$, SI growth rates decrease as $\Lambda_z$ increases from zero, presumably due to stabilization by magnetic tension. 
}

\subsubsection{Streaming instability of Alfv\'{e}n waves}
{ A magnetized} gas can support additional MHD modes: the fast and slow magneto-sonic and Alfv\'{e}n waves. In a dusty media these MHD waves can resonate with the dust-gas drift and produce a variety of instabilities \citep{hopkins18,seligman19,hopkins20}. However, their relevance to PPDs may be limited by non-ideal effects \citep{hopkins18}, a conclusion consistent with our numerical calculations below.  

We demonstrate with Case II that Alfv\'{e}n waves can drive streaming-type instabilities. When rotation is neglected, the dispersion relation for an Alfv\'{e}n wave with a purely vertical background field is  
\begin{align}
\omega_\mathrm{gas}^2 = C_{Az}^2 k_z^2 
\label{Alfven_wave}
\end{align} 
\citep[e.g.][]{ogilvie16}. Using Eq. \ref{Alfven_wave}, Eq. \ref{dimensionless_xdrift}, and Eq. \ref{rdi_cond} the resonance condition between Alfv\'{e}n waves and dust-gas drift becomes
\begin{align}
K_z^2  &= K_x^2\zeta_x^2\beta_z \\
&\simeq 4\st^2 \etatilde^2\beta_z K_x^2 \quad (\epsilon, \st \ll 1),
\label{awsi_cond}
\end{align}
where the second expression applies to small grains in dust-poor disks without a horizontal field. We can therefore expect instabilities in a magnetized, dusty disk  with $K_z\propto K_x$. For $\st=0.1$, $\etatilde=0.05$, $\beta_z=10^4$ we find $K_z \simeq K_x$, while smaller $\beta_z$ reduce the resonant $K_z$ for a given $K_x$. That is, a stronger vertical field produces more vertically elongated disturbances. 

{ 
For Case II we only consider an initially vertical field. If a radial field is also present, Eq. \ref{Alfven_wave} becomes $\omega_\mathrm{gas}^2 = C_{Az}^2 k_z^2 + C_{AR}^2k_x^2$, where $C_{AR}$ is the Alfv\'{e}n speed associated with the radial field. The RDI condition then generalizes to $K_z^2 = K_x^2\zeta_x^2\beta_z\left(1-1/\zeta_x^2\beta_R\right)$, where $\beta_R \equiv C_s^2/C_{AR}^2$, implying the resonant $K_z$ is reduced in magnitude. However, a live radial field will be wound up by the background shear so that $\beta_R\to \infty$ and this reduction eventually vanishes. The azimuthal field generated by the shear  would take no part in axisymmetric RDIs. 
}

{
\subsection{Dynamical effect of dust on the MRI}\label{mri_dust_loading}
The presence of tightly-coupled dust is expected to reduce the Alfv\'{e}n speed since the total density increases with dust-loading, $\rhog \to \rhog(1+\epsilon)$. One can then define 
\begin{align}
    C_{Az,\mathrm{eff}} = \frac{C_{Az}}{\sqrt{1+\epsilon}},\label{eff_alfven}
\end{align}
as the effective vertical Alfv\'{e}n speed of a dusty gas, and similarly for the azimuthal speed. In ideal MHD this reduction increases the most unstable $k_z$ ($\propto 1/C_{Az,\mathrm{eff}}$); while in resistive disks dust-loading reduces growth rates ($\propto C_{Az,\mathrm{eff}}^2$) and the most unstable $k_z$ ($\propto C_{Az,\mathrm{eff}}$), see \cite{sano99} for expressions in the pure gas limit. We demonstrate these effects in \S\ref{awsi_linear} and \S\ref{caseII_resis}.
}

\section{{ Numerical} results} \label{results}


In this section, we solve Eq. \ref{ex_eigen} numerically to obtain the dispersion relation $\sigma=\sigma(k_x, k_z; \etatilde, \beta_{\phi,z}, \Lambda_{\phi,z})$. To connect our results to previous studies, we base our setups on that for the `LinA' ($\st=0.1$, $\epsilon=3$) and `LinB' ($\st=0.1$, $\epsilon=0.2$) SI eigenmodes described in \cite{youdin07}. We normalize growth rates by $\Omega$ and wavenumbers by $\Hgas$. Note that this differs from the conventional wavenumber normalization by $\etatot R$, since we will consider disks with $\etatot=0$. Otherwise, the fiducial value of the reduced pressure gradient $\etatilde = \etatot/\hgas=0.05$. For LinA and LinB the fiducial wavenumbers are then $K_{x,z} = 600$ and $K_{x,z} = 120$, respectively. We apply a lower limit to the growth rates at $s = 10^{-5}\Omega$.   

In Table \ref{table1} we list selected modes in our dusty, magnetized disks. These include SI modes modified by a background accretion flow, SI modes at pressure extrema that are driven by azimuthal drift, and the SI of  Alfv\'{e}n waves. 

To help identify the dominant source of the instabilities uncovered, we also examine the pseudo-kinetic energy of the modes, $U_\mathrm{tot} = \sum_{i=1}^3 U_i$ \citep{ishitsu09,lin21} constructed from the eigenvectors, where $U_1$ corresponds to thermal or pressure contributions, $U_2$ to that from dust-gas drift, and $U_3$ from the magnetic field. In all of the cases examined, $U_1$ is negligible as the modes are nearly incompressible. We thus focus on $U_2$ and $U_3$ (the latter only applicable to Case II). We further decompose $U_2$ into components associated with the background radial and azimuthal drifts, and the drift in the perturbed velocities. Details are given in Appendix \ref{pseudo_energy}. In plots we normalize $U_i$ by $U_\mathrm{tot}$.

\begin{table*}
\caption {Selected unstable modes in dusty, magnetized PPDs. All modes employ $\st = 0.1$. The classic SI modes LinA and LinB of \cite{youdin07} are reproduced for comparison. { The limit of zero initial toroidal and vertical fields are denoted by $\beta_{\phi,z}\to\infty$; while the ideal MHD limit is attained for $\Lambda_{\phi,z}\to\infty$.} Disks initialized with a { purely} horizontal field ($\beta_z = \infty$) do not evolve the induction equation and magnetic perturbations are assumed to be zero. These models possess a non-zero magnetic stress $\alpha_M$ (Eq. \ref{alphaMdef}). Disks initialized with a { purely} vertical field ($\beta_\phi = \infty$) include the full induction equation and Lorentz forces in the perturbed state. \label{table1}
} 
\begin{ruledtabular}
\begin{tabular}{lllllllll}
 Mode & $\epsilon$ & $ K_x, K_z$ & $\etatilde$ &  $\beta_\phi, \beta_z$ & $\Lambda_\phi, \Lambda_z$ &
 $\alpha_M$ &$\sigma/\Omega$ & Comment \\
        \hline
 LinA & $3$ & $600, 600$ & $0.05$ & $\infty, \infty$ & $\infty, \infty$ & 0 & $0.4190 - 0.3480\ii$ & Classic SI \\
 LinAI & $3$ &  $600,600$ &  $0.05$ & $10,\infty$ & $10^{-4},\infty$ & 0.125 &$0.4339 + 0.5671\ii$ & SI with magnetically-induced accretion\\ 
 LinAIeta0 & $3$ &  $5000, 100$ &  $0$ & $10,\infty$ & $10^{-3},\infty$ & 0.0125& $0.1358 + 0.6140\ii$ & SI without pressure gradients\\
 LinAII & $3$ &  $1000, 25$ &  $0.05$ & $\infty,100$ & $\infty,\infty$ & 0 &$0.1248 + 0.4911 \ii$ & SI of Alfv\'{e}n waves \\ 
       \hline
 LinB & $0.2$ & $120, 120$ &  $0.05$ & $\infty, \infty$  & $\infty, \infty$ & 0 & $0.0154 + 0.4998\ii$ & \multirow{4}{*}{As above but in a dust-poor disk} \\
 LinBI & $0.2$ & $120, 120$ &  $0.05$ & $10, \infty$  & $10^{-4}, \infty$  &    0.125 & $0.0243 + 1.1237\ii$ &  \\
 LinBIeta0 & $0.2$ &  $5000, 100$ &  $0$ & $10,\infty$ & $10^{-3},\infty$ & 0.0125 & $0.1377 + 2.4515 \ii$ & \\
 LinBII & $0.2$ &  $1000, 83$ &  $0.05$ & $\infty,100$ & $\infty,\infty$ & 0 &  $0.2215 + 6.7181\ii$ & \\
\end{tabular}
\end{ruledtabular}
\end{table*}

\subsection{Case I: Classic SI in magnetically accreting disks}\label{B_accretion_flow}


We begin by adding a horizontal magnetic field with $\beta_\phi=10$ to LinA and LinB. We vary the drift speeds induced by the field, which remains passive, through $\Lambda_\phi$. { 
Similar values of $\beta_\phi$ have been considered \citep[e.g.][]{krapp18}, although these are rather strong fields compared to that typically found in numerical simulations \citep[e.g.][who find a gas-to-magnetic pressure ratio of order 100]{cui21b}. However, as emphasized in \S\ref{Physical_parameters}, it is the magnetic stress $\alpha_M$ that is relevant to Case I. For nominal values of $\hgas=0.05$, $\beta_\phi=10$, $\Lambda_\phi = 10^{-3}$, we obtain $\alpha_M=0.0125$, which is consistent to non-ideal MHD disk simulations  \citep[e.g.][]{bethune17,bai17,riols20}. 
}

\begin{figure*}
    \centering
    \includegraphics[width=0.5\linewidth]{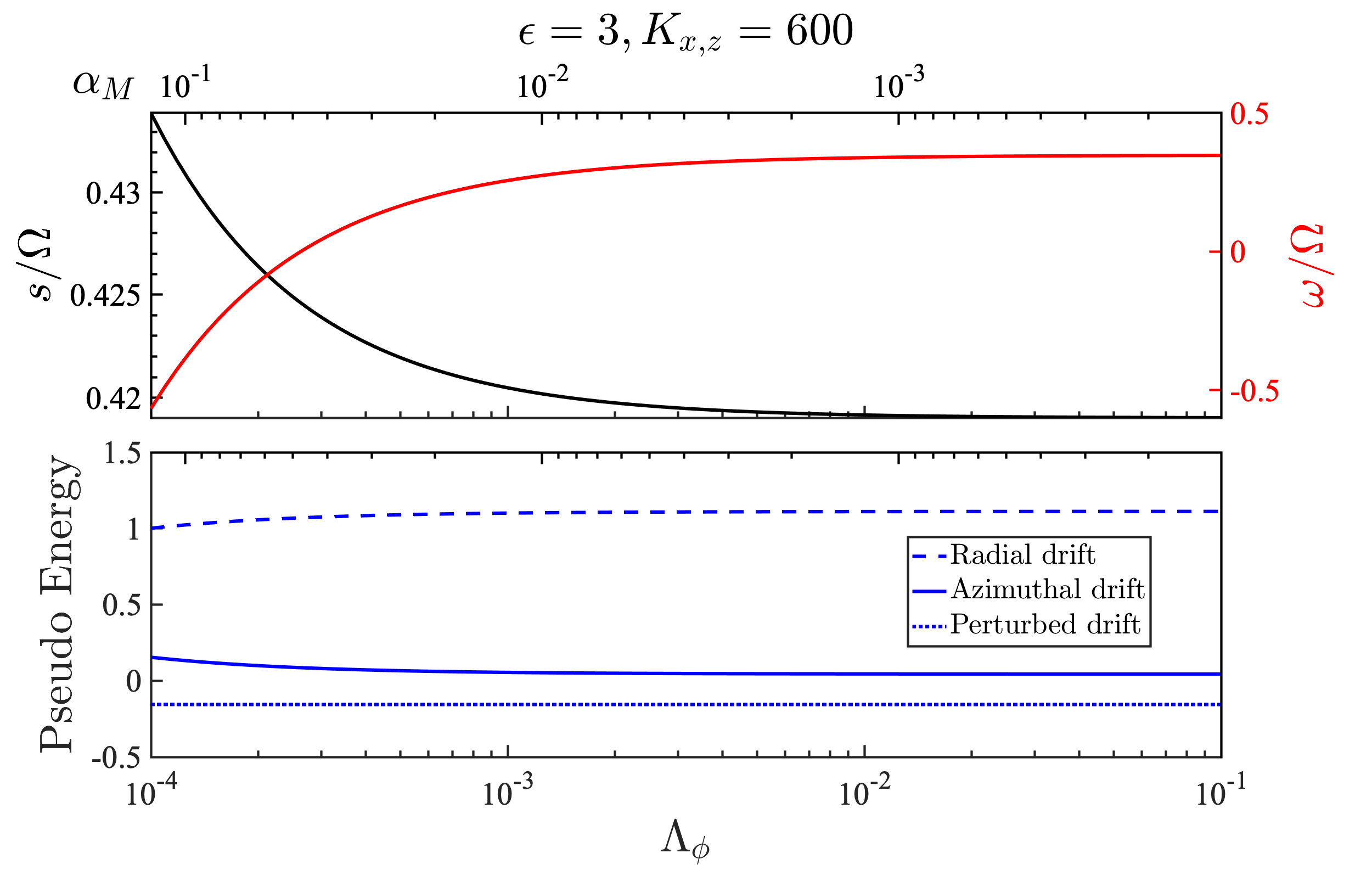}\includegraphics[width=0.5\linewidth]{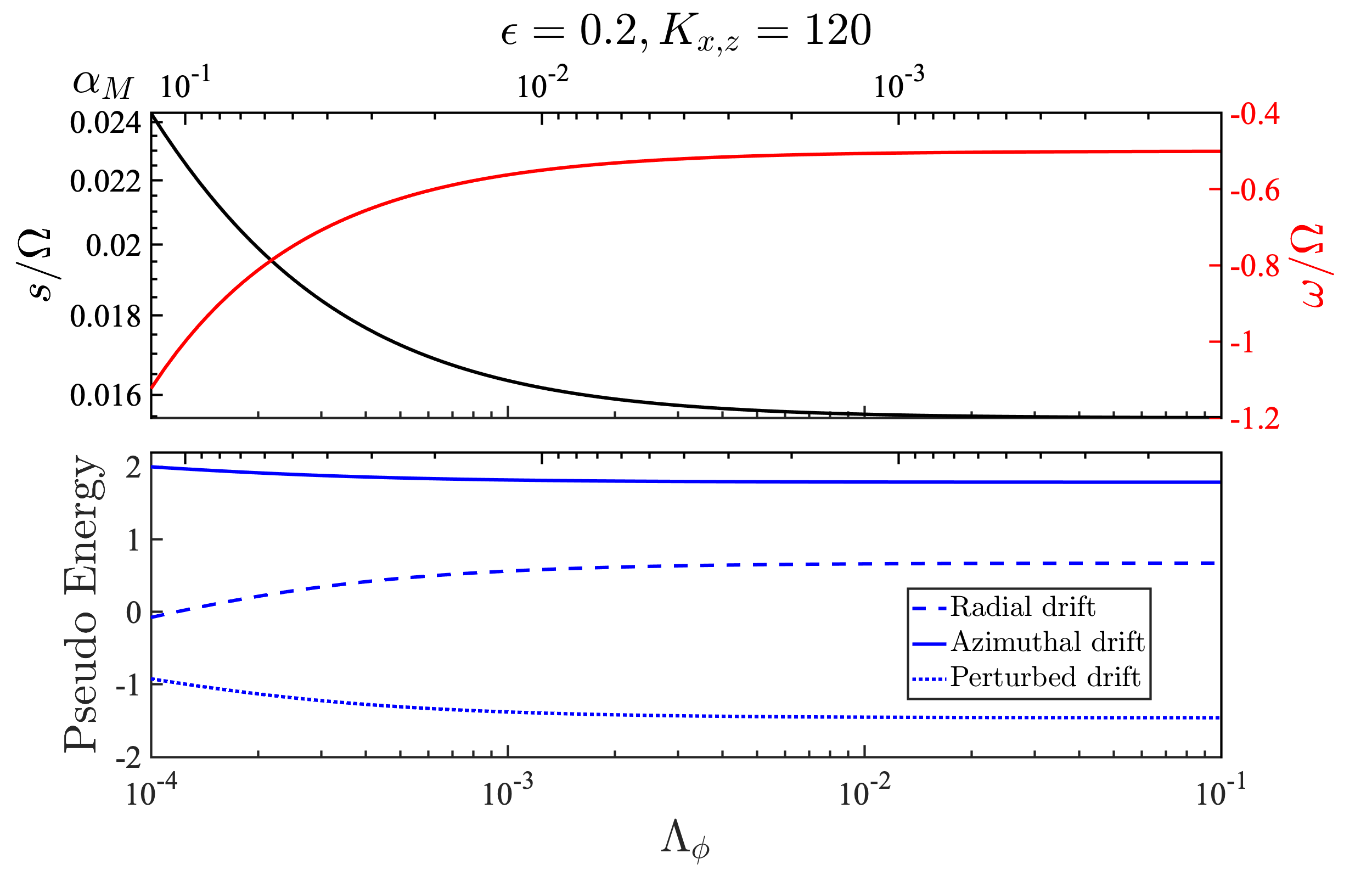}
    \caption{Streaming instability with a background laminar gas accretion flow driven by large-scale, passive horizontal magnetic fields, as a function of the azimuthal Elsasser number $\Lambda_\phi$ for fixed $\beta_\phi =10$, which translates to a range of dimensionless magnetic stresses $\alpha_M$. The grain size is fixed to $\st =0.1$. Top: mode growth rates (black) and oscillation frequencies (red). Bottom: contributions to the modes' pseudo-kinetic energy from the background radial drift (dashed), the background azimuthal drift (solid), and the relative drift in the perturbed velocities (dotted). 
    Left: a dust-rich disk with $\epsilon=3$. Right: a dust-poor disk with $\epsilon=0.2$. 
    \label{caseI_example}}
\end{figure*}

Fig. \ref{caseI_example} shows that SI growth rates increase with decreasing $\Lambda_\phi$, while oscillation frequencies become more negative and can exceed $\Omega$ in magnitude. For LinB, growth rates can increase by $50\%$ for $\Lambda = 10^{-4}$ ($\alpha_M\sim 0.1$) compared to the unmagnetized limit. { Growth rates for other field strengths can be inferred from the equivalent $\alpha_M$ as shown in the figure. For example, if $\beta_\phi=100$ (i.e. a factor 10 increase), then $\alpha_M$ would decrease by a factor of $100$ at fixed $\Lambda_z$ (see Eq. \ref{alphaMdef}). Thus for $\Lambda_z\gtrsim10^{-4}$ the new growth rates would be equal to that with $\beta_\phi=10$ and $\Lambda_z\gtrsim 10^{-2}$, or $\alpha_M\lesssim 10^{-3}$, in which case the magnetically-induced accretion has no effect.}

Recall that the magnitude of the dust-gas radial drift decreases with increasing strength of the magnetic torque, while that of the azimuthal drift increases (\S\ref{dust_drift_magnetized}). This suggests that the increasing growth rates are attributed to the azimuthal drift. We confirm this in the bottom panels of Fig. \ref{caseI_example} by plotting the contributions to the modes' pseudo-kinetic energy. For LinA, modes are driven by the radial drift, but its contribution drops slightly at $\Lambda_\phi = 10^{-4}$ ($\alpha_M\sim 0.1$), whereas that from the azimuthal drift increases. 

On the other hand, for LinB, azimuthal drift has twice the contribution as radial drift even for negligible magnetic stresses; while the radial drift contribution drops noticeably (even becoming slightly negative) for $\Lambda_\phi\lesssim 10^{-3}$ ($\alpha_M\gtrsim 0.01$). It is also for LinB that the increase in growth rates is more pronounced. This indicates that an accretion flow primarily affects SI modes in which the azimuthal drift plays a dominant role.  

We conclude that in dust-poor disks SI will be moderately boosted by an underlying, magnetically-induced accretion flow if the corresponding $\alpha_M\gtrsim 10^{-2}$, while in dust-rich disks the SI only becomes slightly more unstable. The enhancement is limited due to the fact that with a typical $\etatilde$ of $O(\hgas)$, the drifts induced by the magnetic field remain small compared to those directly induced by pressure gradients. However, this picture changes when we consider regions with vanishing $\etatilde$, as explored next. 

\subsection{Case I: SI with vanishing pressure gradients} \label{si_no_pgrad}

Here we examine the SI with vanishing pressure gradients { ($\etatilde\to 0$)}.  
These models can be considered as representing regions near to or at a pressure bump. In this section, we include a small gas viscosity  { and dust diffusion characterized by $\alphass$ (see Appendix \ref{linear_eqns}) to regularize the problem at large wavenumbers.} Otherwise, we find unstable modes { can develop on arbitrarily small scales} with frequencies $|\sigma| \gg \Omega$, which are likely unphysical. { Unless otherwise stated, we set $\alphass = 10^{-9}$.} 

\begin{figure*}
    \centering
    \includegraphics[width=\linewidth]{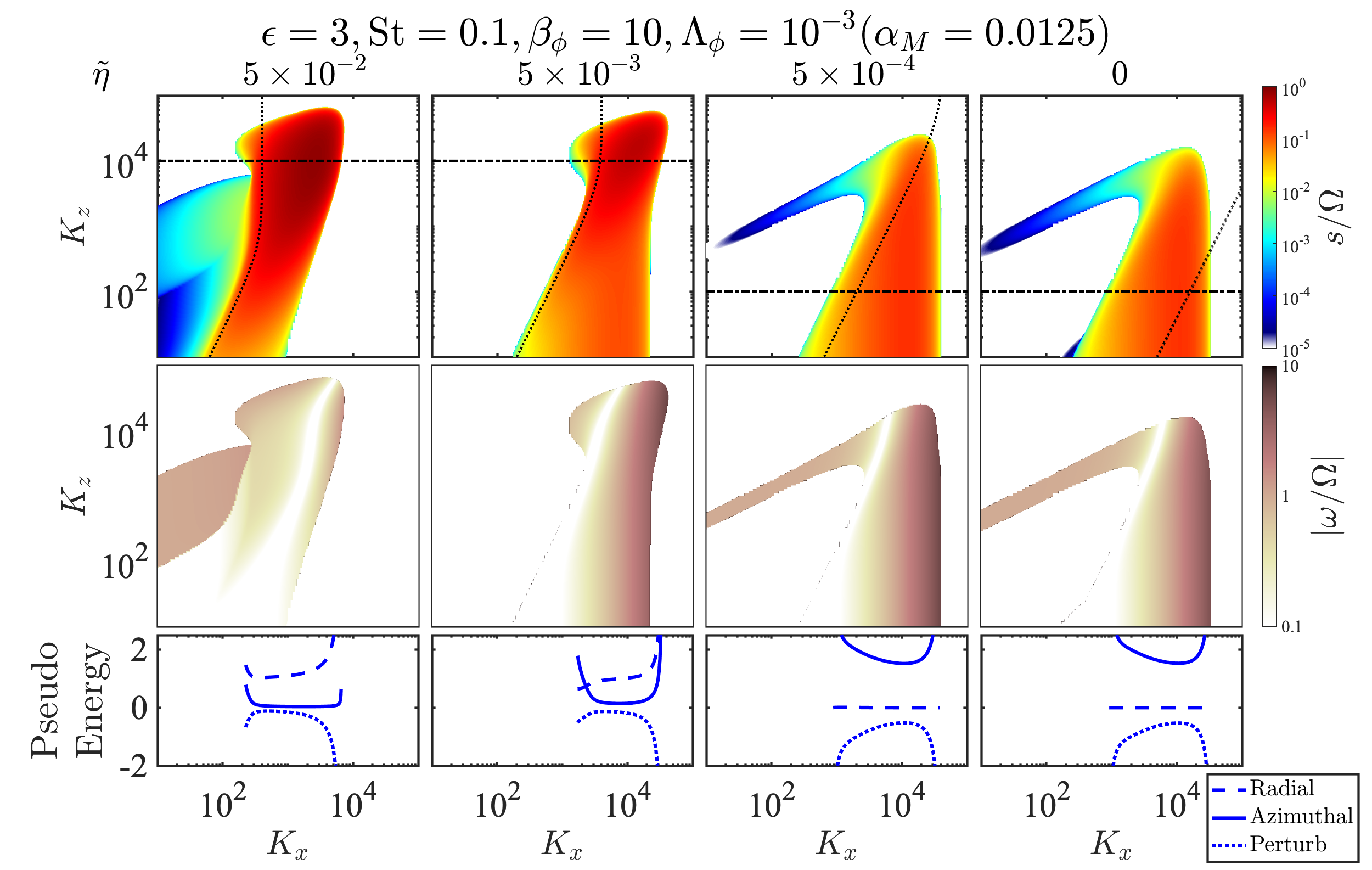}
    \caption{Unstable modes in dusty disks with a background laminar gas accretion flow driven by a passive, horizontal magnetic field. Top and middle: growth rates and oscillation frequencies as a function of wavenumbers, respectively. The dotted line in the top panel corresponds to the RDI condition for the classic streaming instability (Eq. \ref{rdi_si_condition}). Bottom: 
    contributions to the pseudo-energy 
    from the background radial drift (dashed), background azimuthal drift (solid), and perturbed drift velocities (dotted). These are show for a fixed $K_z$ as indicated by the { dash-dotted} line in the top panel and limited to modes with growth rates $>0.01 \Omega$. Left to right: decreasing total radial pressure gradients $\etatilde$, with the rightmost column showing results for vanishing pressure gradient. A small gas viscosity and dust diffusion { with $\alphass=10^{-9}$} is included to regularize the problem at large wavenumbers.}
    \label{caseI_linA_var_eta}
\end{figure*}

In Fig. \ref{caseI_linA_var_eta} we plot unstable modes as a function of wavenumbers and pressure gradients for the magnetized LinA setup. We fix $\beta_\phi = 10$ and $\Lambda_\phi =10^{-3}$, or $\alpha_M=0.0125$. For $\etatilde \geq 0.005$, we find classic SI modes with growth rates up to $\Omega$ for $K_{x} \sim 10^3$--$10^{4}$ and $K_z\sim 10^4$. The most unstable wavenumbers translate to $k_{x}\etatot R \sim 50$ and $k_{z}\etatot R \sim 500$ for $\etatilde = 0.05$ and to about $50$ for $\etatilde = 0.005$. As shown in the pseudo-energy plots, these modes are attributed to the radial  drift.

We find unstable modes change character as $\etatilde$ drops to $\leq 5\times 10^{-4}$. As shown in the right two columns of Fig. \ref{caseI_linA_var_eta}, the instability maintains a dynamical growth rate of $O(0.1\Omega)$ with a characteristic $K_x \sim 10^4$, independent of $K_z$, except at large $K_z$ where the viscous cut-off applies. In fact, instability persists even for $K_z\equiv0$, indicating these modes are one dimensional with negligible perturbations in the gas and dust vertical velocities, as well as that of the gas' radial velocities (see Appendix \ref{azi_si}).  

We remark that for $\etatilde = 5\times10^{-4}$, $K_x=10^4$ corresponds to $k_x\etatot R = 5$. In this sense, these modes have longer radial lengthscales than the standard SI observed for $\etatilde\geq 0.005$. Most notably, however, is that as $\etatilde\to 0$, unstable modes are driven by the azimuthal drift with negligible contributions from the radial drift. Even when there is no pressure gradient whatsoever ($\etatilde = 0$), we find growth rates up to $O(0.1\Omega)$.

\begin{figure*}
    \centering
    \includegraphics[width=\linewidth]{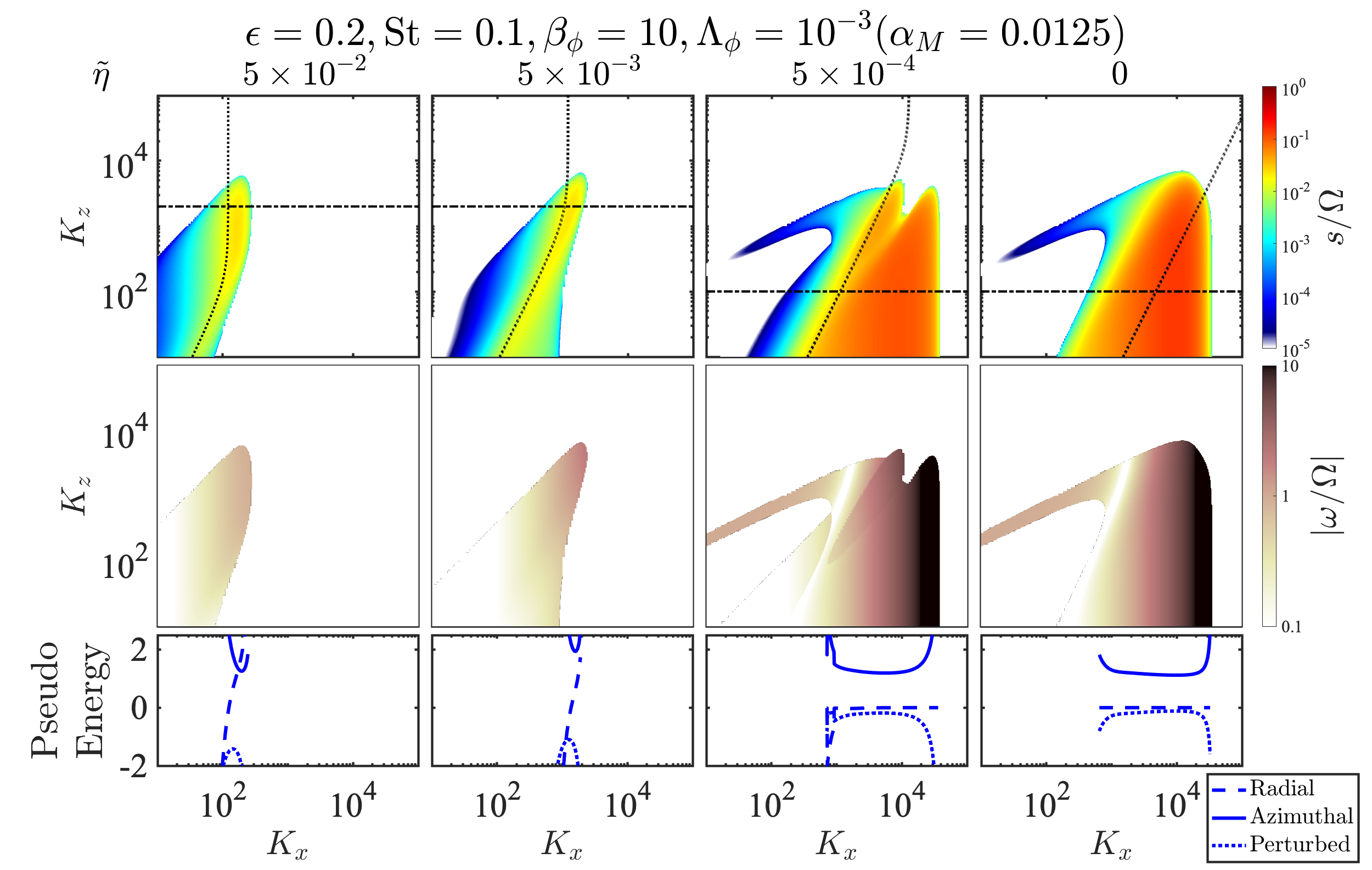}
    \caption{Same as Fig. \protect\ref{caseI_linA_var_eta} but for a dust-to-gas ratio of $\epsilon =0.2$.
    }
    \label{caseI_linB_var_eta}
\end{figure*}

Similarly, in Fig. \ref{caseI_linB_var_eta} we show how modes in the dust-poor LinB setup evolve as the pressure gradient decreases to zero. Again we find the classic SI modes dominate for $\etatilde \geq 0.005$, here corroborated by the coincidence of the most unstable wavenumbers with the RDI resonance condition for the SI as given by Eq. \ref{rdi_si_condition}. 
For $\etatilde = 5\times10^{-4}$, however, the RDI condition no longer predicts the most unstable modes, although a cluster of sub-dominant, classic SI modes are still found around the RDI condition. For $\etatilde \leq 5\times10^{-4}$, we again find the most unstable modes are completely dominated by azimuthal drift and can reach growth rates of $O(0.1\Omega)$. However, for large $K_x$ (here $\sim 10^4$) we find the modes have large oscillation frequencies $|\omega| \gtrsim 10\Omega$, implying that $\left|\taus\omega\right|\gtrsim 1$, which may violate the fluid approximation of dust \citep{jacquet11}.  

{ In Fig. \ref{caseI_no_press_var_beta} we examine how growth rates vary with $\beta_\phi$ at fixed $\Lambda_\phi=10^{-3}$, or more precisely as a function of $\alpha_M$. Here, we set $K_z=0$ and maximize growth rates over $K_x$. Per the simplified model in Appendix \ref{azi_si}, growth rates decrease with $\alpha_M$ and consequently with azimuthal drift. Note that the cut-off at finite $\alpha_M$ is due to dust diffusion, without which growth would persist for decreasing $\alpha_M$ by going to arbitrarily small scales. For $\alphass=10^{-9}$, the instability is quenched when $\alpha_M\lesssim 10^{-3}$ for $\epsilon=3$ and when $\alpha_M \lesssim 10^{-4}$ for $\epsilon=0.2$. Although the corresponding critical magnetic fields of $\beta_\phi\simeq 40$ and $\beta_\phi\simeq100$ are strong, these $\beta_\phi$ values scale as $\hgas\Lambda_\phi^{-1/2}$ (see Eq. \ref{alphaMdef}), and so would increase with increasing $\hgas$, decreasing $\Lambda_\phi$, or both. Thus the instability is more easily realized in highly resistive disks. Larger $\alphass$ requires stronger azimuthal drifts for instability. 


\begin{figure}
    \centering
    \includegraphics[width=\linewidth]{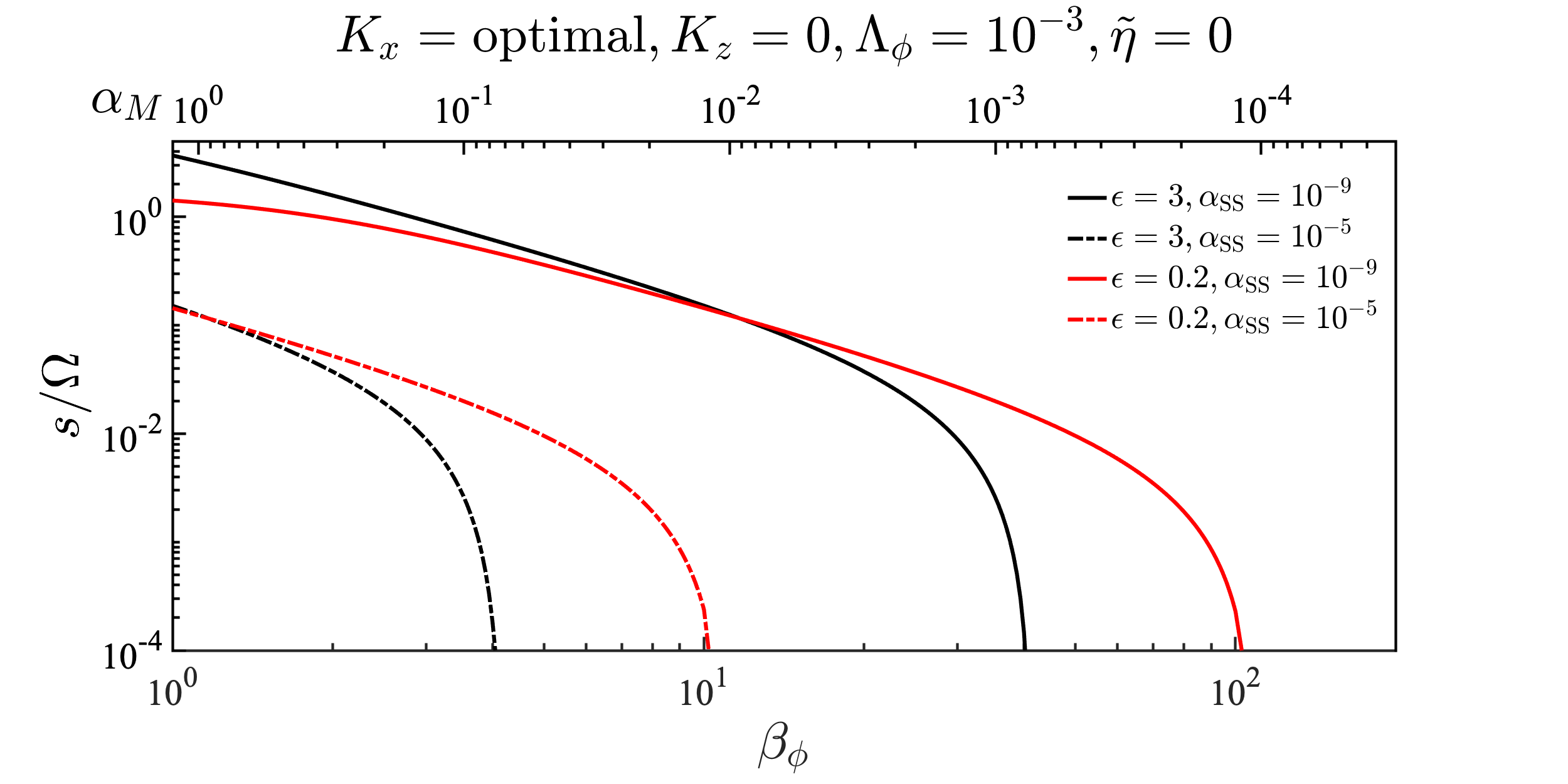}
    \caption{{ Growth rates of the azimuthal drift-driven SI in disks without pressure gradients ($\etatilde=0$) as a function of the plasma beta parameter $\beta_\phi$ at fixed $\Lambda_\phi=10^{-3}$, which corresponds to a range of magnetic stresses $\alpha_M$. Modes have no vertical structure ($K_z=0$) and growth rates are maximized over the radial wavenumbers $K_x$ at each $\beta_\phi$ (or $\alpha_M)$. The black (red) curve show results for a dust-rich (dust-poor) disk with $\epsilon=3$ ($\epsilon=0.2$). The solid (dash-dot) curves are obtained with a viscosity and diffusion coefficient of $\alphass=10^{-9}$ ($10^{-5}$).}
    }
    \label{caseI_no_press_var_beta}
\end{figure}
}

{ The results in this section suggests} that in accreting disks the SI can still operate in regions of weak or even zero pressure gradients, { provided there is sufficient magnetic stress to drive gas accretion with a corresponding $\alpha_M\gtrsim 10^{-4}$--$10^{-3}$}. However, { its nature} differs from the classic SI: here { the SI is driven by} the relative \emph{azimuthal} drift between dust and gas, which is largely induced by the torque acting on gas that is responsible for the underlying accretion flow.

\subsection{Case II: Dust feedback on the MRI and the SI of Alfv\'{e}n waves}\label{awsi_linear}

We now consider disks in the limit of ideal MHD with an initially vertical field and account for Lorentz forces in the perturbed state, i.e. a live field. Recall that a purely vertical field does not modify the background drift velocities from the hydrodynamic limit, since $\beta_\phi \to \infty$ and hence $F_\phi \to 0$ (see Eq. \ref{eqm_vr}--\ref{eqm_wphi}): there is no magnetically-induced accretion flow as in Case I. Here we fix $\etatilde=0.05$. 

We begin by examining how dust-loading affects the MRI { in disks with $\epsilon=0.2$ and $\epsilon=3$. The ensuing MRI turbulence would stir up dust grains, so considering dust-to-gas ratios of order unity may not be self-consistent with a settled dust layer, e.g. \cite{yang18} find $\epsilon\sim 0.3$ under ideal MRI turbulence and solar metallicities (or without feedback). However, order-unity values of $\epsilon$ can be reached at super-solar metallicities because of feedback \citep{yang18} or from radial concentrations by zonal flows \citep{dittrich13} and so are still relevant to explore.} 

\begin{figure*}
    \centering
    \includegraphics[width=\linewidth]{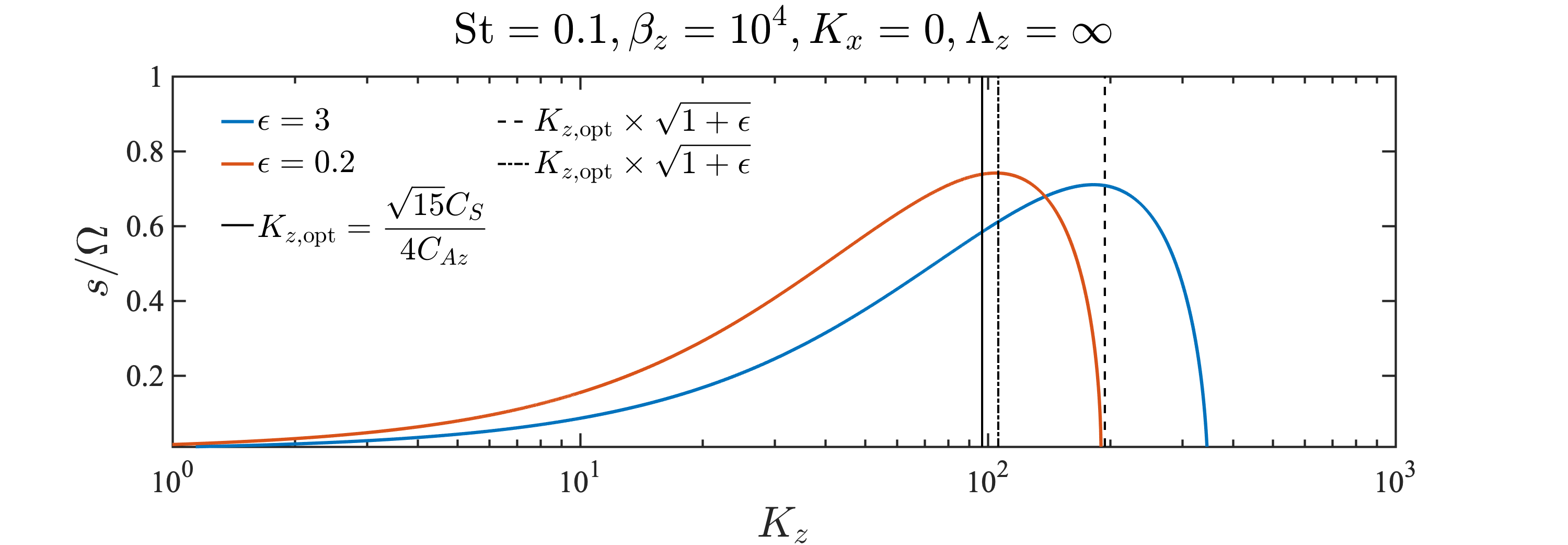}
    \caption{Ideal MRI growth rates in dusty disks threaded by a vertical magnetic field with strength $\beta_z = 10^4$ and dust-to-gas ratio $\epsilon = 3$ (blue) and $\epsilon=0.2$ (brown). The grain size is fixed to $\st=0.1$. The most unstable vertical wavenumber in the pure gas limit is marked by the solid vertical line. Assuming dust-loading reduces the Alfv\'{e}n speed by increasing the total density of the system, the resulting most unstable $K_z$ are 
    marked by the dashed and dash-dotted vertical lines.}
    \label{caseII_ideal_MRI}
\end{figure*}

{ Fig. \ref{caseII_ideal_MRI} show MRI growth rates as a function of $K_z$ at fixed $K_x=0$ and $\beta_z=10^4$.} The maximum growth rate of $0.75\Omega$ is unaffected by { the} dust. However, with increasing $\epsilon$ the MRI shifts to smaller vertical scales. In a pure gas disk the most unstable MRI mode has $k_z = \sqrt{15}C_s/4C_{Az}$ \citep{sano99}. { However, in a dusty gas we expect $C_{Az}\to C_{Az}/\sqrt{1+\epsilon}$ (see \S\ref{mri_dust_loading}}) and thus the most unstable $k_z\propto\sqrt{1+\epsilon}$ increases with dust-loading. This prediction is confirmed in the figure. 

\begin{figure*}
    \centering
    \includegraphics[width=\linewidth]{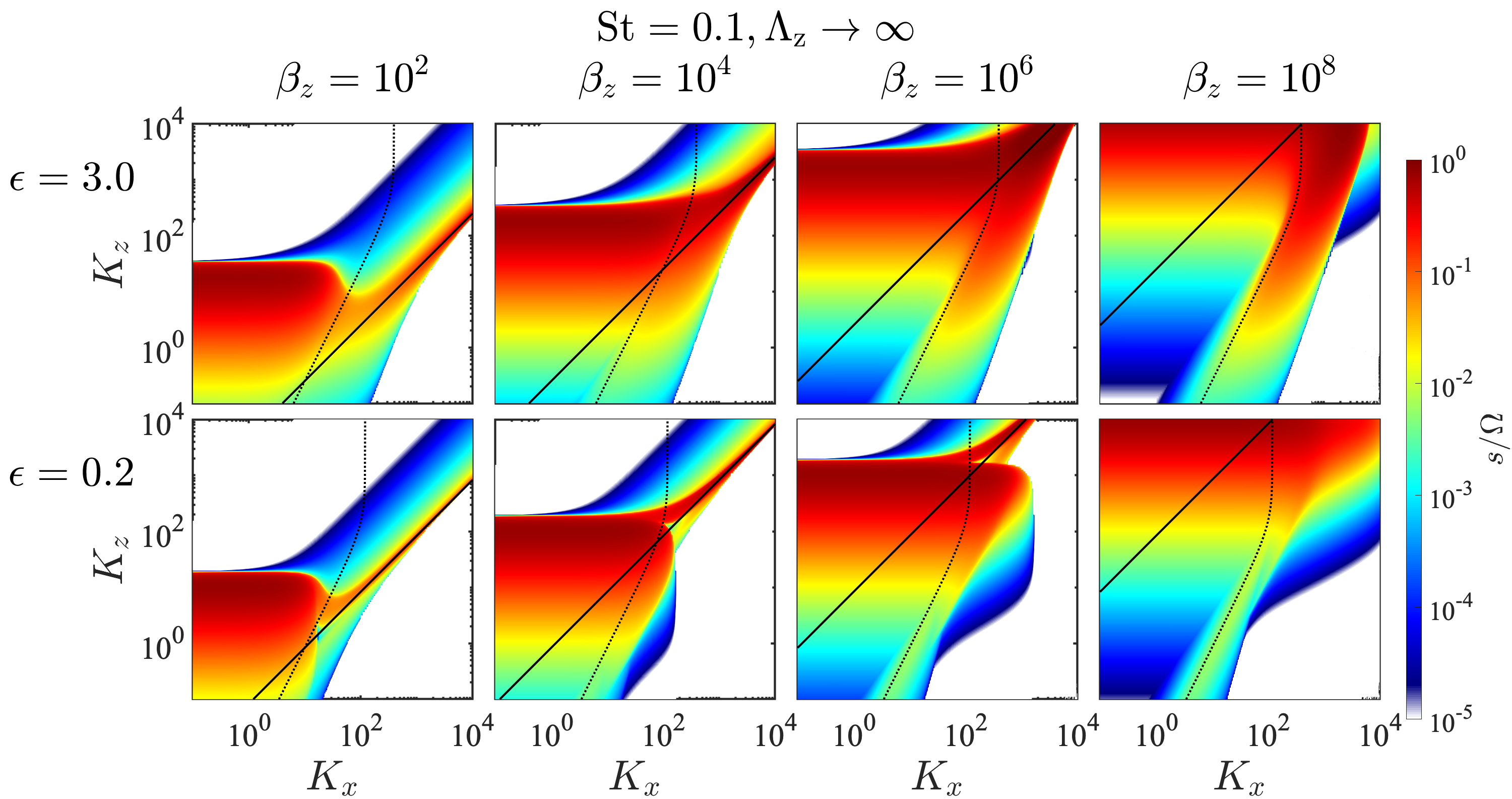}
    \caption{Growth rates of unstable modes in a dusty, actively magnetized disk initialized with a vertical field in the limit of ideal MHD. The grain size is fixed to $\st=0.1$. From left to right: decreasing magnetic field strength (increasing plasma beta parameter $\beta_z$). Upper row: dust-rich disks with $\epsilon=3$. Bottom row: dust-poor disks with $\epsilon=0.2$. Solid lines correspond to the RDI condition between Alfv\'{e}n waves and the dust-gas radial drift (Eq . ~\ref{awsi_cond}). Dotted lines correspond to the RDI condition for the classic streaming instability (Eq. \ref{rdi_si_condition}).  
    }
    \label{Example_CaseII}
\end{figure*}

Next, Fig. \ref{Example_CaseII} show growth rates for the LinA ($\epsilon=3$) and LinB ($\epsilon=0.2$) setups in wavenumber space and for different levels of magnetization. Starting with the 
leftmost column with $\beta_z=100$, the rectangular region with $K_{x,z}\lesssim 20$--$30$ corresponds to the MRI. However, the most prominent feature here is the cluster of new modes around the straight solid line for $K_x\gtrsim 10^2$ ($\epsilon = 3$) and $K_x\gtrsim 20$ ($\epsilon = 0.2$), which corresponds to the resonance condition between Alfv\'{e}n waves and the dust-gas radial drift (Eq. \ref{awsi_cond}), indicating they are a new class of RDI modes unique to dusty, magnetized gas. 

{ These Alfv\'{e}n wave streaming instabilities (AwSI) extend the parameter space of unstable modes to much smaller lengthscales compared to a pure gas disk, wherein only the MRI occurs (see Fig. \ref{Example_MRI_SI}). However, the maximum growth rates of the AwSI modes are typically of  $O(10^{-1}\Omega)$, which is slower than the MRI\footnote{{ We do find AwSI modes that can grow faster than $0.75\Omega$ (i.e. the MRI), but these occur at $K_{x,z}\gtrsim O(10^4)$.}}. Going rightwards to weaker magnetizations, the AwSI modes shift to smaller vertical scales, which is consistent with the resonant condition Eq. \ref{awsi_cond}. 
}

In the top panels of Fig. \ref{Example_CaseII} we also overplot the RDI condition for the classic SI (Eq. \ref{rdi_si_condition}) as the dotted curves. However, for $\beta_z\leq 10^{4}$ we do not find modes to cluster around this resonance. For example, with $\beta_z=100$, at the LinA (LinB) wavenumbers $K_{x,z} = 600$ ($K_{x,z} = 120$) we find growth rates $10^{-4}\Omega$ ($2\times10^{-4}\Omega$), which is much smaller than the classic, unmagnetized SI growth rate of $0.42\Omega$ ($0.015\Omega$). This indicates that the classic SI is suppressed by magnetic perturbations. It is only with an extremely weak field (here for $\beta_z \geq 10^6$) do we observe classic SI modes, but even then the most unstable modes occur at high-$K_z$ and are attributed to the MRI and the AwSI. 

\subsection{Case II: MRI and SI in resistive disks}\label{caseII_resis}

We now add a constant resistivity $\eta_O$ to Case II { to mimic a dead zone in the disk midplane \citep{gammie96}}. As before we first examine the effect of dust-loading on the MRI. { Here, the resulting MRI turbulence would be dampened by Ohmic resistivity. In this case, while grains do not settle much more than in ideal MHD \citep{fromang06, yang18}, 
weakened horizontal diffusion coupled with dust feedback can drive strong radial dust concentrations that lead to order-unity (or larger) values of $\epsilon$ in the dead zone \citep{yang18}\footnote{{ However, note that \citep{yang18} adopt $\Lambda_z\sim 2\times10^{-4}$, which is much smaller than the values we consider.}}. This motivates our consideration of high dust-to-gas ratios.}.

\begin{figure*}
    \centering
    \includegraphics[width=\linewidth]{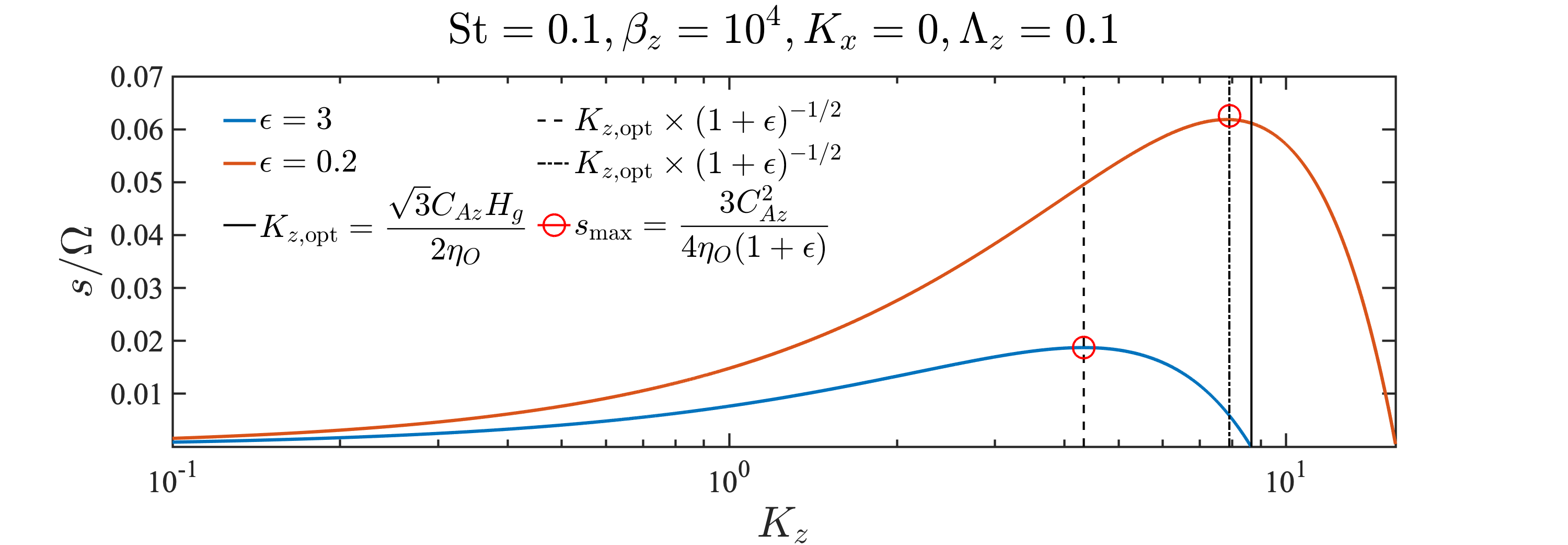}
    \caption{Resistive MRI growth rates in dusty disks threaded by a vertical magnetic field with strength $\beta_z = 10^4$ and dust-to-gas ratio $\epsilon = 3$ (blue) and $\epsilon=0.2$ (brown). The grain size is fixed to $\st=0.1$. The most unstable vertical wavenumber in the pure gas limit is marked by the solid vertical line. Assuming dust-loading reduces the Alfv\'{e}n speed by increasing the total density of the system, the dashed and dash-dotted vertical lines mark the resulting most unstable $K_z$ and the red open circles mark the corresponding maximum growth rates.}
    \label{caseII_nonideal_MRI}
\end{figure*}

Fig. \ref{caseII_nonideal_MRI} shows that increasing $\epsilon$ decreases both the MRI growth rates and the corresponding $K_z$, so modes develop on larger vertical scales in dust-rich disks. This can again be explained by the fact that the effective Alfv\'{e}n speed decreases with $\epsilon$ as the total density of the system increases. In resistive{, gaseous} disks with $\Lambda_z\ll 1$, however, the most unstable MRI mode has $k_z = \sqrt{3}C_{Az}/\eta_O$ with growth rate $s = 0.75C_A^2/\eta_O$ \citep{sano99}. { Again, in a dusty gas $C_{Az}\to C_{Az,\mathrm{eff}}$ (see Eq. \ref{eff_alfven}) so now} $k_z \propto (1 + \epsilon)^{-1/2}$ while $s \propto (1+\epsilon)^{-1}$. These reductions are indeed observed for $\epsilon=3$ compared to $\epsilon=0.2$ in the figure. 

\begin{figure*}
    \centering
    \includegraphics[width=\linewidth]{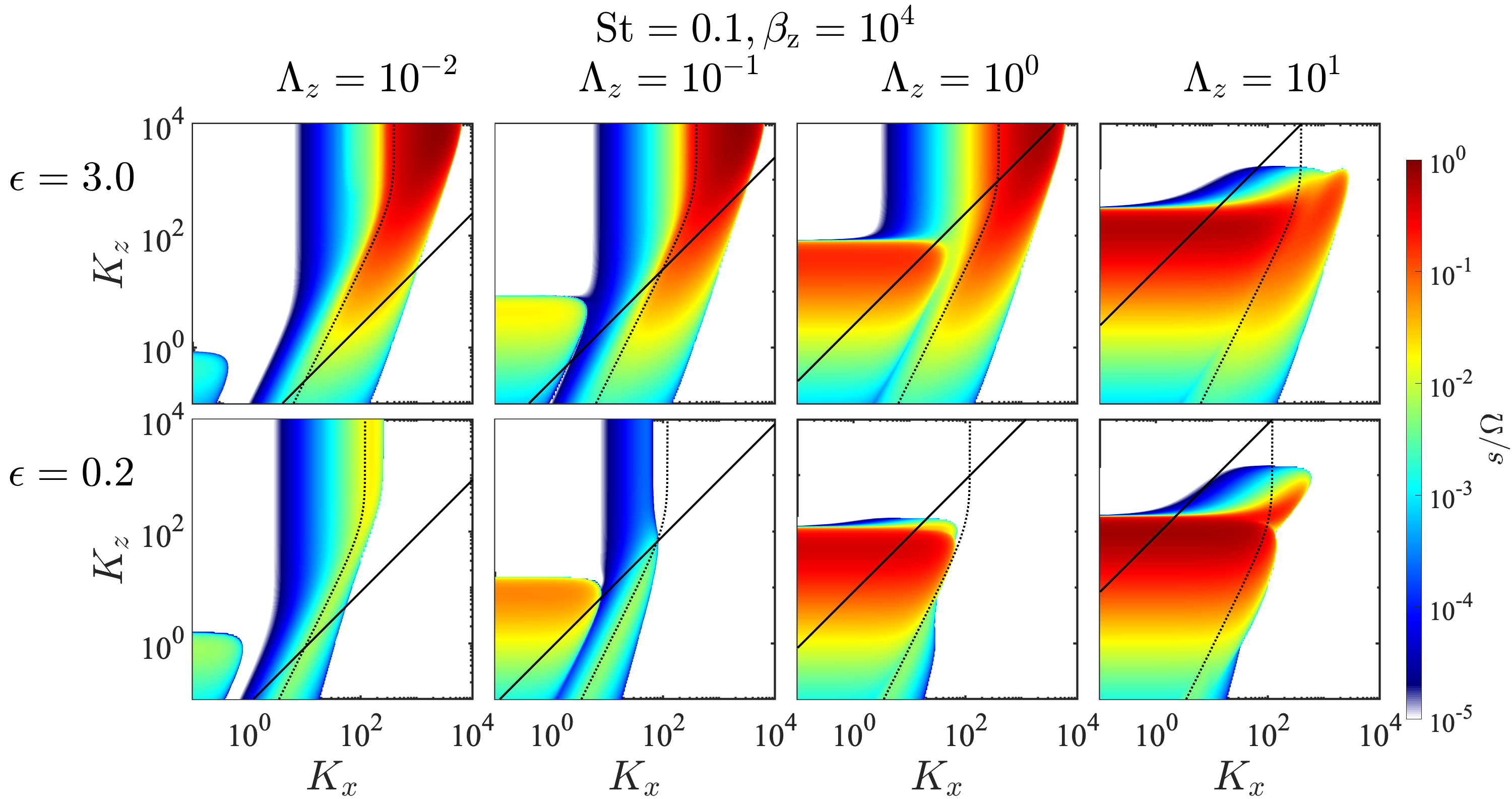}
    \caption{Similar to Fig. \protect\ref{Example_CaseII} but for fixed $\beta_z=10^{4}$ and resistivity decreasing from left to right (increasing Elsasser number $\Lambda_z$). 
    }
    \label{Example_CaseII_resistive}
\end{figure*}

Fig. \ref{Example_CaseII_resistive} show growth rates with fixed $\beta_z = 10^4$ as a function of $K_{x,z}$ and $\Lambda_z$. We find that AwSI modes are easily damped by resistivity: even with $\Lambda_z=10$ there is only a hint of their presence around $K_{x,z}\sim 10^2$ as the `flared' region to the top-right of the MRI modes (this is more pronounced for $\epsilon=0.2$). Note that the RDI condition for the AwSI modes (solid lines) appear to have no relevance to the features in the figure; although coincidentally for $\epsilon=0.2$ and $\Lambda_z=0.1$ they mark the maximum $K_z$ beyond which classic SI modes are strongly stabilized.  

In the $\epsilon=3$ disk the classic SI appears for $\Lambda_z\leq 1$ and its properties are similar in this regime. However, in the $\epsilon=0.2$ disk the SI only appears for $\Lambda_z\leq 0.1$, indicating that slowly-growing SI modes are more easily stabilized by the magnetic field. For $\Lambda_z=0.1$, notice also SI modes are damped for $K_z\gtrsim 10^2$, as mentioned above. 

Thus in resistive, dust-poor disks, there is a lower bound to the scales at which the SI operates, here being $\gtrsim 0.01\Hgas$ in both radial and vertical directions. This is unlike the $\Lambda_z=10^{-2}$ disk, which is effectively unmagnetized, where the SI can operate on much smaller vertical scales as growth rates increase and plateau at large $K_z$.

\begin{figure}
    \centering
    \includegraphics[width=\linewidth]{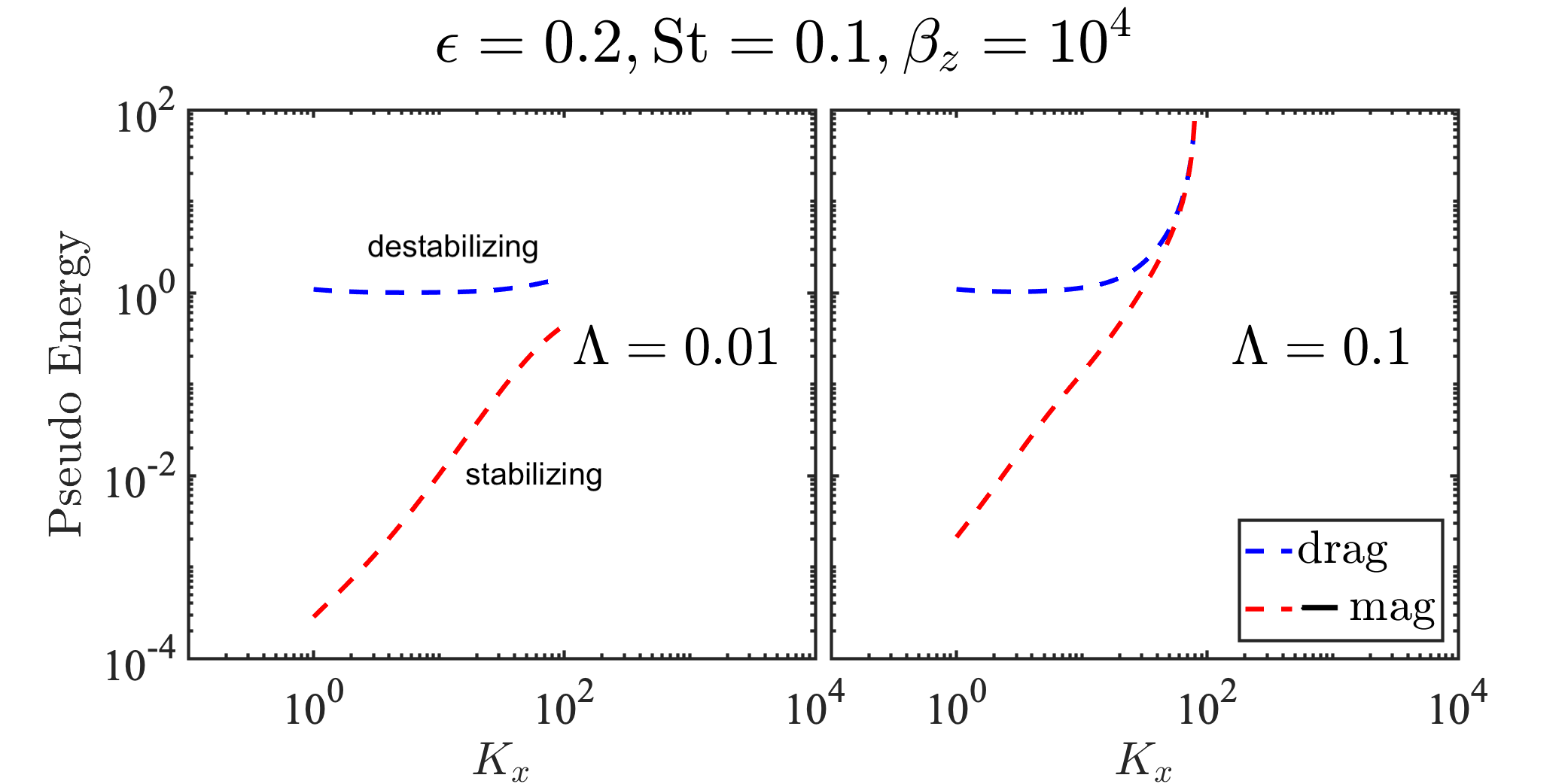}
    \caption{Pseudo-energy decomposition { of} the classic SI in a magnetized disk with $\beta_z=10^4$, $\epsilon=0.2$, and $\st=0.1$. Left: $\Lambda_z=0.01$. Right: $\Lambda_z=0.1$. Blue: the destabilizing contribution from dust-gas drag. Red: { stabilizing} contribution from magnetic perturbations { (the negative of the magnetic pseudo-energy is shown)}. Here, we consider modes with $K_x$ and $K_z$ satisfying the RDI condition for the SI, Eq. \ref{rdi_si_condition}, i.e along the dotted curve in the corresponding panels of Fig. \protect\ref{Example_CaseII_resistive}.}
    \label{caseII_resistive_pseudo_rdi}
\end{figure}

To confirm the magnetic stabilization of the SI, we show in Fig. \ref{caseII_resistive_pseudo_rdi} the pseudo-energy decomposition for the SI modes satisfying the RDI condition for $\epsilon=0.2$ and $\Lambda_z=0.01$, $0.1$. { Note that the resonant $K_z\to\infty$ as $K_x\to 1/|\zeta_x|$ ($\sim 100$ here) according to Eq. \ref{rdi_si_condition}.} As expected drag forces ($U_2$) are destabilizing, but magnetic forces are stabilizing (for ease of comparison we show $-U_3$). The latter effect intensifies with increasing wavenumber, but for $\Lambda_z=0.01$ it is always sub-dominant, so the SI persists. However, for $\Lambda_z=0.1$, both magnetic and drag contributions increase rapidly with $K_{x,z}$, here resulting in a near cancellation and the SI being effectively quenched. { The toy model presented in Appendix \ref{mag_onefluid} also indicate magnetic stabilization at low dust-to-gas ratios for resonant modes with $K_z\gg K_x$.}

\section{Direct simulations}\label{nonlinear}

To verify the new instabilities uncovered above -- namely the azimuthal drift-driven SI without pressure gradients (\S\ref{si_no_pgrad}) and the Alfv\'{en} wave SI (\S\ref{awsi_linear}) -- we also solve the full shearing box equations directly. To this end, we have developed a finite difference code to evolve Eqs. \ref{gas_mass_local}--\ref{dust_mom_local} in their conservative form. We approximate spatial derivatives with a 6$^\mathrm{th}$ order central finite-difference scheme and integrate in time with a 4$^\mathrm{th}$ order Runge-Kutta method. We follow the \textsc{athena} code to treat the source terms related to tidal and Coriolis forces, the large-scale radial pressure gradient, and torques due to a horizontal magnetic field if applicable \citep{stone10}. We integrate the dust-gas drag term explicitly.  

In Appendix \ref{nonlinear_code_test} we test the code by reproducing the standard LinA and LinB SI growth rates, as well as MRI growth rates with and without resistivity. We have found this simple code to be adequate for our primary goal of confirming linear theory. 


\subsection{Numerical setup}

We perform axisymmetric simulations in the ($x,z$) plane but include all three components of the velocity and magnetic fields (the latter in Case II). The domain $x\in[-L_x/2, L_x/2]$ and $z\in[-L_z/2, L_z/2]$ is discretized into $N_x=100$ and $N_z=100$ uniform cells, respectively. To test an eigenmode with wavenumbers $k_x$ and $k_z$, we set the domain size to be one wavelength in each direction, i.e. $L_{x,z} = 2\pi/k_{x,z}$. We adopt $L_x$ as a unit of length and $\Omega^{-1}$ as the unit of time. The velocity normalization is therefore $L_x\Omega\equiv u_\mathrm{norm}$. 
The mass scale is arbitrary for a non-self-gravitating disk, but for convenience, we take $\rhog=1\equiv\rho_\mathrm{norm}$ in the equilibrium. 

We initialize each simulation in steady state consisting of constant densities and drift velocities  as discussed in \S\ref{shear_box}. Following \cite{youdin07}, we add a pair eigenmodes with the same $k_x$ but oppositely signed $k_z$ as a perturbation. The eigenmode amplitude is scaled such that $\dd\rhod = 0.01\rho_\mathrm{norm}$. We apply periodic boundary conditions in $x$ and $z$.

\subsection{Azimuthal drift-driven SI without pressure gradients}


Here, we test for the SI driven by the azimuthal drift induced by magnetic torques from a horizontal field. We adopt the setup of Case I with  $\beta_\phi = 10$ and $\Lambda_\phi  = 10^{-3}$, or equivalently $\alpha_M=0.0125$. We set $\etatilde = 0$ so the equilibrium drift, which is dominated by the azimuthal drift, is entirely due to magnetic torques. For the perturbation we choose $K_x = 5000$ and $K_z=100$, which results in a `tall' shearing box as our domain. We consider $\epsilon=3$ and $0.2$, for which the linearly unstable eigenmodes are labeled `LinAIeta0' and `LinBIeta0' in Table. \ref{table1}. For this test we do not evolve the magnetic field, assuming the  resistivity is sufficiently large such that the field remains passive. 

\begin{figure}
    \centering
    \includegraphics[width=\linewidth]{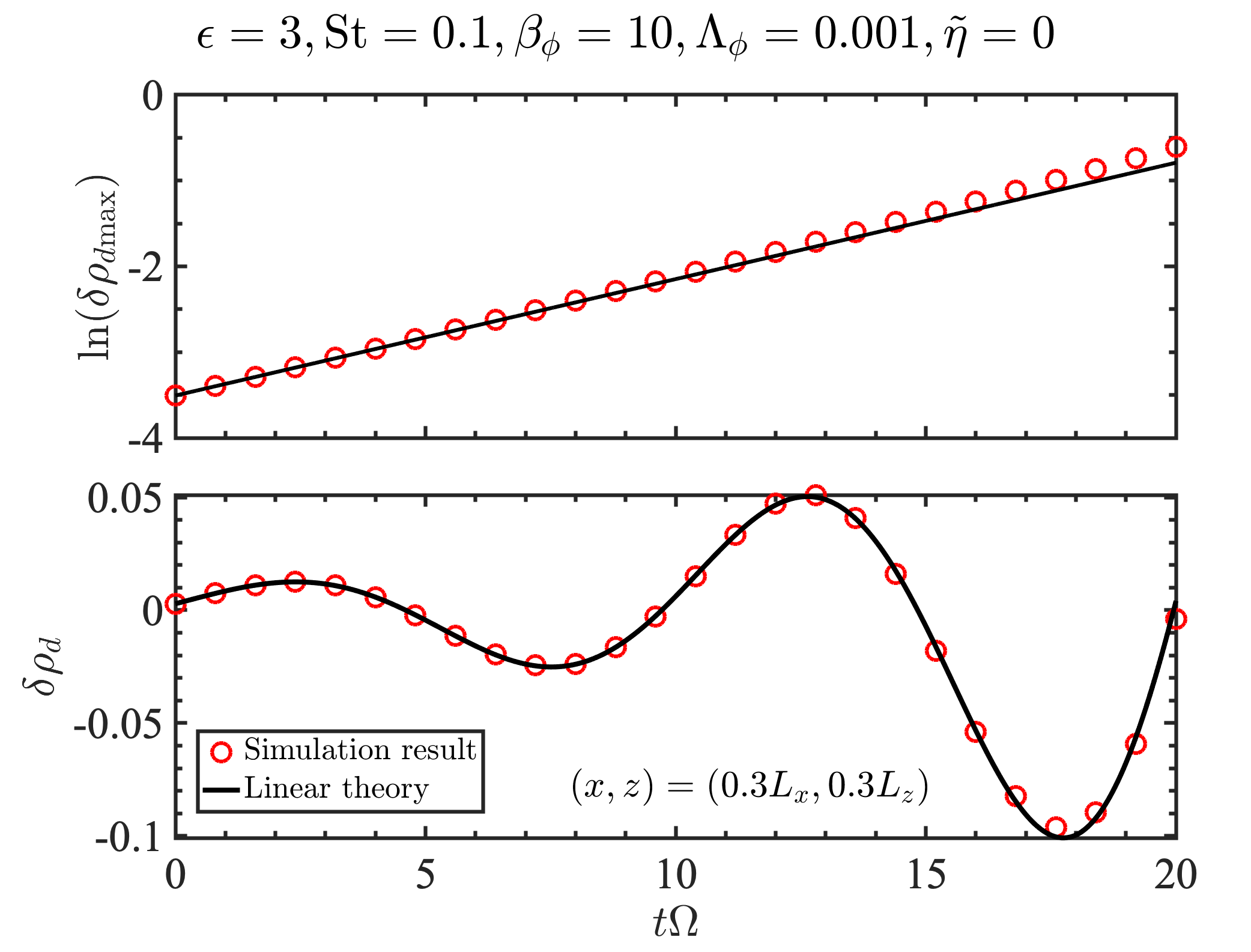}
    \caption{Numerical simulation of the streaming instability without pressure gradients ($\etatilde=0$), driven by the azimuthal drift induced by passive, horizontal magnetic torques with $\beta_\phi=10$, $\Lambda_\phi=10^{-3}$, or equivalently $\alpha_M=0.0125$. The Stokes number (or grain size) $\st$ and dust-to-gas ratio $\epsilon$ are $0.1$ and $3$, respectively.  Top: maximum dust density perturbation in the domain. Bottom: evolution of the dust density perturbation at $(x,z)=(0.3L_x$,$0.3L_z)$. Simulation results are shown in open red circles, while the prediction from linear theory is shown as the solid curves. 
    }
    \label{caseI_simulation}
\end{figure}

\begin{figure}
    \centering
    \includegraphics[width=\linewidth]{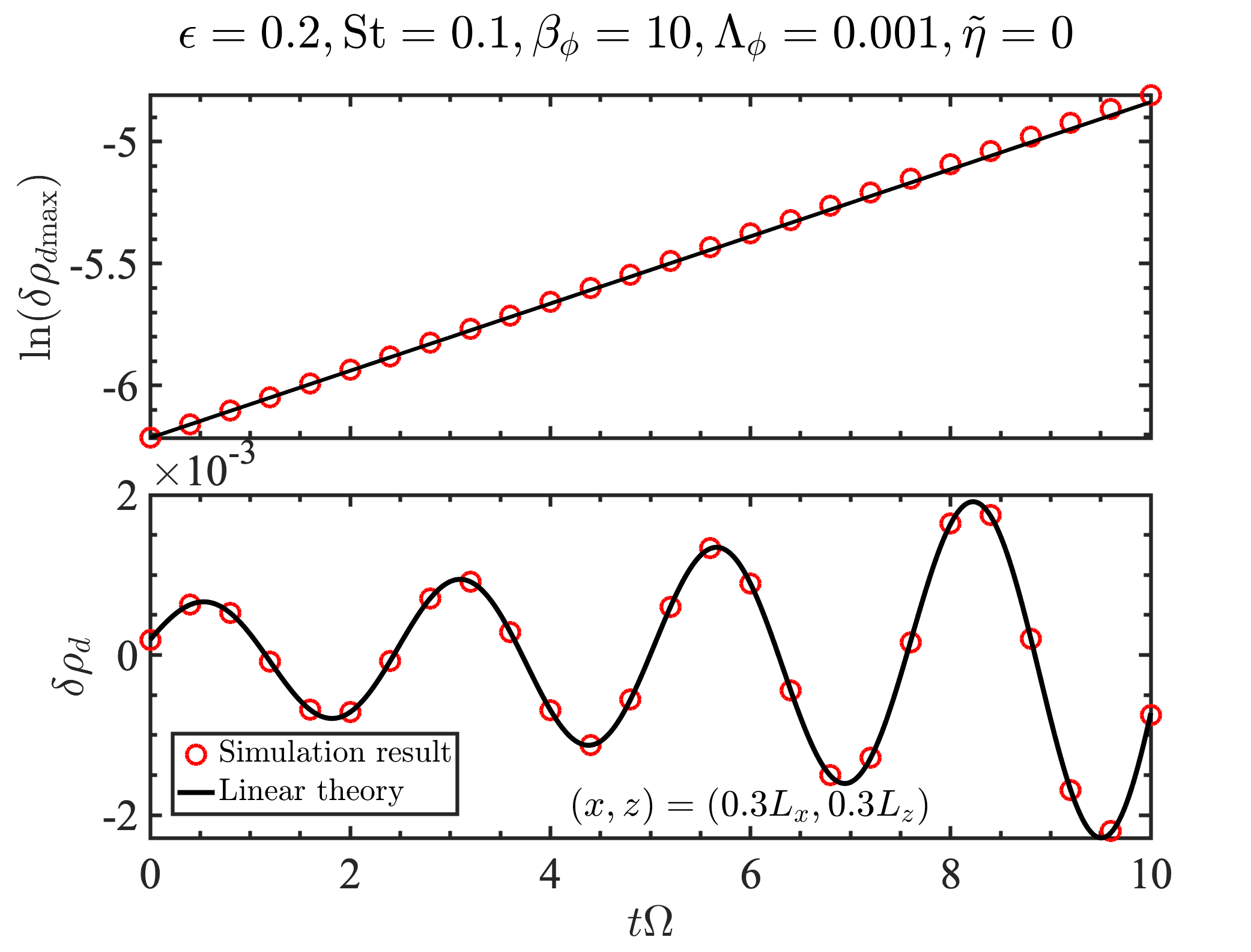}
    \caption{Similar to Fig. \protect\ref{caseI_simulation} except with $\epsilon=0.2$.
    } 
    \label{caseI_simulation2}
\end{figure}

Fig. \ref{caseI_simulation} and \ref{caseI_simulation2} shows the evolution in the perturbed dust density. The upper panel shows the maximum value measured in the domain, while the lower panel shows $w_z$ at a fixed grid point. The measured growth rates $0.1360\Omega$ for $\epsilon=3$ and $0.1380 \Omega$ for $\epsilon=0.2$ are close to that calculated from linear theory ($s=0.1358\Omega$ and $s=0.1377\Omega$, respectively). Furthermore, for $\epsilon=3$ we measure an oscillation period of $T\simeq 10\Omega^{-1}$, which is close to that expected from linear theory as $2\pi/|\omega| =10.233\Omega^{-1}$. Similarly, for $\epsilon=0.2$ we measure $T\simeq 2.6\Omega^{-1}$, compared to the theoretical period of $2.56\Omega^{-1}$. This shows that in dust-poor disks the azimuthal drift-driven SI is highly oscillatory. We find growth rates begin to deviate from linear theory after a few periods. Nevertheless, the close agreement in the early phase of the simulations confirms that the SI can operate without pressure gradients if the gas is subject to passive magnetic torques.  

We note that for large $K_x$ the azimuthal drift-driven SI is insensitive to $K_z$ (see Figs. \ref{caseI_linA_var_eta}-\ref{caseI_linB_var_eta}) and in fact persists with $K_z=0$ (Appendix \ref{azi_si}). Indeed, we have also performed one-dimensional simulations in $x$ only and recovered similar growth rates as above.

\subsection{Alfv\'{e}n wave SI}

Here we adopt the setup of Case II with a live magnetic field to test for the AwSI, which results from a resonance between Alfv\'{e}n waves and the radial dust drift. The induction equation and Lorentz forces are evolved. We consider ideal MHD with an initially vertical field with $\beta_z = 100$ and revert to the nominal pressure gradient $\etatilde=0.05$. The eigenmodes are labeled as `LinAII' and `LinBII' in Table \ref{table1} for $\epsilon=3$ and $0.2$, respectively. We fix $K_x=1000$ but choose $K_z = 25$ for LinAII and $K_z=83$ for LinBII in accordance with the RDI condition given by Eq. \ref{awsi_cond}, i.e. these are resonant wavenumbers. Note that our domain size of one mode wavelength is sufficiently small to exclude the MRI, which operates on larger scales. { However, the MRI
has the larger growth rate (see Fig. \ref{Example_CaseII}), which in reality would dominate and drive rapid disk evolution. The examples here are chosen to confirm the Alfv\'{e}n wave SI.}

\begin{figure}
    \centering
    \includegraphics[width=\linewidth]{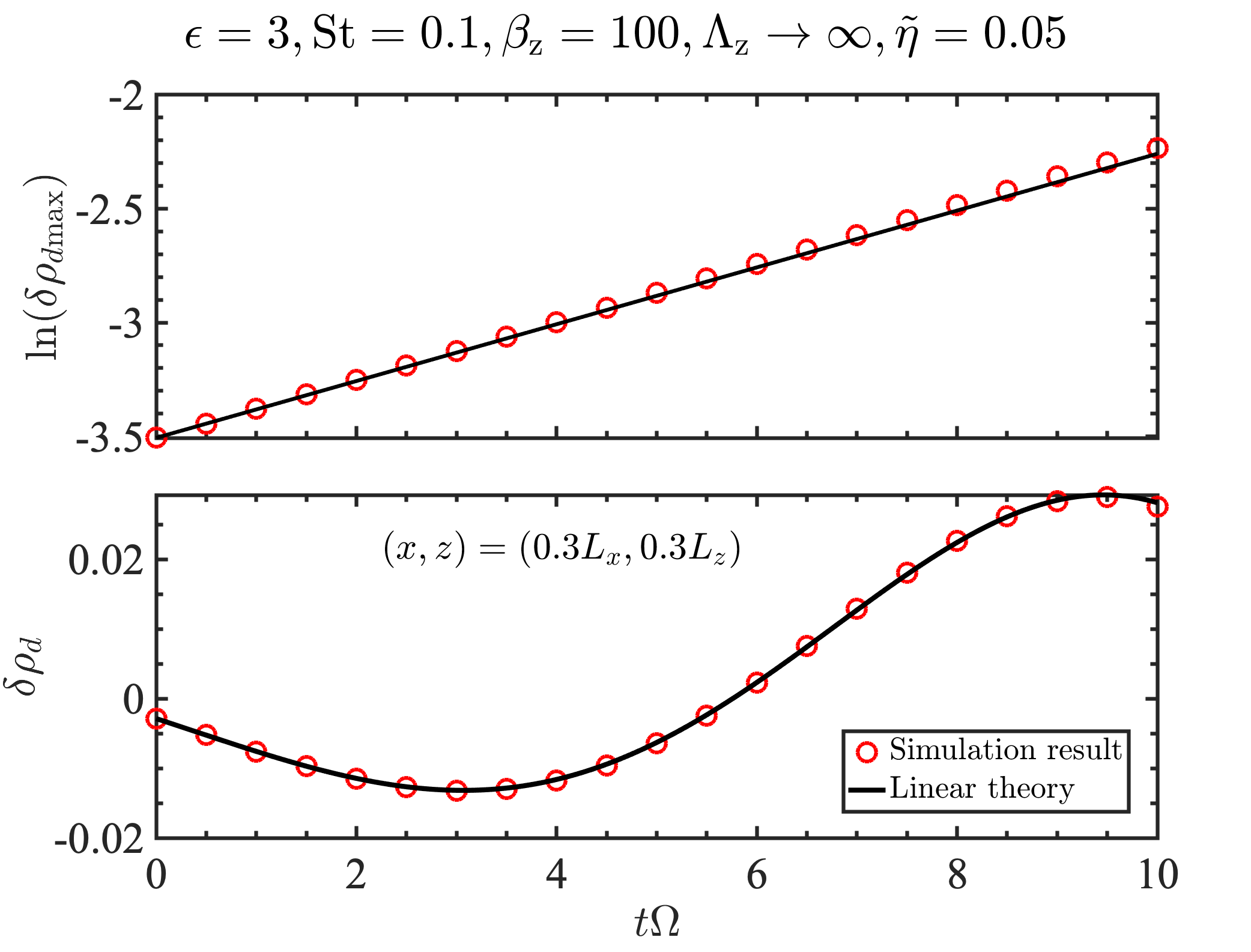}
    \caption{Streaming instability of Alfv\'{e}n waves in a magnetized disk in the limit of ideal MHD. The initial vertical field strength is $\beta_z = 100$ and the reduced pressure gradient is $\etatilde=0.05$. The Stokes number (or grain size) $\st$ and dust-to-gas ratio $\epsilon$ are $0.1$ and $3$, respectively. Top: maximum dust density perturbation in the domain. Bottom: evolution of the dust density perturbation at $(x,z)=(0.3L_x, 0.3L_z)$. Simulation results are shown in open red circles, while the prediction from linear theory is shown as the solid curves. 
    }
    \label{caseII_simulation}
\end{figure}

\begin{figure}
    \centering
    \includegraphics[width=\linewidth]{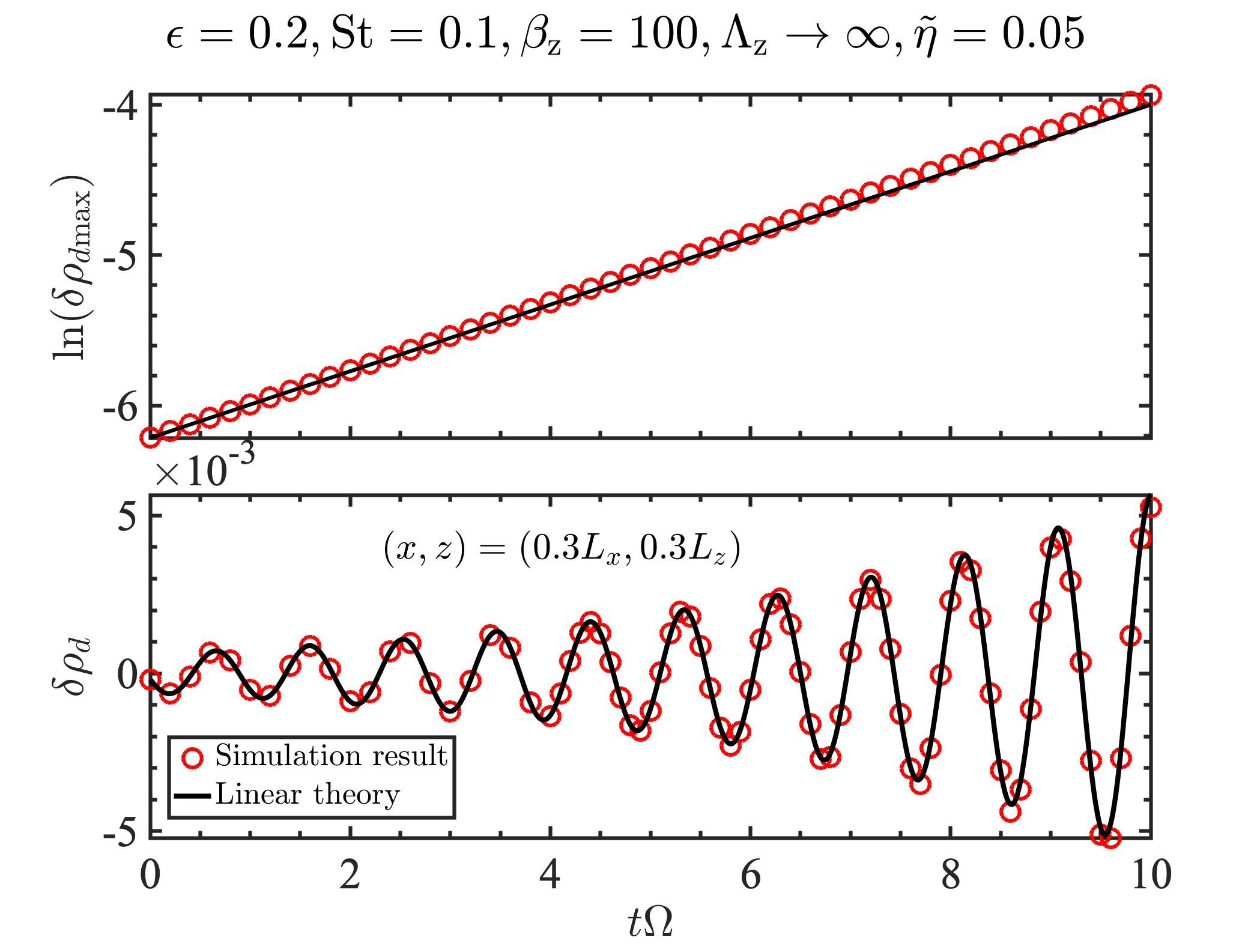}
    \caption{Similar to Fig. \protect\ref{caseII_simulation} except with $\epsilon=0.2$.  
    } 
    \label{caseII_simulation2}
\end{figure}

Figs. \ref{caseII_simulation}-\ref{caseII_simulation2} shows the evolution in the perturbed dust density for the above cases. For LinAII the measured growth rate and oscillation period are $s=0.1250\Omega$ and $T=12.8\Omega^{-1}$; which compares well with the theoretical values of $0.1248\Omega$ and $12.79\Omega^{-1}$. Similarly, for LinBII we measure $s=0.228\Omega$ and $T=0.94\Omega^{-1}$, compared with $0.2215\Omega$ and $0.9353\Omega^{-1}$ from linear theory. 

\section{Discussion}\label{discussion}

\subsection{Planetesimal formation in accreting pressure bumps}\label{discuss_pressure_bump}
We showed that the SI persists in regions of weak or vanishing pressure gradients, provided that there is sufficient azimuthal drift between dust and gas. This suggests that the SI can remain effective inside pressure bumps. 
Although we considered the case of an azimuthal drift resulting from large-scale, horizontal magnetic torques acting on the gas, our results are not limited to this context. The constant azimuthal force $F_\phi$, which represents the background magnetic torque in our models, may also represent other causes of radial gas accretion. For example, as argued by \cite{mcnally17}, if a vertical angular momentum loss (e.g. due to disk winds) resulted in a gas accretion equivalent to that induced by a corresponding $F_\phi$, then our results still apply. 

Using Eq. \ref{azimuthal_si_growth} and Eq. \ref{dimensionless_ydrift}, we can express the growth rate of the azimuthal drift-driven SI at a pressure extremum in terms of the laminar, horizontal stress $\alpha_M$,
\begin{align}\label{nop_si_app}
    \frac{s}{\Omega} = \sqrt{\frac{\epsilon\st}{2(1+\epsilon)^2}\alpha_M \hgas K_x},
\end{align}
where we assumed $\st\ll 1$. We can use this equation to assess the relevance of the SI around pressure bumps, for example in those found by \cite{riols20} in global non-ideal MHD simulations of dusty PPDs. { Recall from Eq. \ref{alphaMdef} that $\alpha_M\propto\beta_\phi^{-2}$ so $s\propto \beta_\phi^{-1}$ when other parameters are fixed.}

In their non-ideal MHD simulations, \citeauthor{riols20} find that gas accretion is largely due to angular momentum removed vertically by a magnetized wind (and therefore occurs near the disk surface), which is a few times larger than that transported radially. The latter is dominated by horizontal magnetic fields with dimensionless stresses of order $10^{-2}$ \citep[see also][]{bethune17}. Hence we consider $\alpha_M\sim 10^{-2}$ below. 

\cite{riols20} also find the spontaneous formation of pressure bumps due to a wind instability \citep{riols19}. In one of their disk models, for a pressure bump at $\sim 20$AU they find $3$mm-sized dust grains (with $\st\sim 0.04$) accumulates into to a ring of width $\sim 0.6\Hgas$ and have settled to a thin layer of thickness $0.25\Hgas$. In the ring, the vertically integrated dust-to-gas ratio (or metallicity $Z$) of all four dust species they considered is about $0.3$. We therefore take $\epsilon\sim 0.3$. We then find $s\sim 10^{-3}\sqrt{K_x}\Omega$ for $\hgas=0.05$, as used in their disk models. The ring width limits $K_x\gtrsim 10$, which implies a growth timescale $\lesssim 50$ orbits. 

Of course, the relative importance between the azimuthal drift-driven SI and the classic, radial drift-driven SI depends on the local pressure gradient, $\etatilde$. It is true that the classic SI formally ceases only where $\etatilde  = 0$ exactly; elsewhere in the pressure bump it can still operate, though on vanishing scales as one approaches the bump center. However, as discussed in \S\ref{dimensionless_drifts}, the torque-induced azimuthal drift is expected to become important when $\etatilde\lesssim \alpha_M\hgas/\st$, which is consistent with numerical results in \S\ref{si_no_pgrad}. We therefore expect a finite region in which the azimuthal drift-driven SI operates. 

To make a crude estimate, we consider a Gaussian pressure bump in the disk midplane, $P\propto e^{-\frac{(R-R_0)^2}{2a^2}}$, where $a$ is the ring width. Neglecting magnetic pressure, the magnitude of the reduced pressure gradient near the bump center is 
\begin{align*}
    \etatilde\simeq \frac{R_0|R-R_0|}{2a^2}\hgas.
\end{align*}
The condition  $\etatilde\lesssim \alpha_M\hgas/\st$ translates to 
\begin{align}
    \frac{|R-R_0|}{\Hgas} < \frac{2\alpha_M}{\st}\frac{a^2}{R_0\Hgas} = \frac{2\alpha_M\hgas}{\st},
\end{align}
where we took $a=\Hgas$ in the second equality. For the above parameters we find azimuthal drift is significant within $|R-R_0|\lesssim 0.025\Hgas$. Hence the longest permissible radial wavelength for the azimuthal drift-driven SI is $0.05\Hgas$, or $K_x\gtrsim 10^2$, for which the growth timescale is $\lesssim 16$ orbits. We can therefore expect rapid instability even close to the bump center.

We emphasize that the SI discussed above requires azimuthal drift. However, it is possible to have a pressure bump without such drift; e.g. those in geostrophic balance, which have been  considered in planetesimal formation simulations \citep[e.g.][]{carrera21a,carrera21b}. In this case, both gas and dust orbit at the Keplerian velocity at the bump center, for which we expect neither the classic nor the azimuthal-drift driven SI. 

{
\subsection{Comparison with secular gravitational instabilities}\label{sgi}

The `Secular Gravitational Instability' \citep[SGI,][]{youdin11,takahashi14,latter17} is another dust-clumping mechanism \citep{tominaga18, pierens21} that may facilitate planetesimal formation \citep{abod19}. Here, dust-gas friction provides an effective `cooling' \citep{lin17} that removes pressure support against gravitational instability  \citep{lin16}\footnote{{ Note that \cite{lin17} erroneously claimed an analogy between SGI and viscous gravitational instabilities. Their Eq. 60 for SGI growth rates is, in fact, distinct from that of the latter \citep[see Eq. 18 of][]{gammie96b}.}}. The SGI does not require large-scale pressure gradients, which also makes it a candidate for planetesimal formation at pressure bumps. It is thus of interest to compare the SGI with the azimuthal drift-driven SI discussed above. 

To this end, in Fig. \ref{sgi_comparison} we calculate SGI growth rates using Eq. 13 of \cite{takahashi14} and compare it with that of the azimuthal drift-driven SI. We consider two disk masses corresponding to Toomre parameters $Q_T=10$ and $100$. A dimensionless dust diffusion coefficient of $\alphass=10^{-9}$ is used without a corresponding gas viscosity as it is neglected in \citeauthor{takahashi14}'s Eq. 13. 

In the $Q_T=10$ disk, the SGI dominates on most scales except the largest ($\sim \Hgas$), which are stabilized by rotation. For $Q_T=100$, the SGI still has the largest overall growth rate, but only dominates for $K_x\gtrsim 60$. These results suggest that the azimuthal drift-driven SI is likely more relevant in low mass disks and at large scales compared to the SGI. When a larger dust diffusion coefficient of $\alphass = 10^{-6}$ is adopted (not shown), we find the instabilities have comparable maximum growth rates ($\sim 10^{-2}\Omega$), but the SI dominates over the SGI for $K_x\lesssim 100$, while both are cut-off for $K_x\gtrsim 200$--$300$. 

Note that the SI model does not include disk self-gravity and the SGI model does not include magnetically-driven azimuthal drifts. A proper comparison between these instabilities in a unified framework is beyond the scope of this paper but should be conducted in the future.

\begin{figure}
    \centering
    \includegraphics[width=\linewidth]{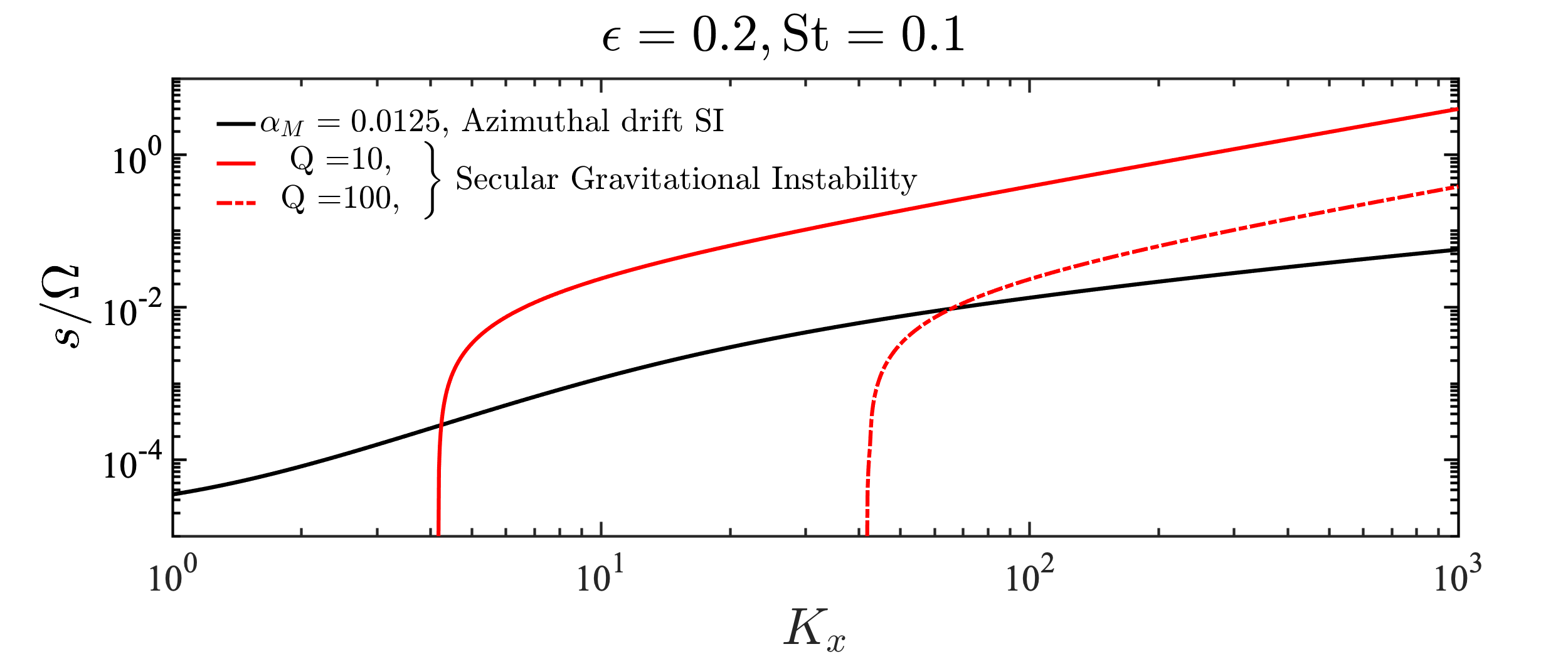}
    \caption{{ Comparison between the SGI (red) and the azimuthal drift-driven SI (black) as a function of radial wavenumbers $K_x$. SGI growth rates are computed from Eq. 13 of \cite{takahashi14} with Toomre parameters $Q_T=10$ (solid) and $Q_T=100$ (dash-dot). The dust-to-gas ratio and Stokes number are fixed to $\epsilon=0.2$ and $\st=0.1$, respectively. Neither disk models have radial pressure gradients ($\etatilde=0$). Modes have no vertical structure, $K_z=0$.}
    } 
    \label{sgi_comparison}
\end{figure}
}

\subsection{SI in off-midplane accreting layers}

Another region of vanishing pressure gradients is off of the disk midplane. To see this, consider a standard Gaussian atmosphere, then the three-dimensional pressure profile is 
\begin{align}
    P_\mathrm{3D}(R,z)  = P(R)\exp{\left(-z^2/2\Hgas^2\right)}. 
\end{align}
The corresponding dimensionless pressure gradient is
\begin{align}\label{eta_3D}
    \eta_\mathrm{3D} = \left(\etatilde - \frac{\hgas}{2}\frac{z^2}{\Hgas^2}\frac{\p\ln\Hgas}{\p\ln R}\right)\hgas.
\end{align}
If $\etatilde > 0$, i.e. a negative midplane pressure gradient, then at some height $\eta_\mathrm{3D}$ will vanish since $\Hgas$ increases with $R$. For $\Hgas \propto R$ and $\etatilde = 0.05$, we find $\eta_\mathrm{3D}\to 0$ as $|z| \to \sqrt{2}\Hgas$. Of course, the dust-to-gas ratio will be small at this altitude due to dust settling, especially for large grains. We may thus only expect small grains there. However, Eq. \ref{nop_si_app} shows that growth rates are not strongly dependent on these parameters. Indeed, even for $\epsilon=0.01$ and $\st=0.01$ we find $s\sim 10^{-4}\sqrt{K_x}\Omega$ (assuming $\alpha_M=0.01$ and $\hgas =0.05$), so for $K_x\sim 100$ the growth timescale $\sim 160$ orbits is still short compared to disk lifetimes. However, a small dust-to-gas ratio may only lead to order-unity turbulent fluctuations rather than dust clumping \citep{johansen07}.

Note that for $z\neq 0$ complications can arise from the `Dust Settling Instability' (DSI): an RDI associated with a resonance between inertial waves and the dust's vertical settling motion \citep{squire18b,krapp20}. However, the DSI requires $K_z\neq 0$, while the azimuthal drift-driven SI can operate for $K_z=0$. We thus expect vertically extended modes to be unaffected by settling. 

{ We remark that the SGI is probably unimportant in the disk layers away from the midplane due to low values of $\epsilon$ and high values of $Q_T$ locally \citep{takahashi14}.} 

\subsection{Dust feedback and the ideal MRI}\label{discuss_feedback_mri}

In a realistically stratified disk and considering ideal MHD, MRI modes can only develop if they fit into the gas disk thickness, $2\Hgas$. Now, the characteristic MRI vertical wavenumber is $k_z\sim \Omega/C_{Az,\mathrm{eff}}$, where the effective Alfv\'{e}n speed $C_{Az,\mathrm{eff}} =  (1+\epsilon)^{-1/2}C_{Az}$ decreases with dust-loading since the total density increases. Then for a given field strength 
$\beta_z$ in the dust-free limit, we require $k_z > \pi/\Hgas$, or 
\begin{align}\label{betaz_crit_ideal}
    \beta_z \gtrsim \frac{10}{1+\epsilon} 
\end{align}
upon dust-loading for the ideal MRI to operate. Thus if the MRI operates without dust, it will continue to operate in the presence of dust, although it will shift to smaller vertical scales. While the maximum growth rate remains at $0.75\Omega$, growth rates for modes with $k_z\lesssim \Omega/C_{Az}$ are somewhat reduced  by dust loading, as shown in Fig. \ref{caseII_ideal_MRI}. 

Recently, \cite{yang18} carried out stratified MHD simulations including dust and found that the particles sediment to a thinner layer when the metallicity $Z$ is increased. They attributed this to the reduction in the effective sound-speed of the dusty gas \citep{lin17}. We point out another possible contributing factor from dust feedback onto the MRI. 

For ideal MHD simulations without particle feedback, \citeauthor{yang18} find a turbulence strength of $\alpha_\mathrm{SS}\sim 10^{-2}$. Using the diffusion model of \cite{dubruelle95}, the expected particle scale height is $\Hd \simeq \sqrt{\alphass/\st}\Hgas$ if $\alphass\ll \st$. For $\st=0.1$,  this gives $\Hd\simeq0.3\Hgas$, as found in the authors' simulations. The most unstable MRI wavelength $\lambda_{z,\mathrm{opt}} \sim 2\pi\Hgas/\sqrt{\beta_z}$. Then for $\beta_z=10^4$ we find $\lambda_{z,\mathrm{opt}} \simeq 0.06\Hgas \ll 2\Hd$. This means that, upon the introduction of dust feedback, MRI modes with vertical lengthscales comparable to the dust layer thickness (i.e. those having $k_z \lesssim \Omega/C_{Az}$, see Fig. \ref{caseII_ideal_MRI}) will be damped, which may promote further settling.  

In practice, however, feedback is only important if the local $\epsilon$ is $O(1)$ or larger, so this effect should only be appreciable at high $Z$. Indeed, \citeauthor{yang18} find  $\Hd$ is only reduced by a factor of $1.5$ for an 8-fold increase in $Z$ to $0.08$ from its canonical value of $0.01$.


\subsection{Dust feedback and the resistive MRI}
In resistive disks the characteristic MRI vertical wavelength is $k_z\sim C_{Az,\mathrm{eff}}/\eta_O$. Thus for fixed $\eta_O$, defined through $\Lambda_z$ in the dust-free limit, requiring the MRI to fit inside the gas disk yields
\begin{align}\label{betaz_crit_resis}
    \beta_z \gtrsim \frac{10}{\Lambda_z^2}(1+\epsilon). 
\end{align}
This condition becomes more difficult to fulfill as $\epsilon$ increases, which reflects the increasing vertical length scale. Furthermore, growth rates are damped on all length scales by dust feedback (Fig. \ref{caseII_nonideal_MRI}).

The result above suggests the possibility of self-sustained dust settling as follows. Consider a dusty disk undergoing resistive MRI turbulence but without dust feedback. For reference, in such a simulation \cite{yang18} measured a $\Hdust = 0.2\Hgas$ for $\st=0.1$ and $\beta_z\sim 10^4$ in quasi-steady state as settling is balanced by turbulent stirring. Switching on feedback weakens the MRI, so the dust grains begin to settle, which further increases $\epsilon$ and weakens the MRI. This positive feedback loop cannot continue indefinitely, however, as the gaseous layers above and below the dusty midplane can still undergo full (resistive) MRI turbulence and stir the particle layer \citep{fromang06}. Nevertheless, we can expect thinner dust layers in higher metallicity disks because the MRI is progressively suppressed. This scenario is similar to that outlined for the Vertical Shear Instability in the presence of dust \citep{lin19}.

A shortcoming of the above discussion (including \S\ref{discuss_feedback_mri}) is that the estimates given in Eqs. \ref{betaz_crit_ideal} and \ref{betaz_crit_resis} assumes a uniform dust-to-gas ratio. In the stratified context, $\epsilon$ could be taken as the average value over the vertical column, in which case $\epsilon\sim Z\ll 1$ and thus feedback should always be negligible in an averaged sense. 

On the other hand, we can apply a similar argument for fitting MRI modes within the dust layer, wherein $\epsilon$ may reach order unity or larger. Doing so, we find the critical $\beta_z\to\beta_z(\Hgas/\Hd)^2$, implying it is difficult to have MRI modes confined to a highly settled dust layer. This does not prevent MRI modes to operate in the gas outside the dust layer, however. Stratified analyses are required to study MRI modes with vertical wavelengths that exceed the dust layer thickness, yet still fit within the gas disk.  

\subsection{Magnetic stabilization of the classic SI}
In \S\ref{awsi_linear}--\ref{caseII_resis} we found that a live vertical magnetic field can suppress the classic SI, which is a stronger statement than MRI modes outgrowing the SI. For ideal MHD, there is no trace of the classic SI for $\beta_z\leq 10^4$. While the classic SI does appear at $\beta_z = 10^{6}$, the most unstable modes are mixed with the MRI. It is only with $\beta_z=10^{8}$, i.e. an extremely weak field, and with $\epsilon=3$, do we find distinct classic SI modes that dominate the system. This observation, together with the fact that we considered rather large grains with $\st=0.1$, which should favor the SI, suggests that the classic SI is sensitive to magnetic stabilization. 

As a consequence, for typical values of $\beta_z=10^4$ adopted for PPDs, the classic SI is only present with sufficient resistivity. 
For $\epsilon=0.2$ we only partially recover the SI with $\Lambda_z\leq 0.1$ (high vertical wavenumbers are still damped), and $\Lambda_z=0.01$ is needed for SI modes to outgrow the MRI. 
At high dust-to-gas ratios, magnetic stabilization is less effective as we find the SI dominates over the MRI with $\Lambda_z\leq 1$ when $\epsilon=3$. However, should such a high $\epsilon$ be attained through dust-settling, then (MRI) turbulence cannot be too strong in the first place, which sets an upper bound on $\Lambda_z$. Therefore, whether or not the SI can occur in magnetized disks is still determined by the value of $\Lambda_z$ below which either the MRI wavelengths exceed the disk scale height (and therefore does not operate), or when SI growth rates exceed the resistive MRI.

\subsection{Dust clumping in MHD-turbulent disks}

At present, the connection between the linear SI and non-linear clumping -- and hence planetesimal formation -- is still unclear (Lesur et al., submitted). Nevertheless, our results suggest that dust clumping directly via the SI may be more difficult in magnetized disks due to magnetic stabilization. This is in addition to stirring by MRI turbulence, which may prevent dust sedimentation. 

On the other hand, planetesimal formation has been successfully simulated in MRI-turbulent disks \citep{johansen07b, johansen11}, but this is not necessarily in contradiction with our results. This is because of the formation of zonal flows or pressure bumps in MRI-turbulent disks  \citep{johansen09b} that can trap dust \citep{dittrich13,xu21}. Similar radial dust concentrations can develop in models with dead zones owing to the weak radial diffusion of particles \citep{yang18}. In either case, the enhanced dust-to-gas ratio in these dust traps should, according to our results, reduce the effect of the MRI and favor further concentration and eventually allow the classic SI to operate, though perhaps not at the exact center of pressure bumps.

\subsection{Relevance of the Alfv\'{e}n wave SI in PPDs}

Our numerical results indicate that the Alfv\'{e}n wave SI requires nearly ideal MHD conditions to operate. This makes their relevance to PPDs questionable, since if ideal MHD conditions are met, for example in the inner disk \citep{flock16}, the MRI would generate vigorous turbulence and overwhelm the Alfv\'{e}n wave SI. 

One possible exception is in the disk atmosphere where $\beta_z$ increases relative to the midplane as the gas density drops. If sufficiently magnetized, i.e. if  $\beta_z\lesssim 10$ (Eq. \ref{betaz_crit_ideal}), then the MRI is stabilized, leaving AwSI modes to survive. As an example, suppose $\beta_z\sim 10^4$ at $z=0$, then at $|z|=4\Hgas$ we find $\beta_z\sim 3$ for a Gaussian atmosphere. Using Eq. \ref{eta_3D}, at this height the reduced pressure gradient parameter is $\etatilde_\mathrm{3D}=\eta_\mathrm{3D}/\hgas \simeq -0.35$, which is much larger in magnitude than the midplane value. For this estimate we assumed $\Hgas = \hgas R$ with $\hgas=0.05$. 

\begin{figure}
    \centering
    \includegraphics[width=\linewidth]{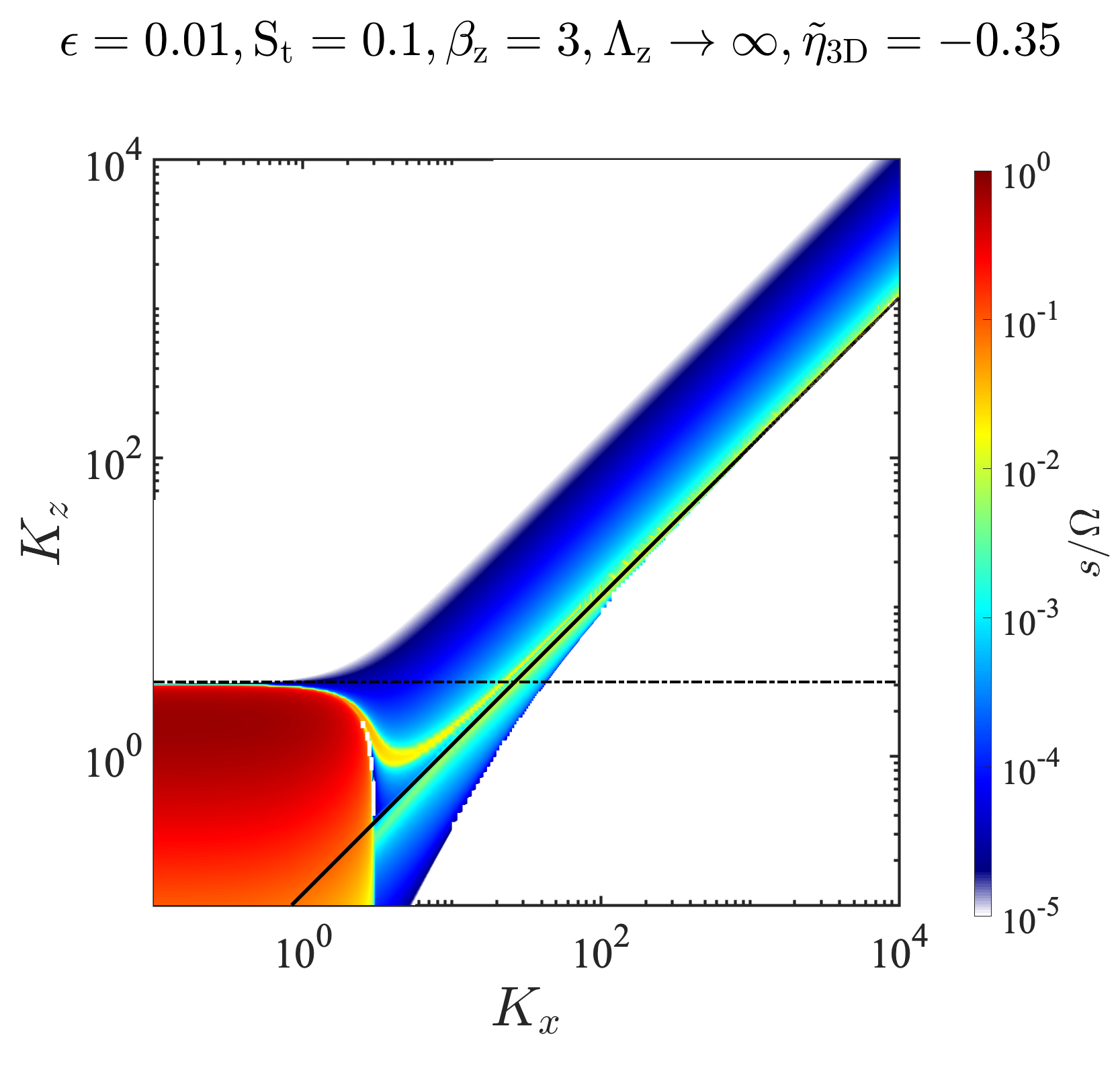}
    \caption{Unstable modes in a strongly magnetized disk atmosphere with $\beta_z=3$ and grains with $\st=0.1$ and $\epsilon=0.01$. Modes with large wavenumbers along the solid line  correspond to Alfv\'{e}n wave streaming instabilities. The horizontal dashed-dotted line ($K_z=\pi$) is the minimum wavenumber for modes to fit within a pressure scale height, which excludes the block of MRI modes (in red) in the lower-left.} 
    \label{awsi_application}
\end{figure}

In Fig. \ref{awsi_application} we plot growth rates for the above example under ideal MHD conditions with $\epsilon=0.01$ and $\st = 0.1$. As expected the MRI is excluded from geometric considerations, but the atmosphere is unstable to the Alfv\'{e}n wave SI with maximum growth rates $\simeq 0.03\Omega$, which may lead to turbulent activity. However, the DSI also operates off of the midplane \citep{squire18b,krapp20}, but is neglected here. How the DSI interacts with 
the Alfv\'{e}n wave SI, or how it is affected by magnetic fields, is beyond the scope of this work.

\subsection{Caveats and outlook}

In the shearing box formalism, the effect of the radial pressure gradient from the global disk is treated through the constant parameter $\etatilde$. We therefore cannot model pressure bumps as a whole. The persistence of the SI with $\etatilde\to0$ only indicates instability localized to regions of vanishing pressure gradients, for example near the center of a pressure bump. A proper stability analysis of pressure bumps requires a variable $\etatilde = \etatilde(x)$. For consistency with global disks, one may also need a variable magnetic torque, $F_\phi = F_\phi(x)$. These generalizations inevitably result in a global eigenvalue problem, i.e. ordinary differential equations in $x$. This should be tackled in a future study. 

Our results demonstrate the potential importance of the azimuthal drift between dust and gas, especially in disks torqued by a laminar magnetic field. According to RDI theory, non-axisymmetric disturbances may grow from a resonance between the azimuthal drift and neutral waves in the gas, for example, Rossby waves \citep{pan20a}. This hypothetical RDI is excluded in our axisymmetric models but should be investigated and compared with the axisymmetric, azimuthal drift-driven SI discussed in this work (which is not an RDI). However, non-axisymmetric modes may not grow exponentially in the shearing box due to differential rotation \citep{johnson05}, so a radially global or cylindrical treatment would be more appropriate. 

We have only considered Ohmic resistivity as the sole non-ideal MHD effect. For passive fields, this is not a limitation since only the effective body force $F_{R,\phi}$ acting on the gas is relevant, regardless of how it arises. For live fields, however, one should also consider  ambipolar diffusion and the Hall effect, which may, in fact, dominate the midplane of PPDs \citep[see][and references therein]{lesur20}. The Hall effect leads to qualitatively new phenomena such as whistler waves and the Hall-shear instability \citep{kunz08}; while ambipolar diffusion can make unstable modes with $k_x\neq0$ dominant \citep{kunz04}. 
The present work should be generalized to study how dust interacts with these modes and instabilities. %

Finally, our results are based on the fluid approximation of dust grains. However, in some calculations, we find modes with high frequency or growth rates such that $|\sigma\taus|\gtrsim O(1)$, which may put the fluid treatment into question \citep{jacquet11}. Ultimately, our results need to be checked against particle-gas simulations \citep[e.g.][]{johansen07}. 

\section{Summary and conclusions}\label{summary}

In this paper, we study the stability of dusty, magnetized protoplanetary disks (PPDs). 
We are mainly motivated by recent models of PPDs that highlight the dominant role of large-scale magnetic fields in driving disk accretion \citep[e.g.]{gressel15,bai17,bethune17} and how this would affect planetesimal formation through the streaming instability \citep[SI,][]{youdin05,johansen07}. We apply standard linear stability analyses and verify key results with direct (magneto-) hydrodynamic (MHD) simulations including dust. 

We extend previous local shearing box models of the SI to account for large-scale magnetic stresses from the global disk, which is treated as a constant, passive torque that modifies the equilibrium dust and gas drift velocities. Here, we find the magnetic torque primarily induces an azimuthal drift between dust and gas, which becomes important in regions where the global radial pressure gradient $\etatilde$ is small. This results in instability even when $\etatilde\to 0$ (in which case the classic SI of \citeauthor{youdin05} ceases to operate). The existence of azimuthal drift-driven SIs suggests that pressure bumps in accreting disks remain viable sites for accelerated planetesimal formation. 

We also consider how a live magnetic field interacts dynamically with dust. Under ideal MHD we find the radial drift between dust and gas can destabilize Alfv\'{e}n waves at sufficiently high wavenumbers. However, these modes are likely overwhelmed by the magneto-rotational instability (MRI) and therefore may be of limited relevance to realistic PPDs, unless the MRI becomes sub-dominant, for example in strongly magnetized disks. In resistive disks, we find dust feedback can stabilize the MRI by reducing the effective Alfv\'{e}n speed of a dusty gas compared to pure gas. Conversely, we find the classic SI can be stabilized by magnetic perturbations, especially at low dust-to-gas ratios. Whether clumping observed in simulations of dusty, magnetized gas can be directly attributed to the SI is, therefore, an open question. 

In a follow-up work, we will use numerical simulations to explore the nonlinear evolution of the azimuthal drift-driven SI, as well as the classic SI and MRI in dusty, magnetized disks. 


\acknowledgements
We are grateful to Xuening Bai and Pinghui Huang for helpful advice on nonlinear simulations. {We thank the anonymous referee for an insightful report that prompted us to compare our results with the SGI.} This research is supported by the Ministry of Science and Technology of Taiwan (grants 107-2112-M-001-043-MY3, 110-2112-M-001-034-, 110-2124-M-002-012-) and an Academia Sinica Career Development Award (AS-CDA-110-M06). { Some numerical simulations were carried out on the TIARA cluster at ASIAA.}

\appendix \label{appendix}

\section{Global steady state}\label{global_density}

We derive the global pressure profile in a steady state disk with velocity fields given by Eqs. \ref{eqm_vr}--\ref{eqm_wphi} at each $R$. We assume $\st$ and $\epsilon$ are both constants. Then the condition for a constant mass flux in dust and gas is equivalent. From Eq. \ref{eqm_vr}, the gas mass flux is 
\begin{align}
RV_R^\prime\rhog = \frac{2 \epsilon \st}{\Delta^2} \etatot \rhog R^2 \OmK + \frac{2\left(\st^2 + \epsilon +1\right)}{\Delta^2}\frac{F_\phi}{\OmK}R\rhog. 
\end{align}
From Eq. \ref{eqm_br}, and \ref{eqm_bphi}, \ref{lorentz_force}, we see that $F_\phi\propto R^{-3/2}/R\rhog$, so the second term on the right hand side is a constant. That is, the magnetic torque induces a constant mass flux (see also Eq. \ref{mgdot}). We therefore require the first term to be constant. This implies, using Eq. \ref{etatot} with  \ref{lorentz_force}, that 
\begin{align}
    \left(\eta\frac{\rhog}{\rhogref} -\mathcal{K}\right)R^{1/2} = \mathcal{C}R_0^{1/2}, \label{const_mass_flux_cond}
    \end{align} 
where 
\begin{align}
    \mathcal{K} \equiv \frac{\hgref^2}{4\beta_\phi},\quad 
    \mathcal{C} \equiv \eta_0 - \mathcal{K}
\end{align}
are dimensionless constants and subscript `0' denotes evaluation at $R_0$. Using the definition of $\eta$ (Eq. \ref{eta_def}), Eq. \ref{const_mass_flux_cond} is an ordinary differential equation for $P$,
\begin{align}
    \frac{1}{2R\OmK^2\rho_\mathrm{g0}}\frac{\p P}{\p R} + \mathcal{K} + \mathcal{C}\left(\frac{R}{R_0}\right)^{1/2}=0.
\end{align}
This can be integrated with the boundary condition $P\to 0$ as $R\to\infty$ to give
\begin{align}
    P(R) = 2\rho_\mathrm{g0}\OmK^2R^2\left[\frac{2}{3}\eta_0\sqrt{\frac{R_0}{R}} + \mathcal{K}\left(1 - \frac{2}{3}\sqrt{\frac{R_0}{R}}\right)\right].
\end{align}
The corresponding gas density profile can be obtained through $\rhog = P/C_s^2$ for an isothermal disk. Note that we require $\eta_0>\mathcal{K}$ to ensure $P>0$ at all radii. This is  satisfied in a thin, weakly magnetized disk wherein $\eta_0$ is $O(\hgref^2)$ and $\beta_\phi\gg 1$.

\section{Linearized equations}\label{linear_eqns}

\begin{figure}
    \centering
    \includegraphics[width=\linewidth]{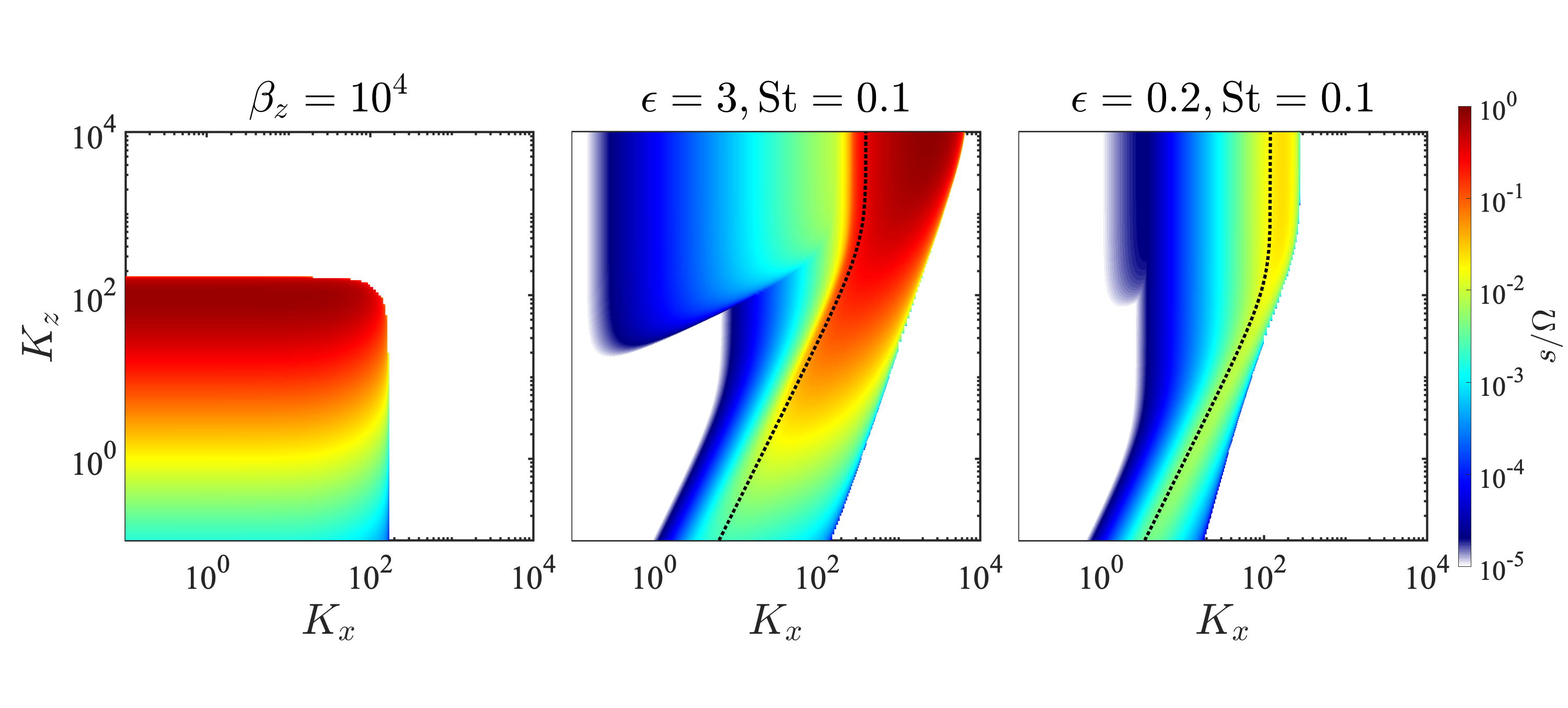}
    \caption{Left: the standard MRI in a pure gas disk with $\beta_z=10^4$ in the ideal MHD limit. Middle: classic SI in an unmagnetized disk with $\epsilon=3$ and $\st=0.1$. Right: similar to the middle panel but with $\epsilon=0.2$. In the SI panels the dotted curves correspond to the RDI condition given by Eq. \ref{rdi_si_condition}.
    }
    \label{Example_MRI_SI}
\end{figure}

We linearize Eqs. \ref{gas_mass_local}--\ref{dust_mom_local} assuming a uniform, vertical background field $B_z$ (Case II). 
We define $W \equiv \dd\rhog/\rhog$, $Q\equiv  \dd\rhod/\rhod$, and $\dd\bm{b} = (\mu_0  \rho_{g})^{-1/2}\dd\bm{B}$. The result is 

\begin{align} 
&\sigma W = -\ii k_x v_x W  - \ii k_{x} \delta v_x - \ii k_{z} \delta v_{z}, \label{lin_gas_mass}\\ 
&\sigma \delta v_x = -\ii k_x v_x \delta v_x + 2 \Omega \delta v_y - \ii k_x C_s^2 W - \frac{\epsilon}{\taus}(w_{x}-v_{x})(W - Q)+ \frac{\epsilon}{\taus}(\delta w_x - \dd v_x)  + \ii C_{Az}(k_z \delta b_x - k_x \delta b_z)  \notag \\ 
& \phantom{-\ii \sigma \delta v_x =}
- \nu k^2 \dd v_x, \label{lin_gas_mom_x}\\
&\sigma \delta v_y = -\ii k_x v_{x} \delta v_y - \frac{\Omega}{2} \delta v_x - \frac{\epsilon}{\taus}(w_{y}-v_{y}) (W-Q) + \frac{\epsilon}{\taus} (\delta w_y - \delta v_y) + \ii k_z C_{Az} \delta b_y - \nu k^2 \dd v_y,  \label{lin_gas_mom_y}\\
&\sigma \delta v_z = -\ii k_x v_{x} \delta v_z   + \frac{\epsilon}{\taus} (\delta w_z - \delta v_z) - \ii k_z C_s^2 W -\nu k^2\dd v_z, \label{lin_gas_mom_z}\\
&\sigma \delta b_x = -\ii k_x v_{x} \delta b_x +  \ii k_z C_{Az} \delta v_x - \eta_O k^2 \delta b_x, \label{lin_mag_x}\\
&\sigma \delta b_y = -\ii k_x v_{x} \delta b_y + \ii k_z C_{Az} \delta v_y -\frac{3}{2} \Omega \dd b_x - \eta_O k^2 \delta b_y, \label{lin_mag_y}\\
&\sigma \dd b_z = -\ii k_x v_{x} \delta b_z - \ii k_x C_{Az} \delta v_x - \eta_O k^2 \delta b_z,  \label{lin_mag_z}\\
&\sigma Q = -\ii k_x w_{x} Q -\ii k_x \delta w_x -\ii k_z\delta w_z - D k^2 (Q-W), \label{lin_dust_mass}\\
&\sigma \delta w_x = -\ii k_x w_{x} \delta w_x + 2 \Omega \delta w_y  - \frac{1}{\taus} (\delta w_x - \delta v_x),\label{lin_dust_mom_x}\\
&\sigma \delta w_y = -\ii k_x w_{x} \delta w_y - \frac{\Omega}{2} \delta w_x  - \frac{1}{\taus} (\delta w_y - \dd v_y), \label{lin_dust_mom_y}\\
&\sigma \delta w_z = -\ii k_x w_{x} \delta w_z - \frac{1}{\taus} (\delta w_z - \delta v_z), \label{lin_dust_mom_z}
\end{align}
where $k^2 = k_x^2 + k_z^2$. Corresponding equations for Case I are obtained by neglecting the magnetic field perturbations and the linearized induction equation (\ref{lin_mag_x}--\ref{lin_mag_z}). { We reproduce the ideal MRI and the classic SI in Fig. \ref{Example_MRI_SI}. In Eqs. \ref{lin_gas_mom_x}--\ref{lin_gas_mom_z} and \ref{lin_dust_mass} we have included an optional gas viscosity term $\propto \nu$ and a corresponding dust diffusion term $\propto D$, which are set to $\alphass C_s\Hgas$ when employed. In \S\ref{si_no_pgrad} we include both diffusion and viscosity, whereas in \S\ref{sgi} and \S\ref{azi_si}) we only include diffusion. Elsewhere, neither effects are included.}    

\section{{ Reduced} model for the SI driven by azimuthal drift}\label{azi_si}

{ We find that the SI can operate without pressure gradients, in which case it is driven by the azimuthal drift between dust and gas (\S\ref{si_no_pgrad}). Here,} we seek to verify this azimuthal drift-driven SI. We consider the setup for Case I, i.e. a purely hydrodynamic model with the magnetic field only modifying the background drift speeds. 

First, we note that this new instability can operate even when $K_z=0$ \citep[contrary to the classic SI, which requires $K_z\neq0$,][]{youdin05}. Henceforth we consider $K_z=0$. Second, as compressibility is negligible in all of the modes examined, we replace the gas continuity equation with the incompressible condition, $\nabla\cdot\bm{v} = 0$. Then the linear perturbations satisfy $k_x\dd v_x + k_z\dd v_z = 0$. Hence for $k_z=0$ the gas has no radial motion, $\dd v_x=0$. Incompressibility also means letting the gas density perturbations $W\to 0$ but $C_s\to \infty$, while the enthalpy perturbation $\delta h \equiv C_s^2W$ remains finite and is determined from the gas' radial momentum equation. Furthermore, the linearized vertical momentum equations for gas and dust can be satisfied with $\delta v_z=\delta w_z =0$. { We include dust diffusion but ignore gas viscosity.} 

The linear eigenvalue problem now only involves the gas' azimuthal momentum equation and the dust's continuity and horizontal momentum equations. In the limit considered, these can be written as 
\begin{align}
\left(\sigma\taus + \mu_\mathrm{g}\right)\dd v_y &= \epsilon\left(w_y - v_y\right)Q + \epsilon\left(\dd w_y - \dd v_y\right),\label{analytic_vy}\\
    \left(\sigma\taus + \mu_\mathrm{d} + \widetilde{D}\right)Q &= -\ii k_x\taus \dd w_x,\\
    \left(\sigma\taus + \mu_\mathrm{d} \right)\dd w_x &= 2 \st \dd w_y - \dd w_x ,\\
    \left(\sigma\taus + \mu_\mathrm{d} \right)\dd w_y &= -\frac{\st}{2}\dd w_x + \dd v_y - \dd w_y\label{analytic_wy},
\end{align}
where 
\begin{align}
\mu_\mathrm{d} \equiv \ii k_x w_x \taus, \quad \mu_\mathrm{g} \equiv \ii k_x v_x \taus, \quad \widetilde{D} \equiv D k_x^2\taus.
\end{align}
These give the dispersion relation
\begin{align}
        \left(\sigma\taus  + \mu_\mathrm{d} + \widetilde{D} \right)\left(\sigma\taus + \mu_\mathrm{g} +\epsilon\right)\left[\left(\sigma\taus + \mu_\mathrm{d} + 1\right)^2 + \st^2\right] = \epsilon\left[\left(\sigma\taus + \mu_\mathrm{d} + \widetilde{D}\right)\left(\sigma\taus + \mu_\mathrm{d} + 1\right) - 2\ii K_x \st^2 \zeta_y\right],\label{analytic_dispersion}
\end{align}
and recall $\zeta_y$ is the dimensionless azimuthal drift (Eq. \ref{dimensionless_ydrift}). The full dispersion relation is a quartic in $\sigma$ and its direct solutions are unwieldy. 

To proceed, we assume modes are slow such that { $|\sigma\taus|
\ll \mathrm{min}(1,\epsilon)$}. We also specialize to the case where radial drift is negligible compared to the azimuthal drift, for example near a pressure extremum ($\etatilde \to 0$), and set $\mu_\mathrm{g} \simeq \mu_\mathrm{d}$. We assume $|\mu_\mathrm{d}| \ll \mathrm{min}(1,\epsilon)$, which can be satisfied for sufficiently small $\st$ at fixed $K_x$. 
The dispersion relation then { reduces} to 
\begin{align}
    \left(\sigma\taus + \mu_\mathrm{d} + \widetilde{D} \right)
\left[\left(\sigma\taus + \mu_\mathrm{d} \right) + \frac{\epsilon\st^2}{1+\epsilon}\right]  =  -\frac{2\ii\epsilon K_x\zeta_y\st^2}{1+\epsilon}.
              \label{analytic_quad}
\end{align}
This quadratic equation for $\sigma\taus + \mu_\mathrm{d}$, and hence $\sigma$, can be solved readily.

{\subsection{Explicit solutions without dust diffusion}
When dust diffusion is negligible ($D=0$),
} a further simplification can be made in the limit $\st^2\ll |\mu_\mathrm{d}|$, which means $K_x$ cannot be too small. { The term $\propto \st^2$ in Eq. \ref{analytic_quad} can then be neglected.} 
{ Seeking} growing solutions, we find 
\begin{align}
   \sigma\taus + \mu_\mathrm{d} \simeq \sqrt{\frac{\epsilon K_x \st^2 \zeta_y}{1+\epsilon}}\times \left(1 - \ii\right).\label{analytic_full}
\end{align}
The growth rate and oscillation frequency are then 
\begin{align}
    &\frac{s}{\Omega} = \sqrt{\frac{\epsilon K_x\zeta_y}{1+\epsilon}},\label{analytic_growth}\\
    &\frac{\omega}{\Omega} = K_x \frac{w_x}{C_s} +  \sqrt{\frac{\epsilon K_x\zeta_y}{1+\epsilon}}.\label{analytic_osc}
\end{align}
(Recall that we defined $\sigma \equiv s -\ii\omega$.)  
We see from Eq. \ref{analytic_growth} that instability is directly powered by azimuthal drift. Furthermore, for azimuthal drifts driven entirely by the magnetic torque, we have $\zeta_y\propto \st$ (Eq. \ref{dimensionless_ydrift} with $\etatilde =0$), thus $s\propto \sqrt{\st}$. Note that both $s$ and $\omega$ are unbounded with increasing $K_x$, but this limit eventually violates the assumption of small $|\mu_\mathrm{d}|$ and $|\mu_\mathrm{g}|$ used to derive Eqs. \ref{analytic_growth}--\ref{analytic_osc}. In any case, the fluid approximation probably breaks down when $|\sigma\taus|\gtrsim 1$ \citep{jacquet11}.  

\begin{figure*}
    \centering
    \includegraphics[width=\linewidth]{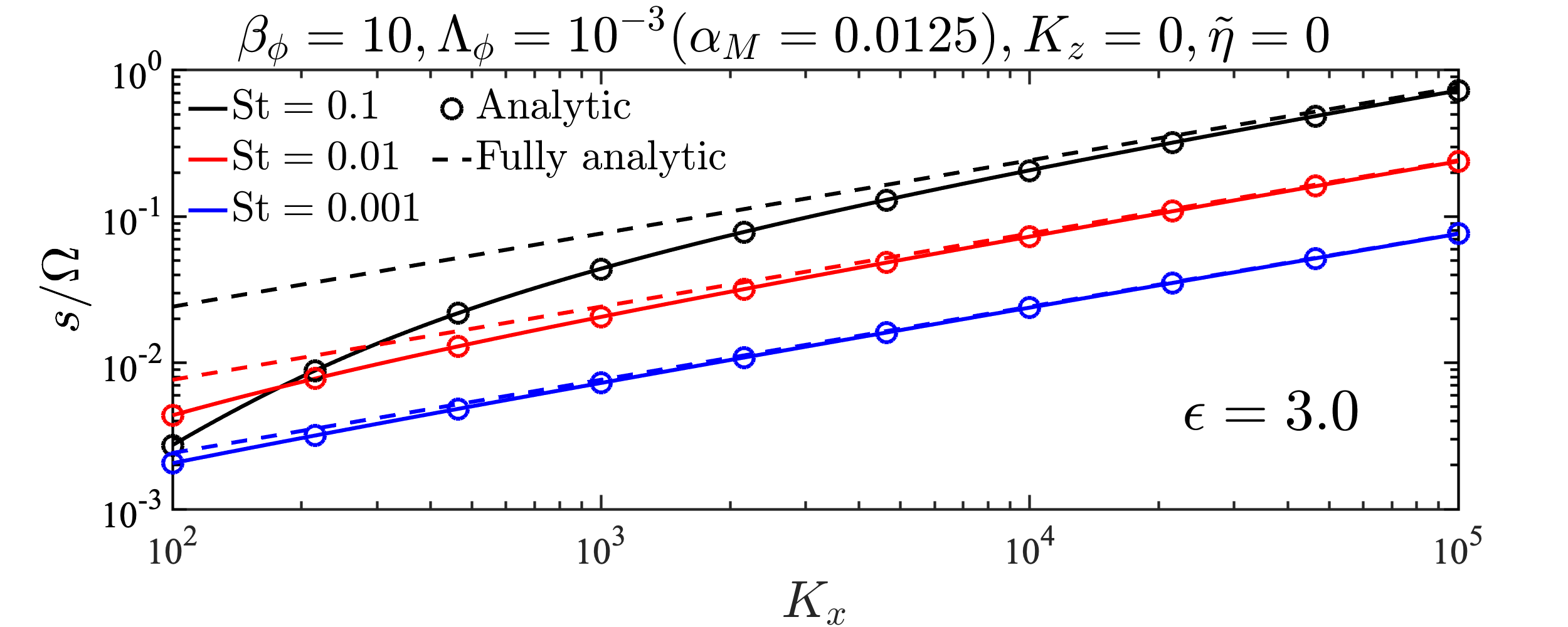}
    \caption{Streaming instability in disks without pressure gradients ($\etatilde = 0$) but including a magnetically-induced accretion flow with a dimensionless Maxwell stress of $\alpha_M = 0.0125$. { No viscosity or diffusion is used ($\nu=D=0$).} Solid curves are computed from the full eigenvalue problem, the circles from the analytic model (Eq. \ref{analytic_quad}), and the dashed curve is the fully analytic solution (Eq. \ref{analytic_growth}). { The dust-to-gas ratio is $\epsilon=3$.} 
    Three Stokes numbers are shown: $\st=0.1$ (black), $\st=10^{-2}$ (red), and $\st=10^{-3}$ (blue). Modes have no vertical structure, $K_z=0$. 
    }
    \label{analytic_test}
\end{figure*}

In Fig. \ref{analytic_test} we test the above model by comparing it with the solution to the full eigenvalue problem. We compute unstable modes with $K_z=0$ in disks with $\etatilde=0$, $\beta_\phi=10$, and $\Lambda_\phi=10^{-3}$ (or $\alpha_M=0.0125$), for a range of stopping times. The analytic model (Eq. \ref{analytic_quad}) accurately reproduces the growth rates, while the fully analytic solutions (Eq. \ref{analytic_growth} works best for small $\st$, large $K_x$, or both. 

{\subsection{Physical interpretation}}

The origin of the instability can be examined through the linearized equations (\ref{analytic_vy}--\ref{analytic_wy}). Under the same approximations as that used to derive Eq. \ref{analytic_full}, { without diffusion} we have 
\begin{align}
    &\delta w_x = 2\st\delta w_y, \label{analytic_wx}\\
    &Q = -\frac{\ii k_x\taus}{\sigma\taus + \mu_\mathrm{d}}\delta w_x.\label{analytic_Q}
\end{align}
The gas and dust azimuthal equations can be combined to give 
\begin{align*}
    (\sigma\taus + \mu_\mathrm{d})\left(\epsilon\dd w_y + \dd v_y \right) = \epsilon \left(w_y - v_y\right) Q - \frac{\epsilon\st}{2}\dd w_x,
\end{align*}
where we have approximated $\mu_\mathrm{g}\simeq \mu_\mathrm{d}$. From Eq. \ref{analytic_wx} it is clear that if $\st^2\ll |\mu_\mathrm{d}|$, then the last term on the RHS can be neglected. Finally, for tightly coupled dust we expect $\delta v_y\simeq \delta w_y$. Hence,
\begin{align}
     (\sigma\taus + \mu_\mathrm{d})\left(1 + \epsilon\right)\delta w_y = \epsilon \left(w_y - v_y\right) Q. \label{analytic_com_vy_reduced}
\end{align}
One can readily check that Eqs. \ref{analytic_wx}, \ref{analytic_Q}, and \ref{analytic_com_vy_reduced} yields the explicit dispersion relation Eq. \ref{analytic_full}. 

The instability mechanism can now be read off Eqs. \ref{analytic_wx}, \ref{analytic_Q}, and \ref{analytic_com_vy_reduced}. Suppose the dust experiences an azimuthal acceleration, moves outward and is slowed down by gas drag (Eq. \ref{analytic_wx}). Then the local dust density increases (Eq. \ref{analytic_Q}), which can be seen by inserting Eq. \ref{analytic_full} to give 
$Q \propto (1-\ii)\delta w_x$ with a positive proportionality constant. For $\zeta_y>0$ this increases the feedback onto the gas that accelerates it in the azimuthal direction as $\delta w_y\propto (1+\ii)Q$, but for tightly coupled dust the latter is  dragged along (Eq. \ref{analytic_com_vy_reduced}). This leads to a positive feedback and hence instability.

{
\subsection{Effect of dust diffusion}
We briefly examine the influence of dust diffusion. Fig. \ref{analytic_test_diff} show growth rates for $\alphass =10^{-9}$, $10^{-8}$, and $10^{-7}$. As expected, diffusion reduces growth rates and sets a minimum scale of instability, which increases with $\alphass$. The analytic model (circles) reproduces results from the full eigenvalue problem with slight deviations at large $K_x$. This is not surprising since the model assumes sufficiently small $\left|\mu_\mathrm{d}\right|\propto K_x$. 

\begin{figure*}
    \centering
    \includegraphics[width=\linewidth]{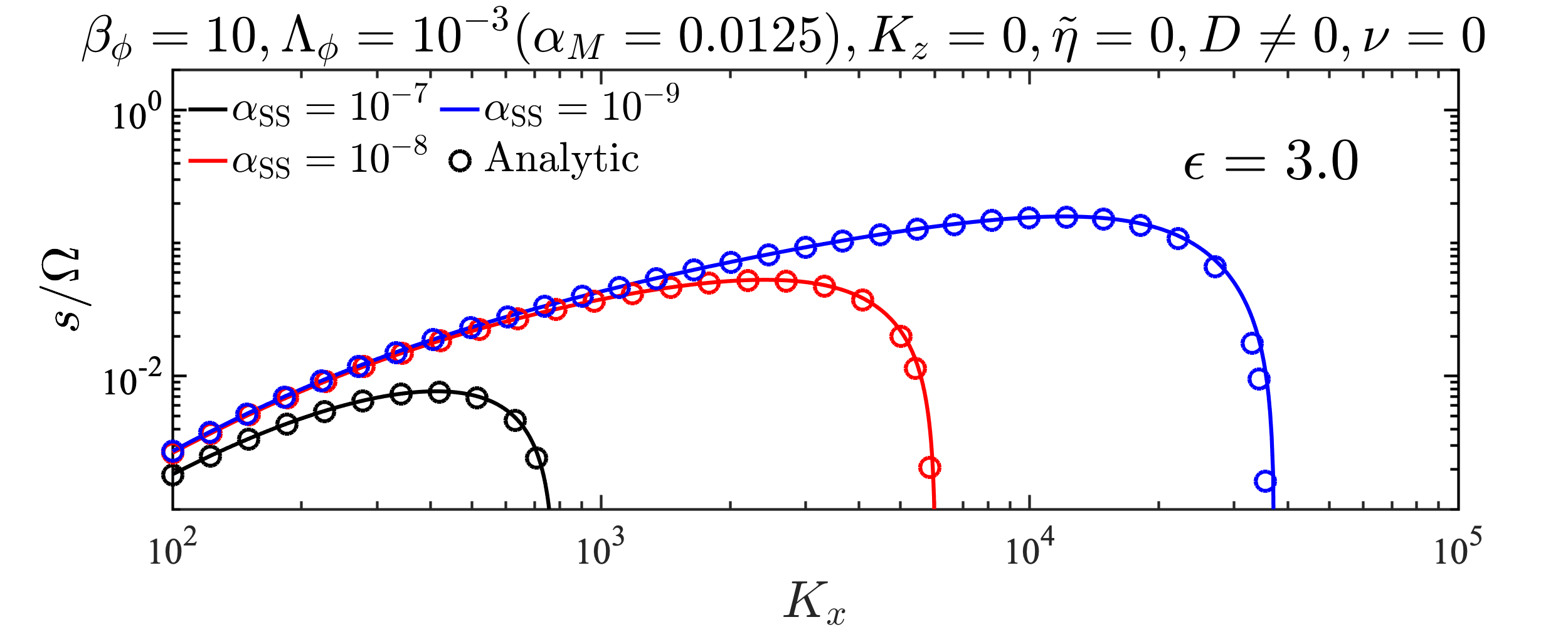}
    \caption{{ Similar to Fig. \protect\ref{analytic_test} but for fixed to $\st=0.1$ and three levels of dust diffusion: $\alphass=10^{-9}$ (blue), $10^{-8}$ (red), and $10^{-7}$ (black). No gas viscosity is included. 
    Solid curves are computed from the full eigenvalue problem and the circles from the analytic model (Eq. \ref{analytic_quad}).} 
    }
    \label{analytic_test_diff}
\end{figure*}
}

{
\section{Single fluid model of a magnetized, dusty gas}\label{mag_onefluid}

Our numerical results show the disappearance of classic SI modes with large $k_z$ as one increases $\Lambda_z$,  even when $\Lambda_z\ll1$ (\S\ref{caseII_resis}, e.g. the bottom left two panels in Fig. \ref{Example_CaseII_resistive}). This is curious because, for large $k_z$, Ohmic diffusion should render magnetic fields ineffective. To check this result, we analyze a one-fluid model of a dusty, magnetized gas following the framework developed by \cite{laibe14}. We assume the gas is incompressible ($\nabla\cdot\bm{v}=0$) and $\rhog$ is constant. We neglect gas viscosity and dust diffusion.  


In the single fluid approach, one works with the total density 
\begin{align}
    \rho \equiv \rhog + \rhod,
\end{align}
and the center of mass velocity 
\begin{align}
    \bm{u} = \fgas\bm{v} + \fdust\bm{w},
\end{align}
where $\fgas = 1/(1+\epsilon)$ is the gas fraction and $\fdust=\epsilon/(1+\epsilon)$ is the dust fraction. We define the dust-gas drift as $\Delta\bm{u}\equiv \bm{w} - \bm{v}$. Then $\bm{v} = \bm{u} - \fdust\Delta\bm{u}$ and $\bm{w} = \bm{u} + \fgas\Delta\bm{u}$. 

For sufficiently small grains, dust follows the gas with a correction from the differential force between dust and gas. In our case, this arises from pressure gradients and Lorentz forces that only acts on the gas. As a result, dust grains drift relative to the gas with
\begin{align}
\Delta\bm{u} = \taus\left[\frac{\nabla P}{\rho} - 2 \eta R \Omega^2 \fgas \hat{\bm{x}}    
- \frac{1}{\mu_0\rho}\left(\nabla\times\bm{B}\right)\times\bm{B}
\right]\label{terminal_velocity}
\end{align}
\citep{johansen05,fromang06}. For clarity, we have dropped the subscript zero to indicate evaluation of $\eta$, $R$, and $\Omega$ at the reference radius $R_0$.

Eq. \ref{terminal_velocity} is also known as the terminal velocity approximation  \citep[TVA,][]{youdin05,lovascio19}. While the TVA simplifies the modelling of dusty gas considerably, its shortcomings were recently analyzed by \cite{pan21}, where it was shown that the TVA underestimates linear growth rates when $\epsilon\ll 1$, $k_z\gg k_x$, or both, and is particularly significant at resonant wavenumbers (i.e. those satisfying Eq. \ref{rdi_si_condition}) -- regimes that we will consider below. The remedy to the TVA's deficiency involve additional contributions to $\Delta\bm{u}$, see \cite{pan21} for details. We leave this to future work and proceed with the caution that the following discussion is aimed at capturing  qualitative effects, rather than producing a quantitative replacement of the two-fluid treatment. 

The gas incompressibility condition, dust continuity equation, and the center-of-mass momentum equation for the dust-gas mixture are:  
\begin{align}
    &\nabla\cdot\bm{u} = \nabla\cdot\left(\fdust\Delta\bm{u}\right),\\
    &\frac{\p\epsilon}{\p t} + \nabla\cdot\left(\epsilon\bm{u}\right) = -\nabla\cdot\bm{u},\\
    &\frac{\p\bm{u}}{\p t} + \bm{u}\cdot\nabla\bm{u} = -\frac{1}{\rho}\nabla P + 2 \eta r \Omega^2 \fgas \hat{\bm{x}} + 2\Omega u_y \hat{\bm{x}} - \frac{\Omega}{2}u_x\hat{\bm{y}} + \frac{1}{\mu_0\rho}\left(\nabla\times\bm{B}\right)\times\bm{B}.
\end{align}
See also \citet[][Appendix B]{lin21} for the case of a compressible, unmagnetized gas. Strictly speaking, one should express the induction equation in terms of $\bm{u}$ and $\Delta\bm{u}$. We neglect this complication and set $\bm{v}\to\bm{u}$ in Eq. \ref{Faraday_local}, so the induction equation retains the form as in the two-fluid model. See \cite{fromang06} for a similar treatment. This approximation is applicable when $\fdust\ll 1$, but to connect to hydrodynamic results, we shall not take this formal limit until later. The equilibrium disk consists of $u_x = u_z= 0$, $u_y = -\eta r \Omega^2\fgas$ and constant $P$, $\rho$, $\epsilon$. The initial magnetic field $\bm{B} = B_z \hat{\bm{z}}$  with a constant $B_z$. We ignore passive magnetic torques, so this setup corresponds to Case II discussed in the main text. 

\subsection{Dispersion relation}

The linearized one-fluid equations are
\begin{align}
    &\nabla\cdot\dd\bm{u} = - \ii k_x \taus \mathcal{F} \frac{(1-\epsilon)}{(1+\epsilon)^2}\delta\epsilon - \fdust \taus k^2\frac{\dd P}{\rho} - \fdust\taus k^2 \frac{B_z}{\mu_0\rho}\dd B_z,\label{mag_onefluid_lin_mass}\\
    &\sigma \delta \epsilon = -(1+\epsilon)\nabla\cdot\dd\bm{u},\\
    &\sigma \dd u_x  = -\ii k_x \frac{\dd P}{\rho} -\mathcal{F}\frac{\dd\epsilon}{1+\epsilon} + 2\Omega\dd u_y
    -\frac{B_z}{\mu_0\rho}\left(\ii k_x\delta B_z - \ii k_z \dd B_x\right),\\
    &\sigma\dd u_y = -\frac{\Omega}{2}\dd u_x + \ii k_z \frac{B_z}{\mu_0\rho}\delta B_y,\\
    &\sigma\dd u_z = -\ii k_z\frac{\delta P}{\rho},\\
    &\widetilde{\sigma} \delta B_x = \ii k_z B_z \dd u_x,\\
    &\widetilde{\sigma} \delta B_y = \ii k_z B_z \dd u_y - \frac{3}{2}\Omega\dd B_x,\\
     &\widetilde{\sigma} \delta B_z =-\ii k_x B_z \dd u_x,\\
\end{align}
where $\mathcal{F} = 2\eta R \Omega^2\fgas$ and $\widetilde{\sigma} = \sigma + \eta_O k^2$. We have also used $\nabla\cdot\dd\bm{B} = 0$ in Eq. \ref{mag_onefluid_lin_mass}. The last two terms in Eq. \ref{mag_onefluid_lin_mass} reflect the tendency for dust to concentrate at maxima in the \emph{total} pressure. These equations yield the dispersion relation 
\begin{align}
    &\widetilde{\sigma}^2\left[
    \underbrace{\fdust\taus \sigma^4}_\text{neglect} + \sigma^3 + \taus\left(\fdust\Omega^2 - \ii k_x \mathcal{F}\fgas\right)\sigma^2 + \frac{k_z^2}{k^2}\Omega^2 \sigma - \ii k_x \taus \mathcal{F}\fgas\Omega^2\left(1-\epsilon\right)\frac{k_z^2}{k^2}
    \right]\notag\\  
    &+k_z^2C_{\mathrm{A}z\mathrm{,eff}}^2\left\{
    \left(2\sigma\widetilde{\sigma}+k_z^2C_{\mathrm{A}z\mathrm{,eff}}^2\right)
    \left(\fdust\taus\sigma^2 + \sigma -\ii k_x  \taus  \mathcal{F}\fgas \right)
    -3\Omega^2\left[\fdust\taus\sigma^2 + \sigma\frac{k_z^2}{k^2} - \ii k_x  \taus \mathcal{F}\fgas\left(1-\epsilon\right)\frac{k_z^2}{k^2}\right]\right\}=0.\label{magSI_onefluid_disp}
\end{align}
Recall the effective Alfv\'{e}n speed $C_{Az\mathrm{,eff}} = C_A/\sqrt{1+\epsilon}$. When $B_z\to 0$, only the first term survives, which is equivalent to the dispersion relation for the SI as derived by \citet[][their Eq. 97]{lin17}, see also \cite{jacquet11}. \citeauthor{lin17} argued that the term  $\fdust\taus\sigma^4$ should be neglected to avoid spuriously growing epicycles, although recently \cite{jaupart20} and \cite{pan21} showed that epicycles can indeed be unstable. Nevertheless, we neglect this term to focus on classic SI modes. When $\taus\to 0$, Eq.  \ref{magSI_onefluid_disp} is equivalent to the dispersion relation for the MRI in resistive, gaseous disks \citep[][their Eq. 22]{sano99} with the Alfv\'{e}n speed reduced by perfectly-coupled dust.

We now make further simplifications. We assume $B_z\neq 0$, large resistivity ($\Lambda_z\ll 1$), and consider modes with $k_z\gg k_x$. Then $k_z/k\simeq 1$ and  $\widetilde{\sigma}\simeq \eta_Ok_z^2$. Note that $\eta_O$ is parameterized through $\Lambda_z$ and $C_{Az}$ in the pure gas limit (Eq. \ref{def_elsa}). In terms of the dimensionless frequency $\hat{\sigma} = \sigma/\Omega$, Eq. \ref{magSI_onefluid_disp} reduces to  
\begin{align}
&\hat{\sigma}^3 + \left(\fdust\st - 2\ii K_x \st \etatilde \fgas^2\right)\hat{\sigma}^2 + \hat{\sigma} -2\ii K_x \st \etatilde \fgas^2\left(1-\epsilon\right) + \fgas\Lambda_z\left(2\hat{\sigma}+\fgas\Lambda_z\right)\left(\fdust\st\hat{\sigma}^2 + \hat{\sigma} - 2\ii K_x \st \etatilde \fgas^2\right)\notag\\
&-\frac{3\beta_z}{K_z^2}\fgas\Lambda_z^2\left[
\fdust\st\hat{\sigma}^2 + \hat{\sigma} - 2\ii K_x \st \etatilde \fgas^2\left(1-\epsilon\right)
\right]=0.\label{magSI_onefluid_disp_simple}
\end{align}

In Fig. \ref{magSI_onefluid_compare} we compare SI growth rates obtained from Eq. \ref{magSI_onefluid_disp_simple} to that from the full treatment. We take $\beta_z=10^4$, $\etatilde=0.05$, $\st=0.1$, and consider $\epsilon=[0.01$,\, $0.2$,\, $3]$ with increasing $\Lambda_z$ for larger $\epsilon$ as the impact of magnetic fields is smaller in dustier disks. 
For $\epsilon < 1$ and $\Lambda_z=0$, the persistence of instability at $K_x\gtrsim 200$--$300$ in the analytic model is an artifact of the TVA \citep{pan21}. Introducing $\Lambda_z>0$ leads to a reduction in growth rates  but is over-estimated by the analytic model. The analytic model does improve with smaller $\epsilon$: for $\epsilon=0.01$ the cut-off at $K_x\simeq 90$ is similar to the full model ($K_x\simeq 100$). 

For completeness, we present a dust-rich case in Fig. \ref{magSI_onefluid_compare} with $\epsilon=3$, which is actually inconsistent with taking the center-of-mass velocity in the induction equation. Despite this, the analytic model performs well for $K_x\gtrsim 300$, including the slight drop in growth rates as $\Lambda_z$ increases. In fact, for $\Lambda_z=1$, analytic growth rates are similar to the full model for all $K_x$ values considered. The full model shows that growth rates for $K_x\lesssim 50$ decrease with $\Lambda_z$; while growth rate increase for $50 \lesssim K_x\lesssim 300$. On the other hand, the analytic model always predicts an increase for $K_x\lesssim 300$. Resolving these discrepancies probably requires accounting for dust-gas drift in the induction equation.

\begin{figure*}
    \centering
    \includegraphics[width=\linewidth]{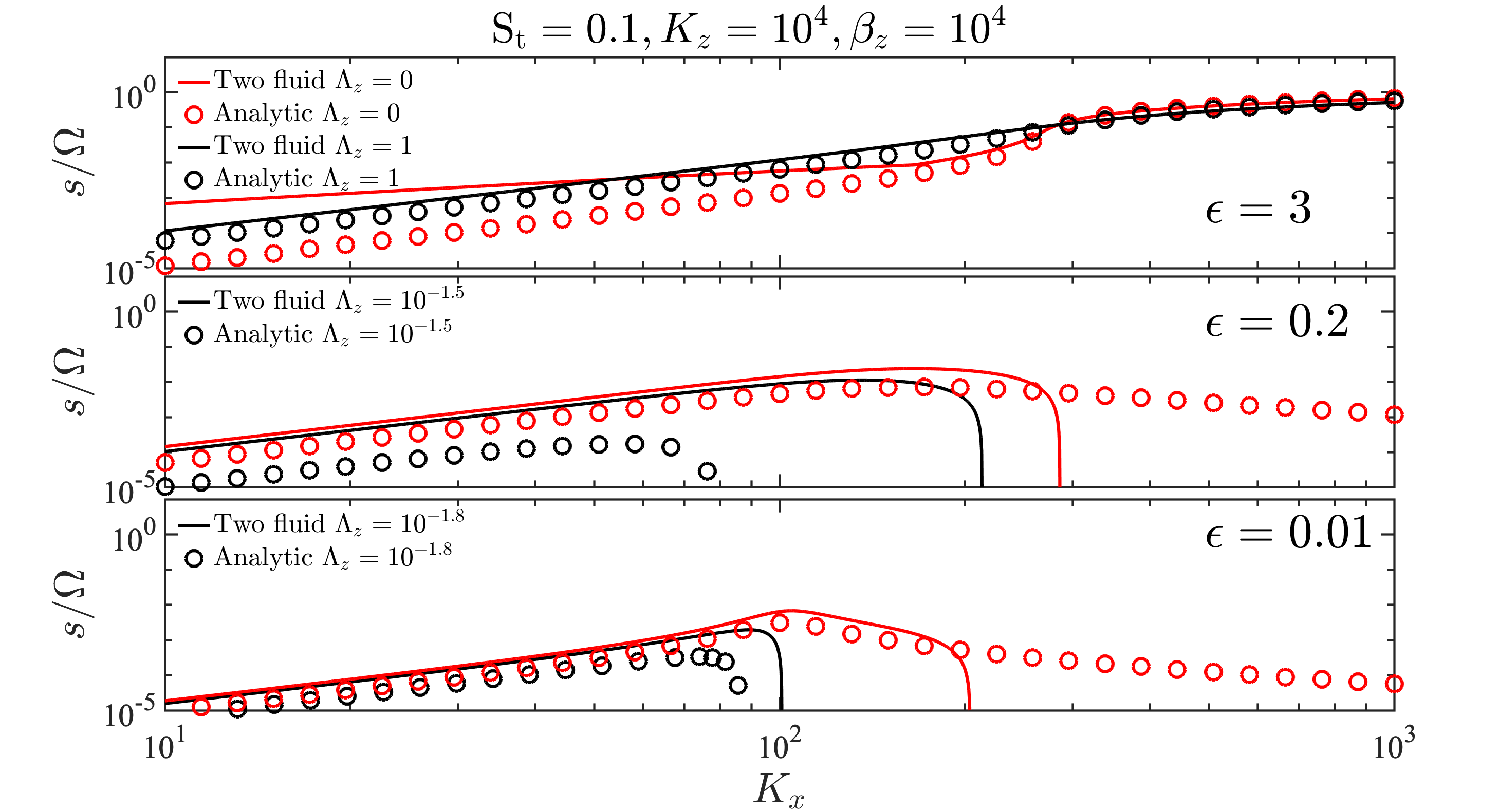}
    \caption{{ SI growth rates in effectively unmagnetized disks ($\Lambda_z=0$, red) compared to magnetized disks ($\Lambda_z>0$, black). The disk parameters are $\beta_z=10^4$, $\etatilde=0.05$, and $\st=0.1$. Modes have fixed $K_z=10^4$. Solid lines are obtained from the full equations while circles are obtained from the reduced dispersion relation Eq. \ref{magSI_onefluid_disp_simple}. 
    Dust-to-gas ratios from top to bottom: $\epsilon=3$, $0.2$, and $0.01$. }
    }
    \label{magSI_onefluid_compare}
\end{figure*}

\subsection{Magnetic stabilization of the classic SI}

We can evaluate $\left.\frac{\p\hat{\sigma}}{\p\Lambda_z}\right|_{\Lambda_z=0}$ to examine the effect of introducing magnetic perturbations to the SI. Note that for large resistivity and $K_z$ we do not expect complications from the MRI as it should remain suppressed. Denoting the solution in the limit of $\Lambda_z\to 0$ as $\hat{\sigma}_*$, we have  
\begin{align}
     \left[3\hat{\sigma}_*^2 + 2\hat{\sigma}_*\left(\fdust\st - 2\ii K_x \st\etatilde \fgas^2\right) +   1\right]\left.\frac{\p\hat{\sigma}}{\p\Lambda_z}\right|_{\Lambda_z=0} 
     + 2\fgas\hat{\sigma}_*\left(\fdust\st\hat{\sigma}_*^2 + \hat{\sigma}_* - 2\ii K_x \st\etatilde \fgas^2 \right) = 0.
    \label{magsi_analytic_resonant}
\end{align}
We make use of series solutions for $\hat{\sigma}_*$ (in $\st$) as developed by \cite{youdin05,jacquet11,jaupart20}. In terms of dimensionless variables and for $K_z\gg K_x$, Eqs. 28--29 of \cite{jacquet11} give 
\begin{align}
    \hat{\sigma}_* = \left(2K_x\st\etatilde\fgas\right)\left(\fdust - \fgas\right)\left[\left(2K_x\st\etatilde\fgas\right)\left(\fdust - \fgas\right)\fdust\st-\ii\right].
\end{align}

We finally set $\fdust\ll 1$ so $\fgas\simeq 1$ and specialize to radial wavenumbers such that $K_x^2\zeta_x^2 = 1$. Recall from Eq. \ref{dimensionless_xdrift} that $\zeta_x$ is the dimensionless radial drift. According to RDI theory, at this $K_x$ the resonant $K_z\to\infty$,  see Eq. \ref{rdi_si_condition}. These regimes are also where magnetic stabilization is most apparent in Fig. \ref{Example_CaseII_resistive}. Then $2\st\etatilde K_x = 1$ and $\hat{\sigma}_* = \fdust \st + \ii$. Inserting these into Eq.  \ref{magsi_analytic_resonant}, 
we find the growth rate  $\hat{s}=\real{\hat{\sigma}}$ decreases with increasing $\Lambda_z$,
\begin{align} 
    \left.\frac{\p\hat{s}}{\p\Lambda_z}\right|_{\Lambda_z=0} = - \frac{2}{3}\left(\fdust\st\right)^2 < 0,  
\end{align}
where we have used $\fdust\st\ll 1$. Both finite drag and feedback are necessary for magnetic stabilization ($\fdust\st\neq 0$) in the limits considered.  
}

\section{Pseudo-energy decomposition}\label{pseudo_energy}

Following \cite{ishitsu09} and \cite{lin21} we construct a pseudo-kinetic energy associated with a linear mode from Eqs. \ref{lin_gas_mom_x}--\ref{lin_gas_mom_z} and Eqs. \ref{lin_dust_mom_x}--\ref{lin_dust_mom_z}:  
\begin{align} 
& U_{\rm{tot}} \equiv  \epsilon (|\delta w_x|^2 + 4 |\delta w_y|^2 + |\delta w_z|^2) + (|\delta v_x|^2 + 4 |\delta v_y|^2 + |\delta v_z|^2) =  \sum_{i=1}^{3} U_i  , \label{pseudo_tot}
\end{align}
where 
\begin{align} 
& s U_1 = k_x C_s^2 {\rm Im} (W\delta v_{x}^*) +  k_z C_s^2 {\rm Im} \it(W\delta v_{z}^*), \label{pseudo_thermal}\\ 
& s U_2 = - \frac{\epsilon}{\taus}\left\{(v_x-w_x) {\rm Re} \left[(Q-W) \delta v_x^*\right] + 4 (v_y-w_y) {\rm Re}  \left[(Q-W) \delta v_y^*\right] + |\delta v_x - \delta w_x|^2 + 4 |\delta v_y - \delta w_y|^2 + |\delta v_z - \delta w_z|^2\right\}, \label{pseudo_drag} \\
& s U_3 = -k_z C_{Az} {\rm Im} (\delta b_x \delta v_{x}^*) + k_x C_{Az} {\rm Im} (\delta b_z \delta v_{x}^*) - 4 k_z C_{Az} {\rm Im} (\delta b_y \delta v_{y}^*). \label{pseudo_mag}
\end{align}
are contributions from pressure forces, dust-gas drag, and magnetic forces, respectively. We further decompose $U_2$: 
\begin{align}
    &U_2 = U_{2x} + U_{2y} +  U_{2\delta},\\
    &U_{2x} = - \frac{\epsilon}{s\taus}(v_x-w_x) {\rm Re} \left[(Q-W) \delta v_x^*\right],\\
    &U_{2y} = -\frac{4\epsilon}{s\taus}(v_y-w_y) {\rm Re}\left[(Q-W) \delta v_y^*\right],\\
    &U_{2\delta} = -\frac{\epsilon}{s\taus}\left(|\delta v_x - \delta w_x|^2 + 4 |\delta v_y - \delta w_y|^2 + |\delta v_z - \delta w_z|^2\right).
\end{align}
We associate $U_{2x}$ and $U_{2y}$ with the radial and azimuthal drifts in the background, and $U_{2\dd}$ with the relative drift in the perturbed velocities. Note that $U_{2\dd}$ is always stabilizing.

\section{Code tests}\label{nonlinear_code_test}
\begin{figure*}
    \centering
    \includegraphics[width=\linewidth]{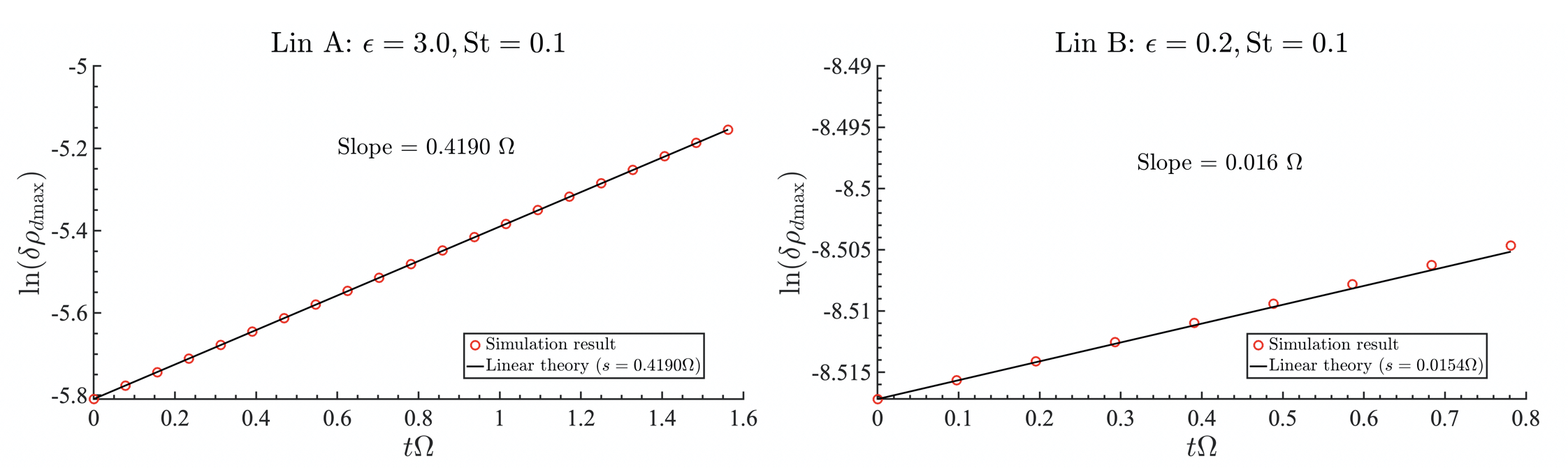}
    \caption{Linear growth of the LinA (left) and LinB (right) SI eigenmodes in an unmagnetized, dusty disk.}
    \label{nonlinear_code_test_SI}
\end{figure*}

To test our finite difference code used in \S\ref{nonlinear}, we first reproduce the classic SI eigenmodes LinA and LinB as described in \cite{youdin07} and summarized in Table \ref{table1}. For this test the magnetic field is switched off. Fig. \ref{nonlinear_code_test_SI} shows the evolution of the maximum dust density perturbation in the domain, which shows a good agreement between the growth measured in the simulation and the expected growth rate calculated from linear theory. We remark that for LinB the measured growth rate in $\delta\rho_d$ is slightly higher than the theoretical value, but for $\delta w_z$ we do measure the same growth rate of $s=0.0154\Omega$ as in linear theory. 

\begin{figure*}
    \centering
    \includegraphics[width=\linewidth]{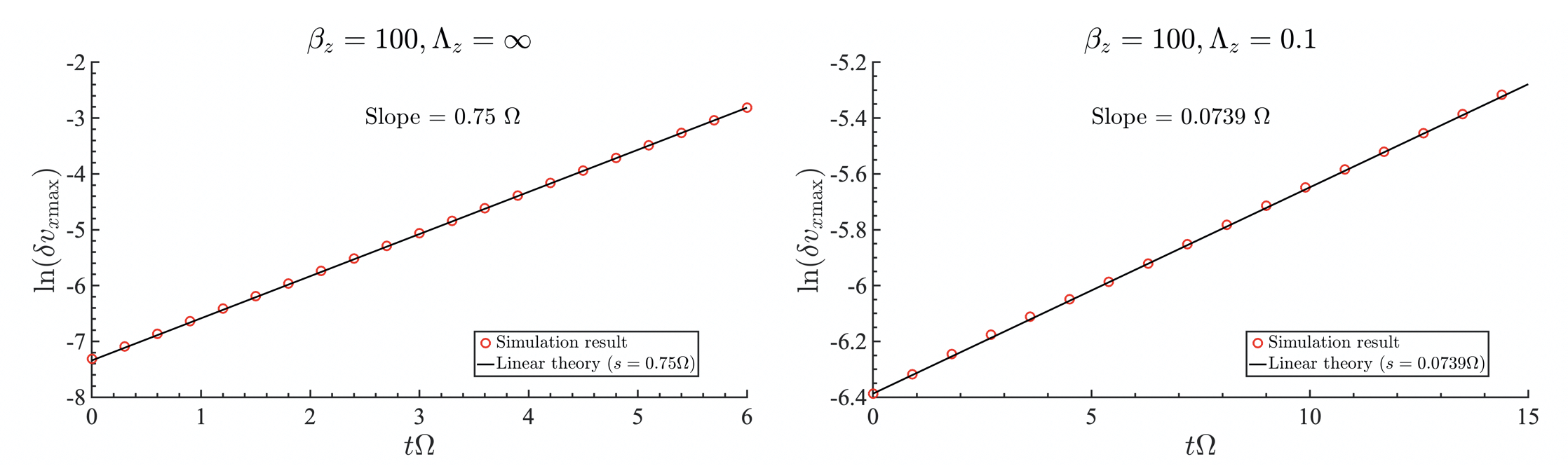}
    \caption{Linear growth of the MRI in the ideal (left) and resistive (right) regimes without dust.}
    \label{nonlinear_code_test_MRI}
\end{figure*}

Similarly, we reproduce the standard MRI with a purely vertical field in Fig. \ref{nonlinear_code_test_MRI}. For this test we disable the dust component and set $\etatilde=0$. We fix $\beta_z=100$ , $K_x=0$, and choose the most unstable $K_z=5\sqrt{15}/2$ ($K_z=\sqrt{3}/2$) in the ideal (resistive cases) according to linear theory. For ideal MHD we obtain a linear growth rate of $0.75\Omega$ as expected. For the non-ideal case, we set  $\Lambda_z = 0.1$ and obtain a growth rate of $0.074\Omega$, which is close to the analytic value of $0.75\Lambda_z\Omega$ in the limit $\Lambda_z\ll 1$. \citep{sano99}.

\bibliographystyle{aasjournal}
\bibliography{ref}

\end{document}